\documentclass[%
 reprint,
 amsmath,amssymb,
 aps,
 prx,longbibliography,
 superscriptaddress
]{revtex4-2}

\usepackage{graphicx,xcolor}
\usepackage{dcolumn}
\usepackage{bm}
\usepackage[percent]{overpic}
\usepackage{physics}
\usepackage{mathtools}
\usepackage{hyperref}
\usepackage{booktabs}
\usepackage{longtable}
\usepackage[pdftex]{pict2e}
\usepackage{upgreek}
\usepackage{braket}
\usepackage{orcidlink}
\usepackage[normalem]{ulem}

\hypersetup{colorlinks=true,%
            urlcolor=blue,%
            citecolor=blue,%
            linkcolor=blue}

\newcommand{\DUKE}{Duke Quantum Center, Departments of Physics and Electrical and Computer Engineering, Duke University, Durham, North Carolina 27708, USA}
\newcommand{\CORNELL}{School of Applied and Engineering Physics, Cornell University, Ithaca, NY 14853.}
\newcommand{\JQI}{Joint Quantum Institute, NIST/University of Maryland, College Park, Maryland 20742}
\newcommand{\QUICS}{Joint Center for Quantum Information and Computer Science, 
\\
NIST/University of Maryland, College Park, Maryland 20742}

\begin{document}
\title{
String-Breaking Dynamics in Quantum Adiabatic and Diabatic Processes
}
\author{Federica~Maria~Surace~\orcidlink{0000-0002-1545-5230}}
\affiliation{Department of Physics and Institute for Quantum Information and Matter,
California Institute of Technology, Pasadena, California 91125, USA}
\author{Alessio~Lerose~\orcidlink{0000-0003-1555-5327}}
\affiliation{Rudolf Peierls Centre for Theoretical Physics, Clarendon Laboratory, Oxford OX1 3PU, United Kingdom}
\affiliation{Institute for Theoretical Physics, KU Leuven, Celestijnenlaan 200D, 3001 Leuven, Belgium}

\author{Or~Katz~\orcidlink{0000-0001-7634-1993}}
\affiliation{\DUKE}
\address{\CORNELL}

\author{Elizabeth~R.~Bennewitz
}
\affiliation{\QUICS}
\affiliation{\JQI}
\author{Alexander~Schuckert~\orcidlink{0000-0002-9969-7391}
}
\affiliation{\QUICS}
\affiliation{\JQI}
\author{De~Luo~\orcidlink{0000-0002-4842-0990}}
\affiliation{\DUKE}
\author{Arinjoy~De}
\affiliation{\DUKE}
\affiliation{\JQI}
\author{Brayden~Ware}
\affiliation{\QUICS}
\affiliation{\JQI}
\author{William~Morong}
\thanks{Current address: AWS Center for Quantum Computing, Pasadena, California 91125, USA. Work done prior to joining AWS.}
\affiliation{\QUICS}
\affiliation{\JQI}
\author{Kate~Collins}
\affiliation{\QUICS}
\affiliation{\JQI}
\author{Christopher~Monroe}
\affiliation{\DUKE}
\author{Zohreh~Davoudi~\orcidlink{0000-0002-7288-2810}}
\affiliation{\QUICS}
\affiliation{Department of Physics and Maryland Center for Fundamental Physics, University of Maryland, College Park, Maryland 20742}
\author{Alexey~V.~Gorshkov}
\affiliation{\QUICS}
\affiliation{\JQI}

\date{\today}

%%%%%%%%%%%%%%%%%%%%%%%%%%%%%%%%%%%%%%%%%%%%%%%%%%%%%%%%%%%%%%%%%%%%%%%%%%%%%%%%%%%%%%%%%%%%%%%%%%%%%%%%%%%%%%%%%%%%%%%%%%%%%%%%%%%%%%%%%%%%%%%%%%%%%%%%%%%%%%%%%%%%%%%%%%%%%%%%%%%%%%%%%%%%%%%%%%%%%%%%%%%%%%%%
\begin{abstract}
Confinement prohibits isolation of color charges, e.g., quarks, in nature via a process called \emph{string breaking}: the separation of two charges results in an increase in the energy of a color flux, 
visualized as a string, connecting those charges. Eventually, creating additional 
charges is energetically favored, hence breaking the string. Such a phenomenon can be probed in simpler models, including quantum spin chains, enabling enhanced understanding of string-breaking dynamics. 
A challenging task is to understand how string breaking occurs as time elapses, in an out-of-equilibrium setting.
This work establishes the phenomenology of dynamical string breaking induced by a gradual increase of string tension over time. It, thus, goes beyond instantaneous quench processes and enables tracking the real-time evolution of strings in a more controlled setting. We focus on domain-wall confinement in a family of quantum Ising chains. 
Our results indicate that, for sufficiently short strings and slow evolution, 
string breaking can be described by the transition dynamics of a two-state quantum system akin to a Landau-Zener process. For longer strings, a more intricate spatiotemporal pattern emerges: the string breaks by forming a superposition of bubbles (domains of flipped spins of varying sizes), which involve highly excited states. We finally demonstrate that string breaking driven only by quantum fluctuations can be realized in the presence of sufficiently long-ranged interactions. This work holds immediate relevance for 
studying string breaking 
in quantum-simulation experiments.
\end{abstract}
%%%%%%%%%%%%%%%%%%%%%%%%%%%%%%%%%%%%%%%%%%%%%%%%%%%%%%%%%%%%%%%%%%%%%%%%%%%%%%%%%%%%%%%%%%%%%%%%%%%%%%%%%%%%%%%%%%%%%%%%%%%%%%%%%%%%%%%%%%%%%%%%%%%%%%%%%%%%%%%%%%%%%%%%%%%%%%%%%%%%%%%%%%%%%%%%%%%%%%%%%%%%%%%%

\maketitle

%%%%%%%%%%%%%%%%%%%%%%%%%%%%%%%%%%%%%%%%%%%%%%%%%%%%%%%%%%%%%%%%%%%%%%%%%%%%%%%%%%%%%%%%%%%%%%%%%%%%%%%%%%%%%%%%%%%%%%%%%%%%%%%%%%%%%%%%%%%%%%%%%%%%%%%%%%%%%%%%%%%%%%%%%%%%%%%%%%%%%%%%%%%%%%%%%%%%%%%%%%%%%%%%
\section{Introduction}
Unlike electrons that can be detached from atoms, isolated quarks have never been observed in nature. The absence of isolated quarks stems from the confining nature of quantum chromodynamics (QCD), the theory of the strong force~\cite{Greensite:2011zz}. In the case of static charges, the energy stored in a color flux tube connecting those charges, namely a string, grows linearly with the string length. At a critical distance, the energy threshold is passed for the production of new charges, which screen the original static charges, in a process known as string breaking~\cite{NAMBU1979372}.
While QCD studies of confinement and string breaking have been carried out  
using the nonperturbative method of lattice QCD~
\cite{bali1998glueballs,aoki1999static,pennanen2000string,bali2000static,duncan2001string,bernard2001zero,kratochvila2003observing,bali2005observation}, they have so far been limited to probing these phenomena in static, equilibrium settings. Since string breaking is conjectured to be a primary mechanism behind the hadronization of color charges in the aftermath of high-energy particle collisions~\cite{andersson1983parton}, a detailed understanding of string breaking in dynamical, nonequilibrium settings will be of significant value~\cite{brambilla2014qcd,bauer2023quantum}.
As current computational methods generally become inadequate in probing real-time QCD processes, the focus has shifted to studies of simpler models, including spin models and field theories in lower spatial dimensions,
so as to offer insights into dynamical mechanisms governing string breaking~\cite{Hebenstreit2013,Hebenstreit2014,Kuhn2015,Buyens2016,Pichler2016,Kormos:2017aa,SALA2018,Spitz2019,ch2019confinement,Lerose2020,Magnifico2020realtimedynamics,Milsted22,Lee2023}. Notably, recent interest surrounds quantum simulators' potential to emulate and study string breaking in such models~\cite{Lerose2020,Verdel19_ResonantSB,Surace2020,Verdel2023,batini2024particle}. 

Indeed, certain Ising spin models in 1+1 dimensions are known to exhibit confinement of domain walls~\cite{McCoyWu,DELFINO1996469,Kormos:2017aa,Liu2019},
i.e., the elementary excitations of the Ising chain in the ordered phase, which can be interpreted as \emph{charges}~\footnote{The term ``charge'' can be understood from the operation of ``gauging'' the global $\mathbb{Z}_2$ symmetry of the Ising model, i.e., promoting the global symmetry to a local one by introducing redundant degrees of freedom. In the lattice gauge theory obtained from this procedure, domain walls play the role of charges. See, for example, Refs.~\cite{balian1975gauge,zhang2018quantum,Lerose2020,borla2020gauging,Surace_2021,exp1} for an explicit derivation of the mapping between the Ising chain and the Ising lattice gauge theory.}. Nearest-neighbor interactions control the mass of individual domain walls, whereas interactions beyond nearest neighbors induce an attractive potential between domain walls, causing them to form bound states, or \emph{mesons}. 
If interactions decay sufficiently slowly with distance, this potential diverges at large distances: the domain walls are confined, meaning that it is impossible to isolate a single domain wall with any finite amount of energy~\cite{McCoyWu,Liu2019}.
If interactions decay quickly with distance, a confining potential can still be induced by a longitudinal field. Bound states formed by pairs of domain walls have been observed experimentally, and their low-lying spectrum was deduced
from nonequilibrium dynamics after a quantum quench~\cite{tan2021domain}. However, the simple notion of a confining potential growing unbounded with distance is only valid in the limit of suppressed charge fluctuations, e.g., for infinitely massive charges.  When the mass is finite, instead, as the distance between the two domain walls grows, the potential energy eventually overcomes the threshold for producing a new pair of domain walls, 
which \emph{screen} the confining potential. The resulting state is a pair of mesons, which can be easily separated, since the meson-meson interaction 
can be shown to be 
suppressed at large distances. Similarly to QCD, this string breaking manifests as the saturation of the potential between two static charges beyond a critical distance~\cite{
Kuhn2015,Buyens2016}.

The experimental observation of domain-wall bound states~\cite{tan2021domain} raises a natural question: can one observe string breaking in spin quantum simulators? The simplest scenario to create nonequilibrium conditions is an instantaneous quantum quench: a string, i.e., a pair of static domain walls at a distance, is set to evolve under the confining Hamiltonian, producing a state with nonvanishing overlap with broken string states. This process was 
first
investigated theoretically in the regime of resonant string breaking~\cite{Verdel19_ResonantSB,Verdel2023}. Recently string breaking in quench processes has been realized for an Ising model, i.e., a (1+1)D lattice gauge theory, in a trapped-ion quantum simulator~\cite{exp1}, for a non-Abelian lattice gauge theory in (1+1)D on an IBM quantum processor \cite{Ciavarella2024}, and for lattice gauge theories in (2+1)D
in arrays of superconducting qubits~\cite{Cochran2024} and Rydberg atoms~\cite{Gonzalez2024}. Nonetheless, inducing string breaking in adiabatic and controlled diabatic processes allows for an enhanced theoretical understanding of string breaking in real-time dynamics. These controlled processes, for instance, can reveal potential universal dynamics, and share similarities with the physical setup of an expanding string generated by a collision of charges and bound states in high-energy experiments. Our work develops a theoretical foundation for, and numerical analysis of, string-breaking dynamics in adiabatic and controlled diabatic processes. As will be demonstrated in 
an upcoming paper~\cite{exp2},
recent advances in trapped-ion systems enable realizing our proposed beyond-quench string-breaking processes in present-day experiments.

We focus on string breaking in the presence of a pair of static charges in a quantum Ising chain
with either exponentially decaying or power-law decaying interactions. This choice is inspired by the capabilities of modern trapped ion experiments, which can implement both types of interactions~\cite{exp1}. We study the time-evolved state of the region enclosed between the static charges as the string tension is slowly increased in time through the breaking point.
We find two different regimes. For short strings, the breaking materializes as a collective flip of all the spins in the region, adiabatically following the lowest-energy state. This case can be interpreted as the creation of two dynamical charges at the edges of the region, such that the static charges are completely screened. We describe this breaking using two-state transition dynamics within the Landau-Zener formalism. As the corresponding energy gap is exponentially small in the string length, for long strings, such a collective flip 
involves diabatic transitions which excite low-energy many-body states, unless the variation of the string tension is prohibitively slow.
For a ramping speed that is independent of 
the string length, we find that the string breaking manifests in the formation of finite \emph{bubbles} of flipped spins. We study how bubble nucleation depends on the ramp speed and establish that the typical size of the bubbles decreases with the speed. A similar evolution has previously been studied in the regime of small transverse fields and long evolution times~\cite{Sinha2021}. This analysis demonstrates the value of controlled diabatic processes in characterizing the evolution of a string toward a superposition of broken strings.

Our analysis establishes a direct connection between the string-breaking dynamics and the crossing of a first-order quantum phase transition, offering a novel perspective on the real-time behavior of confined systems. We identify concrete observables and dynamical protocols that  quantitatively capture string breaking in real-time. Our results show that identifying static string breaking for long strings from real-time dynamics can be challenging, as achieving adiabaticity is generally difficult. We, therefore, provide pathways for studying string-breaking dynamics in the absence of adiabaticity, which turned out to present rich dynamics; with associated diagnostics that are accessible in experiment. Our findings in such diabatic settings may have implications for other processes, such as string breaking induced by collisions, although establishing a precise connection is left for future work. Moreover, our analysis reveals an unexpected scaling law for non-equilibrium dynamics across the string-breaking transition, as well as novel phenomena arising from long-range interactions, such as the emergence of string breaking driven purely by quantum fluctuations.

The remainder of this paper is structured as follows. In Sec.~\ref{sec:sb}, we introduce the model and the string-breaking setup. In Sec.~\ref{sec:dyn}, we focus on the model with exponential decay of interactions. We first study static string breaking by analyzing how the energy spectrum depends on the string tension. We then study the dynamics of the string as the string tension is linearly increased in time. In Sec.~\ref{sec:longrange}, we discuss the case of power-law decaying interactions and show how string breaking can occur upon increasing the strength of quantum fluctuations at constant string tension. We conclude in Sec.~\ref{sec:conclusions} with a discussion of the results, and with considerations for experimental realization. Three Appendices offer further details on our setup, models, and assumptions.

%%%%%%%%%%%%%%%%%%%%%%%%%%%%%%%%%%%%%%%%%%%%%%%%%%%%%%%%%%%%%%%%%%%%%%%%%%%%%%%%%%%%%%%%%%%%%%%%%%%%%%%%%%%%%%%%%%%%%%%%%%%%%%%%%%%%%%%%%%%%%%%%%%%%%%%%%%%%%%%%%%%%%%%%%%%%%%%%%%%%%%%%%%%%%%%%%%%%%%%%%%%%%%%%
\section{Quantum Ising chain and string-breaking setup}
\label{sec:sb}
Consider a quantum Ising chain described by the Hamiltonian
\begin{equation}
\label{eq:Ising}
H_\text{Ising} = -\sum_{i<j} J_{i,j} \sigma^z_i \sigma^z_j-g\sum_{j} \sigma_j^x-h\sum_j \sigma_j^z.
\end{equation}
The ferromagnetic Ising coupling $J_{i,j}>0$ is a function of the distance $|j-i|$ between spins $i$ and $j$, $g$ is the transverse-field strength, and $h$ is the longitudinal-field strength. 
For $h=0$, the model has a second-order quantum phase transition between a ferromagnetic and a paramagnetic phase at a critical value $g=g_{*}$.
We will focus on sufficiently small values of $g$, such that, for $h=0$, the system is in the ferromagnetic phase. In the limit of vanishing $g$, the ground state of the system for $h>0$ is the maximally polarized state along the $+\hat z$ direction. 
Furthermore, ferromagnetic domain walls, pictorially denoted as $\dots\uparrow\uparrow\uparrow\downarrow\downarrow\downarrow\dots$ and $\dots\downarrow\downarrow\downarrow\uparrow\uparrow\uparrow\dots$, are the elementary excitations 
for $h=0$, 
where $\uparrow$ and $\downarrow$ denote spins pointing along the positive and negative $\hat z$ directions, respectively. In the ferromagnetic phase, away from the $g=0$ limit, the spins are 
no longer maximally polarized
but the description in terms of domain walls is still valid. 
The interpretation of domain walls as charges can be made more rigorous by mapping the quantum Ising chain to a $1+1$-dimensional lattice gauge theory~\cite{Lerose2020}: upon this mapping, domain walls become particles and domains of $\downarrow$ spins are electric-flux strings.

\begin{figure}[t!]
    \centering
    \includegraphics[width=\linewidth]{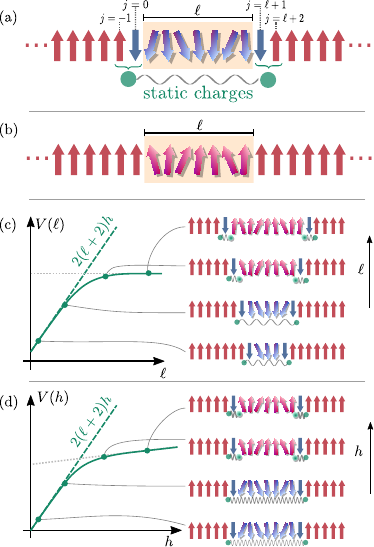}
    \caption{(a) Schematic illustration of a ``string-like'' ground state in the presence of static charges. The static charges (in green) are domain walls separated by a string of length $\ell+2$. Only the $\ell$ spins between the static charges 
    (yellow region) are dynamical. Red and blue colors are used for static spins polarized along $+\hat z$ and $-\hat z$, respectively, 
    while a
    red (blue) gradient illustrates the dynamical spins with net positive (negative) $\hat z$ magnetization. 
    (b) Schematic illustration of the ground state of $\ell$ dynamical spins without static charges. The potential is computed as the energy difference between the ground states with and without static charges. (c) 
    The saturation of the potential $V(\ell)$ for large $\ell$ indicates the presence of string breaking in the model. The potential saturates when the string potential $V(\ell)$ overcomes the energy needed to produce two additional charges, thus breaking the string.
    (d) Another indication of string breaking is the sudden change of slope of the potential (from $2(\ell+2)$ to $4$) as a function of string tension $h$ for a fixed string length $\ell$. The slopes noted correspond to $g=0$ but they remain good approximations for $g\neq 0$.
    }
    \label{fig:1}
\end{figure}

Realizing the paradigmatic string-breaking thought experiment requires imposing \emph{static} charges as well. This can be done by freezing a pair of neighboring spins in the domain-wall configuration $\uparrow\downarrow$ and another pair in the domain-wall configuration $\downarrow\uparrow$. The domain walls are fixed at a distance $\ell+2$ from each other. The $\ell$ dynamical spins located between the frozen pairs
are allowed to evolve, while the external spins located beyond the static charges are kept static. This simplified arrangement with non-dynamical external spins does not qualitatively alter the string-breaking scenario \cite{Verdel19_ResonantSB,exp1}, as discussed in Appendix \ref{app:ext}. We will consider a configuration like the one shown in Fig.~\ref{fig:1}(a): $\ell$ spins (with $j=1,\dots \ell$) are dynamical (i.e., acted upon by the transverse field); one external spin on each edge ($j=0$ and $j=\ell+1$) is polarized along $-\hat z$, and all the other external spins ($j<0$ and $j>\ell+1$) are polarized along $+\hat z$. The string is considered to be the state consisting of only two domain walls---the ones involving the static edge spins. Therefore, the internal spins in the string are polarized along $-\hat z$. This state may or may not be the ground state of the model, as will be discussed shortly.

The external spins have no dynamics:  their only effect is to induce an effective longitudinal field $h_j^\text{eff}$ on the dynamical spins. Explicitly, the Hamiltonian governing the $\ell$ dynamical spins, including the effective longitudinal field $h_j^\text{eff}$ generated by the external spins and the homogeneous field $h$, reads
\begin{equation}
\label{eq:H}
H = -\sum_{1\le i<j\le \ell} J_{i,j} \sigma^z_i \sigma^z_j-g\sum_{j=1}^\ell \sigma_j^x+\sum_{j=1}^\ell (h_j^{\mathrm{eff}}-h)\sigma^z_j,
\end{equation}
where
\begin{equation}
\label{eq:Beff}
    h_j^\text{eff}=J_{0,j}+J_{j,\ell+1}-\sum_{i=-\infty}^{-1} J_{i,j}-\sum_{i=\ell+2}^{+\infty} J_{j,i}
\end{equation}
represents the site-dependent coupling of a dynamical spin $j$ to the static spins.

%%%%%%%%%%%%%%%%%%%%%%%%%%%%%%%%%%%%%%%%%%%%%%%%%%%%%%%%%%%%%%%%%%%%%%%%%%%%%%%%%%%%%%%%%%%%%%%%%%%%%%%%%%%%%%%%%%%%%%%%%%%%%%%%%%%%%%%%%%%%%%%%%%%%%%%%%%%%%%%%%%%%%%%%%%%%%%%%%%%%%%%%%%%%%%%%%%%%%%%%%%%%%%%%
\section{String breaking with short-range interactions}
\label{sec:dyn}
For concreteness, in this section,  
we focus on the case of exponentially decaying interactions of the form
\begin{equation}
\label{eq:Jij}
    J_{i,j}\equiv J_{|j-i|}=e^{-(|j-i|-1)/\xi},
\end{equation}
with $\xi>0$. We set our units such that the nearest-neighbor interaction is $J_{1}=1$. This interaction type is inspired by condensed-matter systems whose underlying interaction is short-range~\cite{Dutta_2015}. Such interactions can be experimentally realized in a native manner in trapped-ion quantum simulators~\cite{Nevado2016,schuckert2023observation,katz2024observing}, photonic meta-materials~\cite{Zhang2023}, and other quantum simulators whose synthetic interaction can be programmed efficiently~\cite{periwal2021programmable,Lee2016}.  
The phase diagram of the Ising model with such interactions is qualitatively the same as that of the standard Ising model with nearest-neighbor interactions. 
With this choice of couplings, the effective field in Eq.~\eqref{eq:Beff} has the form
\begin{equation}
\label{eq:Beffsr}
    h_j^\mathrm{eff}=\frac{1-2e^{-1/\xi}}{1-e^{-1/\xi}}\left(e^{-(j-1)/\xi}+e^{-(\ell-j)/\xi}\right).
\end{equation}
The effective field $h_j^\text{eff}$, thus, is positive (negative) for all $j$ if $\xi<1/\log(2)$ ($\xi>1/\log(2)$).
In the following, we will assume $\xi<1/\log(2)$, such that the positive effective field $h_j^\text{eff}$ favors polarization along the $-\hat{z}$ direction and the string is stable for $h=0$. 
In all the numerical simulations, we will set $\xi=1$.

%%%%%%%%%%%%
\subsection{String-breaking statics}
Before studying string breaking in a dynamical, nonequilibrium setting, we analyze the statics of the string and the broken string in the short-range model to gain an intuition for states and their properties.  In the short-range model, the longitudinal field $h>0$ induces a linear potential (proportional to $h\ell$) between the static charges, which are now confined in composite excitations (mesons, pictorially represented as $\dots\uparrow\uparrow\downarrow\downarrow\downarrow\uparrow\uparrow\dots$). In this section, we derive the analytic form of the confining potential, and of a residual potential among mesons in a broken string, in the absence of a transverse field, and further analyze these potentials numerically when a nonvanishing transverse field is introduced.

%%%%%%%%%%%%
\subsubsection{Case $g=0$}
\label{subsec:h0}
\begin{figure*}[t!]
    \centering
    \includegraphics[width=\textwidth]{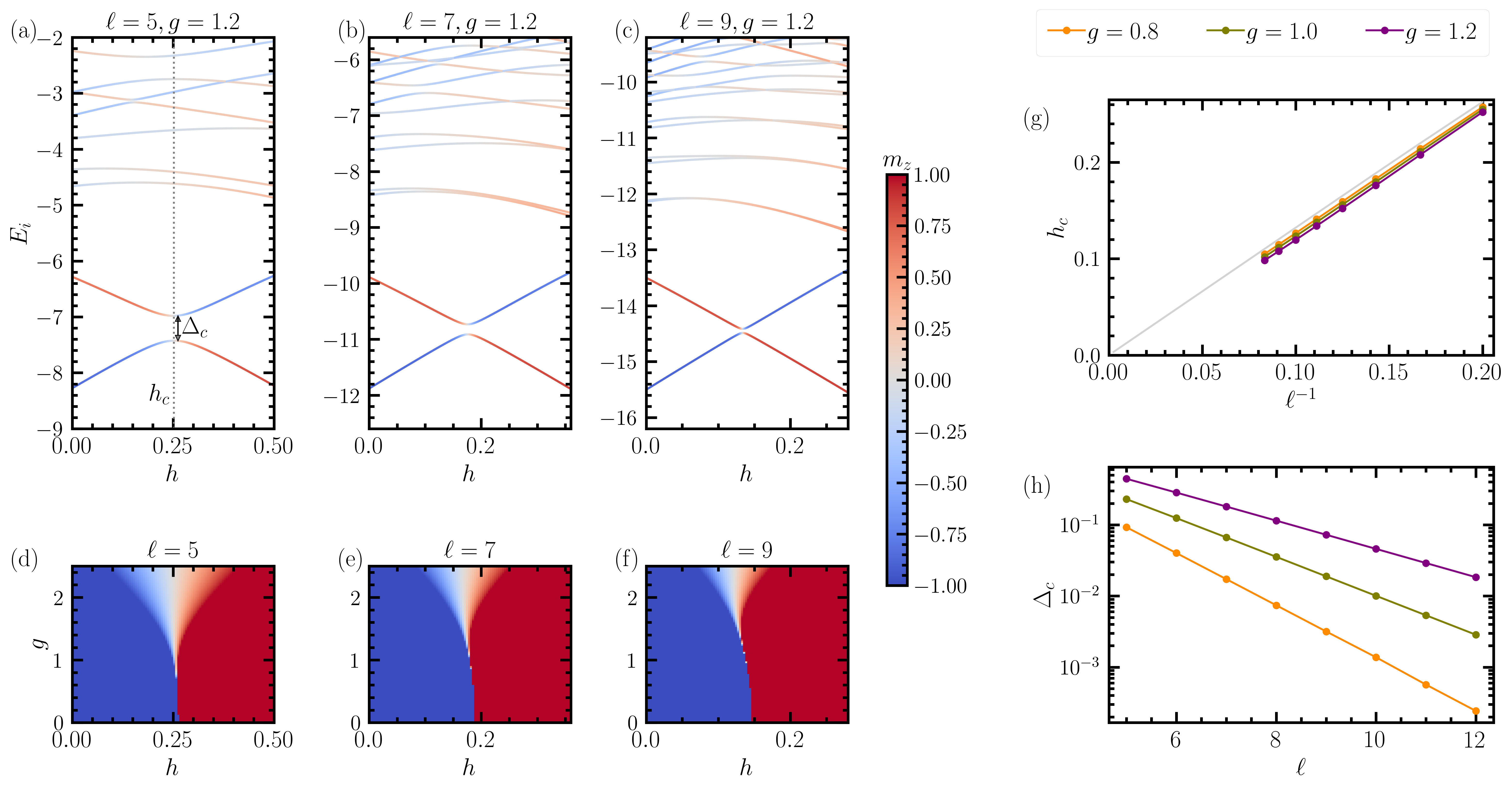}
    \caption{ (a,b,c) Lowest energy eigenvalues, $E_i$, of the Hamiltonian in Eq.~\eqref{eq:H} with short-range coupling $J_{i.j}$ given in Eq.~\eqref{eq:Jij} and effective longitudinal field in Eq.~\eqref{eq:Beffsr} (with $\xi=1$), as a function of the homogeneous longitudinal field $h$ for $g=1.2$ and strings of lengths $\ell=5,7,9$. The color indicates the expectation value of the magnetization along the $\hat{z}$ direction for each energy eigenstate. The two lowest levels are well separated from the rest of the spectrum. They correspond to the string (in blue) and broken-string (in red) configurations and undergo an avoided level crossing at $h\approx h_c(\ell)$. (d,e,f) Magnetization of the ground state as a function of $g$ and $h$ for strings of lengths $\ell=5,7,9$. (g) The static string-breaking point  $h_c$, corresponding to where the minimum gap $\Delta_c$ between ground and the first excited state occurs, as a function of $\ell^{-1}$ for different values of $g$. Even for fairly large values of $g$ (up to $g=1.2$), $h_c$ shows rather small deviations from the $g=0$ 
    case (gray line) obtained in Eq.~(\ref{eq:Bc}). (h) Gap $\Delta_c$ at the avoided crossing as a function of $\ell$. The gap is exponentially small in $\ell$ and grows with $g$. A fit to the data gives $\Delta_c \propto (g/1.88)^\ell$.}
    \label{fig:spectrum_sc}
\end{figure*}
In order to understand how the string can break as one introduces a confining potential $h\neq 0$, it is useful to first focus on the case $g=0$. The Hamiltonian is diagonal in the $z$ basis and the energy difference between the string  state, labeled by `$s$' (with all dynamical spins polarized along $-\hat z$) and the broken string, labeled by `$bs$' (with all dynamical spins polarized along $+\hat z$) can be computed as
\begin{align}
\label{eq:DeltaE}
    E_{bs}-E_{s}&=2\sum_{j=1}^\ell h_j^\text{eff}-2\ell h \nonumber\\
    &=4\frac{(1-2e^{-1/\xi})(1-e^{-\ell/\xi})}{(1-e^{-1/\xi})^2}-2\ell h.
\end{align}
Equation~(\ref{eq:DeltaE}) shows that there is a value $h_c(\ell)$ such that, for $h>h_c(\ell)$,  the broken string has a lower energy than the string. We will refer to this value as the {\it string-breaking point}, which 
reads
\begin{equation}
\label{eq:Bc}
    h_c(\ell)= 2\frac{(1-2e^{-1/\xi})(1-e^{-\ell/\xi})}{(1-e^{-1/\xi})^2} \, \ell^{-1}.
\end{equation}
Since $h_c(\ell)$ decreases as $\ell^{-1}$, in the limit of large $\ell$, a vanishingly small string tension is sufficient to break the string.

Finally, to argue that $h_c(\ell)$ is the string-breaking point, one also needs to check that all the other states in the $z$ basis have an energy larger than the string configuration for $h<h_c(\ell)$. This check is provided in Appendix \ref{app:bubbles}. Note that this is enough to prove that the broken string state is the ground state for $h>h_c(\ell)$ \footnote{Following Appendix \ref{app:bubbles}, we can show that at the string-breaking point $h=h_c$, one has $E_s=E_{bs} < E_\mathrm{other}$, where $E_\mathrm{other}$ is the energy of an arbitrary configuration with $n<\ell$ spins polarized along $-\hat z$. For a larger $h>h_c$, the difference $E_\mathrm{other}-E_{bs}$ can only be larger, since it differs from the $h=h_c$ value by a quantity $2n(h-h_c)$.}.

For the simple case of $g=0$, one can also write an explicit expression for the potential for the string and broken-string states, defined with respect to the energy of the configuration without static charges (shown in Fig.~\ref{fig:1}(b)):
\begin{align}
    V_s(\ell, h)&=2h(\ell+2)+2\sum_{i=-\infty}^{-1}\sum_{j=0}^{\ell+1}J_{i,j}+2\sum_{i=0}^{\ell+1} \sum_{j=\ell+2}^{\infty}J_{i,j} \nonumber\\
    &=2h(\ell+2)+a_s+b_s e^{-\ell/\xi},
\end{align}
and
\begin{align}
\label{eq:Vbs}
    V_{bs}(\ell, h)&=4h+2\sum_{j\in \mathbb{Z}\setminus\{0,\ell+1\}}(J_{0,j}+J_{\ell+1,j}) \nonumber\\
    &=4h + a_{bs}+b_{bs}e^{-\ell/\xi},
\end{align}
where $a_s= 4(1-e^{-1/\xi})^{-2}$, $b_s=-4(e^{1/\xi}-1)^{-2}$, $a_{bs}=8(1-e^{-1/\xi})^{-1}$, and $b_{bs}=-4$.

This calculation shows how (static) string breaking is associated with a sudden change of the slope of the potential computed for the lowest-energy configuration: the slope changes from $dV/dh = 2 (\ell + 2)$ for the string to $dV/dh = 4$ for the broken string.
Note that, for $g=0$, the Hamiltonian is diagonal in the $z$ basis, and there is no matrix element between the string and broken-string states. Therefore, to observe string breaking in real-time dynamics, one has to consider a finite transverse field $g\neq 0$.

%%%%%%%%%%%%
\subsubsection{Case $g\neq 0$}
For $g\neq 0$, the spectrum cannot be computed analytically, but one can use intuition from the $g=0$ case to understand the general features of the low-energy spectrum as a function of $h$.
Moreover, for relatively small system sizes (up to $\ell \lesssim 16$), the spectrum can be computed numerically using exact diagonalization. As we will show, all the key signatures of interest in our work are already clearly visible in this regime, making larger system sizes unnecessary for establishing our main conclusions. 

For $g=0$, the string and the broken-string configurations cross at $h=h_c(\ell)$. For small $g$, we expect that the main effect of the transverse field is to open a gap at the crossing of the lowest two energy levels, i.e., the crossing turns into an avoided crossing. As shown in Fig.~\ref{fig:spectrum_sc}(a-c), this picture is true even for fairly large values of $g$ (but small enough
to remain in the ferromagnetic phase).
The minimum gap as a function of $\ell$ for a fixed $g$ occurs for a value $h_c(\ell, g)$ that is very close to the one in Eq.~(\ref{eq:Bc}), which was computed for the $g=0$ case [Fig. \ref{fig:spectrum_sc}(g)]. We refer to $h_c(\ell, g)$ as the (static) string-breaking point. Following the ground state for different values of $h$, the properties of the state change abruptly as one crosses the string-breaking point $h_c(\ell, g)$: the magnetization, for example, rapidly changes sign, as shown by the colormap in Fig.~\ref{fig:spectrum_sc}(a-c). This abrupt change is shown in Fig.~\ref{fig:spectrum_sc}(d-f) as a function of $g$ and $h$ for different values of $\ell$: the change is less abrupt as $h$ is varied across the string-breaking point when $g$ is larger and $\ell$ is smaller. For such values, the gap at the avoided level crossing is larger. 

Comparing the spectrum for different values of $\ell$, we observe that the minimum gap $\Delta_c$ where the avoided crossing takes place is suppressed very rapidly upon increasing $\ell$. 
This 
rapid decrease can be understood as follows: the gap is given by the matrix element between the string and the broken-string state; these two states are macroscopically different (intuitively, all $\ell$ spins have to be flipped to connect them), and hence this matrix element is exponentially small in $\ell$. The plot in Fig.~\ref{fig:spectrum_sc}(h) confirms the exponential scaling of the gap for $g<g_*$: $\Delta_c\propto e^{-c\ell}$, where $c$ depends on $g$ roughly as $c\approx -\log (g/1.88)$. This shows that the gap can vary by many orders of magnitude as $\ell$ changes.

The exponential closing of the gap and the sudden change in the nature of the ground state across the string-breaking point is a witness for a first-order phase transition. In the phase diagram of $H_\mathrm{Ising}$ [Eq.~(\ref{eq:Ising})], the line $h=0$ (for $g$ smaller than its critical value) corresponds to a first-order phase transition, with a discontinuous change of magnetization between $h<0$ and $h>0$. In our string-breaking setup, the system size is finite, and the boundary conditions corresponding to the presence of external charges 
shift the crossing point from $h=0$ to $h=h_c(\ell, g)$. The transition point of the mixed-field Ising chain is recovered in the thermodynamic limit since $h_c(\ell)$ vanishes as $\ell^{-1}$ for large $\ell$.

% %%%%%%%%%%%%%%%%%%%%%%%%%%%%%%%%%%%%%%%%%%%%%%%%%%%%%%%%%%%%%%%%%%%%%%%%%%%%%%%%%%%%%%%%%%%%%%%%%%%%%%%%%%%%%%%%%%%%%%%%%%%%%%%%%%%%%%%%%%%%%%%%%%%%%%%%%%%%%%%%%%%%%%%%%%%%%%%%%%%%%%%%%%%%%%%%%%%%%%%%%%%%%%%%

\subsection{String-breaking dynamics}
The energy of the string is proportional to both the length and the string tension. In the conventional notion of string breaking, the potential between the static charges is tracked as a function of the distance between them [Fig.~\ref{fig:1}(c)]. Here, we choose to focus on a different protocol: instead of increasing the distance between the static charges, we increase the string tension, i.e., the strength of the confining field $h$, while keeping the distance fixed [Fig.~\ref{fig:1}(d)]. This choice is particularly convenient for experimental implementations, where system sizes are typically limited. Moreover, the string tension is an external parameter that can be controlled and smoothly changed in time while keeping the external charges static. Changing the position of the charges dynamically instead involves significant technical challenges.

As one increases the string tension, the result is analogous to increasing the string length: the energy of the string becomes larger than the energy needed to produce additional charges (domain walls) and the string breaks. 
In this protocol, the potential 
does not saturate as $h$ is increased past the string-breaking point, but keeps growing with a smaller $\ell$-independent slope ${dV}/{dh}=4$ [dotted line in Fig.~\ref{fig:1}(d)]. The reason is that the masses of the two mesons that are created when the string is broken grow linearly with $h$:  for large $h$, the spins are $z$-polarized, and the mass of a meson is the energy needed to flip a single spin, i.e., $M_\text{mes}\approx 2h +4\sum_{j=1}^\infty J_{0,j} =2h+a_{bs}/2$, where $a_{bs}$ is defined as in Eq.~(\ref{eq:Vbs}).

To observe string breaking, expected to occur at a value $h_c$ of the string tension, we imagine a protocol where the system is prepared in the ground state for $h=0$ and finite $g$~\footnote{The initial state considered here is the ground state at finite transverse field $g$. Experimentally, this can be prepared with high fidelity with a quasi-adiabatic ramp of the transverse field from $g=0$, since the gap remains rather large along the ramp.}, which happens to be a string state. Then the longitudinal field is slowly turned on in time~\footnote{To make the time dimensionality of $\tau$ explicit, one can reintroduce the energy scale $J_{1}$, set to $1$ earlier for simplicity, and define $h(t)=J_{1}\, t/\tau$.}
\begin{equation}
    h(t)=t/\tau.
\end{equation}

If the speed $\tau^{-1}$ is sufficiently small, then the system follows the ground state adiabatically, and magnetization, as a function of $h$, changes abruptly when $h\approx h_c$. However, with the exception of very short strings, the gap at $h\approx h_c$ is extremely small, and an adiabatic protocol is practically unattainable. 
Naively, this implies that probing string breaking at $h\approx h_c$ is not possible. However, we will show below 
that string breaking can 
be probed even for long strings 
when accounting for the nontrivial diabatic component of the evolution. In this case, the string-breaking process observed in real-time evolution will be rather different from that for the (perfectly adiabatically prepared) ground state.

\begin{figure}[t!]
    \centering
\includegraphics[width=\linewidth]{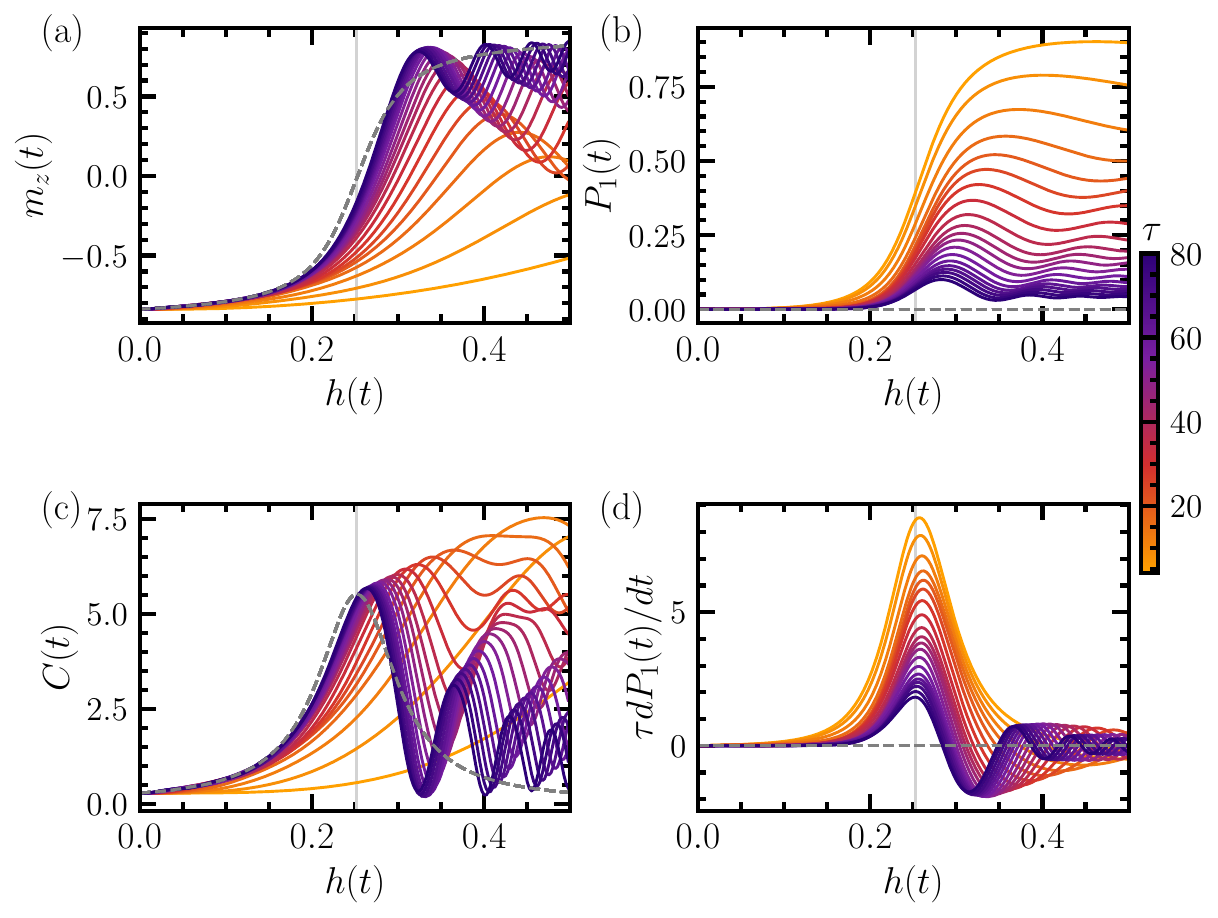}
    \caption{Time evolution as a function of $h(t)=t/\tau$ for different values of $\tau$. The dashed gray lines represent the adiabatic limit $\tau\rightarrow \infty$. The length of the string is $\ell=5$, and the transverse field is $g=1.2$. (a) Magnetization along $\hat{z}$ [Eq.~(\ref{eq:mx})]. (b) Population $P_1$ of the (instantaneous) first excited state [Eq.~(\ref{eq:Pn})]. (c) Connected spin-spin correlator [Eq.~(\ref{eq:C})]. (d) $dP_1/dh=\tau dP_1/dt$. In all plots, the value of $h_c = 0.252$ where the gap is minimal is indicated as a gray vertical line. 
    }
    \label{fig:ramp_sc}
\end{figure}
%

%%%%%%%%%%%%
\subsubsection{Observables across a longitudinal-field ramp}
\label{subsec:obsramp}
To probe string breaking in the diabatic regime, we will first focus on the following protocol.
We evolve from $h=0$ at time $t=0$ to $h=h_\mathrm{f}$ at time $t=\tau h_\mathrm{f}$. We choose $h_\mathrm{f}\simeq 2h_c$, so that we can characterize the evolution close to the avoided crossing. Figure~\ref{fig:ramp_sc} displays various observables as a function of $h(t)$ for different values of $\tau$. We consider the average magnetization along $\hat{z}$ [Fig.~\ref{fig:ramp_sc}(a)], 
\begin{equation}
\label{eq:mx}
    m_z(t)=\frac{1}{\ell}\sum_{j=1}^\ell \braket{\Psi(t)|\sigma_j^z| \Psi(t)}, 
\end{equation}
and the populations (i.e.\ occupation probabilities) of the instantaneous eigenstates [Fig.~\ref{fig:ramp_sc}(b)],
\begin{equation}
\label{eq:Pn}
    P_n(t)= \left|\braket{\Psi(t)|\psi_n(h(t))}\right|^2,
\end{equation}
where $\ket{\Psi(t)}$ is the state of the system at time $t$ and $\ket{\psi_n(h)}$ is the $n$-th eigenstate of the Hamiltonian $H(h)$ (ordered from lowest to highest energy). In particular, $n=0$ is the ground state and $n=1$ is the first excited state.
As shown in Fig.~\ref{fig:ramp_sc}(a) for $\ell=5$ and $g=1.2$, the magnetization grows when $h$ is close to $h_c$, indicating that the string is (partially) breaking. For sufficiently large values of $\tau$, the magnetization changes sign and starts to oscillate around a positive value. The growth is faster for large $\tau$, where the evolution is closer to an adiabatic one. The deviation from an adiabatic evolution can also be observed in Fig.~\ref{fig:ramp_sc}(b). For large $\tau$, the evolution is slow and the population $P_1$ of the (instantaneous) first excited state remains small for the entire evolution. For sufficiently small $\tau$, instead, this population can reach $P_1\approx 1$: this is the case of a completely diabatic evolution, where the system essentially remains in the string state.

To further characterize the evolution, and, in particular, to find independent methods to estimate $h_c$ from real-time evolution, it is useful to consider additional observables. We define the following connected correlator
\begin{equation}
\label{eq:C}
    C=\sum_{i=1}^\ell\sum_{j=i+1}^\ell (\braket{\sigma_i^z\sigma_j^z}-\braket{\sigma_i^z}\braket{\sigma_j^z}).
\end{equation}
For adiabatic evolution, we expect $C$ to have a sharp peak at $h\approx h_c$, where the ground state is a coherent superposition of string and broken-string states, and to be small for all other values of $h$. As shown in Fig.~\ref{fig:ramp_sc}(c), for large values of $\tau$, the correlator $C$ has a large peak at $h\approx h_c$, in agreement with the expectation from the adiabatic case. 
Residual oscillations for $h>h_c$ 
indicate that the system remains in a superposition of  string and broken-string states for some time after the avoided crossing. The value of $h_c$ can be estimated by extrapolating the value of $h$ of the first local maximum of $C$ for $\tau\rightarrow \infty$. 

Another method to estimate $h_c$ from real-time dynamics is to look for the maximum of the derivative $dP_1/dh$, as shown in Fig.~\ref{fig:ramp_sc}(d). This method, however, has the disadvantage that, since the instantaneous eigenstate $\ket{\psi_1(h)}$ is not known for $g\neq 0$, it is not easy to measure $P_1$ experimentally. The correlator $C$ can instead be directly measured in present-day experiments, since these experiments can typically sample from the distribution of bitstring configurations of the quantum state of interest.

%%%%%%%%%%%%
\subsubsection{Landau-Zener approach}
\label{subsec:LZ}
\begin{figure*}[t!]
    \centering
\includegraphics[width=\textwidth]{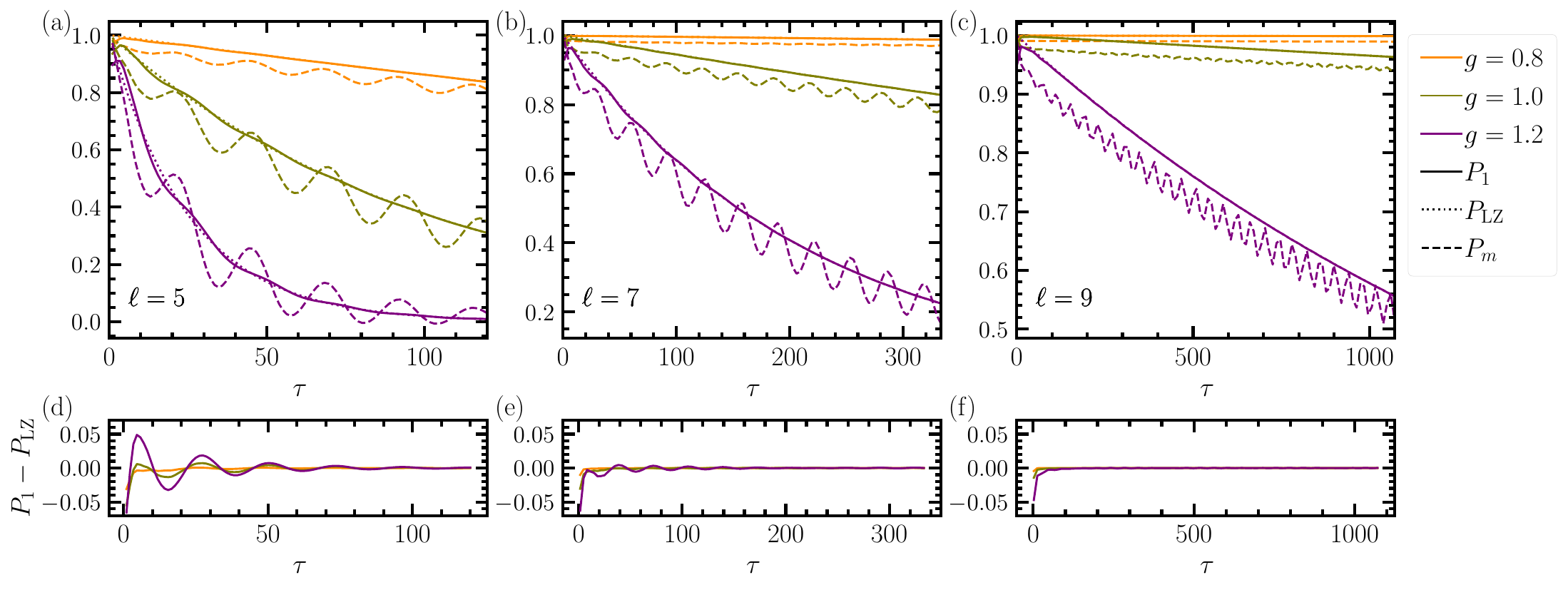}
    \caption{The final values of probabilities after the ramp from $h=0$ to $h=h_\mathrm{f}$ for (a) $\ell = 5$, (b) $\ell = 7$, and (c) $\ell = 9$, t (here, $h_\mathrm{f}=0.5, 0.36, 0.28$ for $\ell=5,7,9$, respectively). Solid lines: population $P_1$ of the first excited state.  Dotted lines: population $P_\text{LZ}$ from the Landau-Zener formula in Eq.~(\ref{eq:LZ}). Where not visible, the dotted lines overlap with the solid lines. Dashed lines: population $P_m$ defined from the magnetization [see Eq.~(\ref{eq:Pm})]. Difference $P_1-P_{\rm LZ}$ for (d) $
    \ell=5$, (e) $\ell=7$ and (f) $\ell=9$. }
    \label{fig:LZsc}
\end{figure*}
The evolution discussed earlier can be quantitatively understood using a Landau-Zener approach (see Ref.~\cite{Pelissetto2020} for a similar discussion). The string and broken-string states can be approximated by a two-level system described by the Hamiltonian
\begin{equation}
\label{eq:HLZ}
    H_\text{LZ}(t)=\begin{pmatrix}
\ell m_* (h(t)-h_c)& \Delta_c/2\\
\Delta_c/2 &-\ell m_* (h(t)-h_c)
    \end{pmatrix}.
\end{equation}
The effective magnetization $m_*$, the crossing point $h_c$, and the gap $\Delta_c$ at the crossing are obtained by fitting $E_1-E_0$ as a function of $h$ close to its minimum to the functional form $E_1-E_0=2\sqrt{\ell^2m_{*}^2(h-h_c)^2+(\Delta_c/2)^2}$, where $E_0$ and $E_1$ are the energies of the ground and the first excited states of the original Ising Hamiltonian, respectively. 
This model is expected to be accurate as long as the two lowest states are separated from the other states in the spectrum of $H$ by a large gap (compared to $\sqrt{m_*\tau^{-1}}$), such that the other states are not excited during the evolution. As shown in Fig.~\ref{fig:spectrum_sc}(a-c), a large gap is present for the range of parameters considered here. We also check in Appendix \ref{app:data} that the population of other levels is smaller than $~10^{-2}$ for the range of $\tau$ considered ($\tau \ge 4.0$ for $\ell=5$, $\tau \ge 5.6$ for $\ell=7$, and $\tau \ge 7.1$ for $\ell=9$).

The Landau-Zener formula \cite{Zener1932} predicts that the population of the excited state at $t=+\infty$ after evolving it from the ground state at $t=-\infty$ has the form
\begin{equation}
\label{eq:LZ}
    P_\mathrm{LZ}=\exp\left(-\frac{\pi \Delta_c^2 \tau}{4\ell m_*}\right).
\end{equation}
In Fig.~\ref{fig:LZsc}, we compare this expression with the value of $P_1$ at the end of the evolution (when $h=h_\mathrm{f}$) as a function of $\tau$: despite the finite time interval of the ramp in the calculation of $P_1$, we find excellent agreement between the two. This agreement shows that the Landau-Zener prediction from a simple two-level system is able to capture the string-breaking dynamics of the much more complex many-body Hamiltonian.

The excited state $\ket{\psi_1(h_\mathrm{f})}$ is not analytically known for $g\neq 0$. One way to measure the population $P_1$ experimentally is to adiabatically turn off the transverse field $g$, in order to map $\ket{\psi_1(h_\mathrm{f})}$ to the first excited state in the $\hat{z}$ basis. This approach can be challenging because the system may lose coherence during such an adiabatic ramp. 
As an alternative approach, it is useful to consider a more accessible quantity, $P_m$, defined as
\begin{equation}
\label{eq:Pm}
    P_m = \frac{m_z^0+m_z^\mathrm{f}}{2m_z^0},
\end{equation}
which can be directly obtained by measuring the final magnetization. Here, $m_z^0$ and $m_z^\mathrm{f}$ are the values of the magnetization at the beginning and the end of the linear ramp, respectively. For small $g$, the $z$ magnetization is approximately diagonal in the Landau-Zener basis used in Eq.~(\ref{eq:HLZ}), with diagonal elements $\mp m_*$. Within this approximation, the initial and final states have magnetization $m_z^0\approx-m_*$ and $m_z^\mathrm{f}\approx P_0 m_*+P_1(-m_*)=m_
*(1-2P_1)$. By plugging these expressions in Eq.~(\ref{eq:Pm}), we find that, in this regime, $P_m$ is a good approximation of the final $P_1$. In Fig.~\ref{fig:LZsc}, we compare $P_m$ with $P_1$ and the Landau-Zener prediction. We find that $P_m$ captures the overall decay, even for large values of $g$, but exhibits additional oscillations as a function of $\tau$.

%%%%%%%%%%%%
\subsubsection{Potential}

As mentioned before, string breaking is triggered for a fixed $\ell$ by increasing $h$: a sharp suppression of the growth of the potential is expected between two static charges as a function of $h$ for $h>h_c$. In the dynamical setup of interest here, we need a suitable definition of the potential to probe such an effect. At a time $t$ of the linear ramp, we define the potential $V(t)$ as the difference between the energy $E(t)=\braket{\Psi(t)|H(t)|\Psi(t)}$ with the Hamiltonian defined in Eq.~\eqref{eq:H} and the ground-state energy of the state in the absence of the static charges, $E^{\rm{vac}}(t)$. Explicitly,
\begin{equation}
\label{eq:V}
    V(t)=E(t)-E^{\rm{vac}}(t
    )+4h(t).
\end{equation}
Here, we have included $4h(t)$, the contribution to the energy difference coming from the longitudinal field applied to the static spins at positions $j=0$ and $\ell+1$. In this equation and in the numerical results, we have dropped some additional terms, coming from interactions between static spins, that do not depend on $t$, and only contribute a constant shift to the potential. 
The energy $E^{\rm{vac}}(t)$ is defined as the ground-state energy of the Hamiltonian in Eq.~\eqref{eq:Beff} with $h=h(t)$ but with $h_j^\text{eff}$ replaced with $h_j^\text{vac}$, where
\begin{equation}
    h_j^\text{vac}=-\sum_{i=-\infty}^{0} J_{i,j}-\sum_{i=\ell+1}^{+\infty} J_{i,j}.
\end{equation}
This state corresponds to the vacuum configuration in Fig.~\ref{fig:1}(b).
\begin{figure}[t!]
    \centering
    \includegraphics[width=\linewidth]{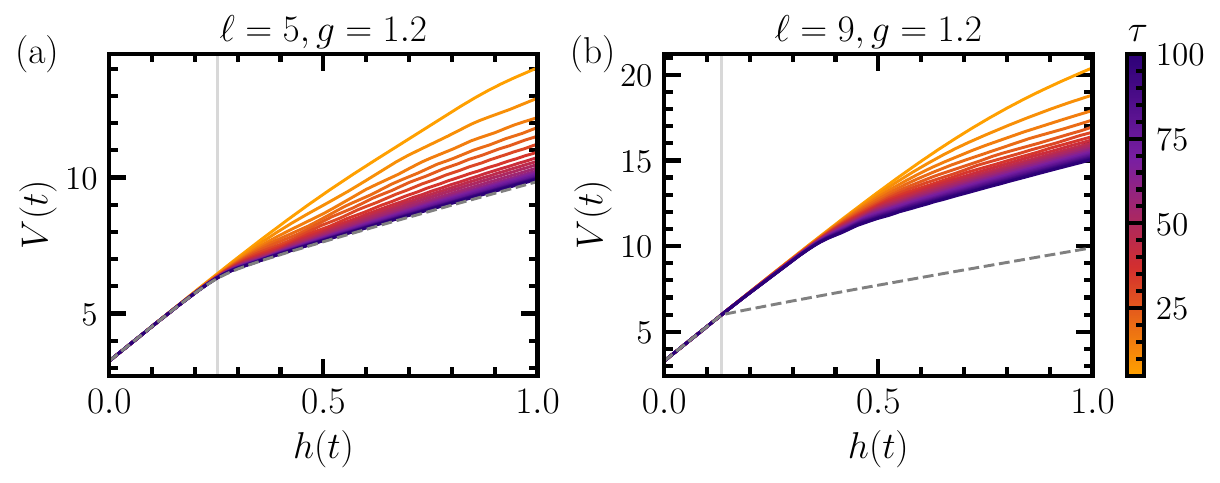}
    \caption{Potential of the static charges in the dynamical protocol, as defined in Eq.~(\ref{eq:V}), for different values of $\tau$ in the range $5.0$ to $100.0$ and for (a) $\ell=5$ and (b) $\ell=9$. The dashed line is the value of $V$ in the ground state (i.e., in the adiabatic limit $\tau\rightarrow \infty$), and the vertical line indicates the values of $h_c$: in (a), $h_c=0.25$ and in (b), $h_c=0.13$. }
    \label{fig:Vdyn}
\end{figure}

Figure~\ref{fig:Vdyn} displays the potential $V(t)$ throughout the evolution from $h=0$ at time $t=0$ to $h_\text{f}=1$ at time $t=\tau$.
As shown in Fig.~\ref{fig:Vdyn}(a), for $\ell=5$, the potential grows linearly with $h$ with almost no dependence on $\tau$ for $h<h_c$, while for $h>h_c$, significant deviations are observed among potential values for different $\tau$. For small $\tau$ (large diabaticity), the probability of string breaking is small, 
and the potential exhibits an approximately linear growth, similar to the potential for $h < h_c$. 
For large $\tau$ (closer to adiabaticity), on the other hand, the growth is greatly suppressed for $h>h_c$, and the potential extracted dynamically approximates the static potential computed from the ground state of $H$ in Eq.~\eqref{eq:H}; 
the probability of string breaking is large, and the potential tends to grow with a smaller slope at large $h$. This behavior can be understood by comparing these results with Fig.~\ref{fig:LZsc}(a) and Eq.~(\ref{eq:LZ}): in the range of $\tau$ considered in Fig.~\ref{fig:Vdyn}(a), the population of the first excited state goes from the large value, $P_\text{LZ}\approx 0.82$ (for $\tau=5$), to the very small value, $P_\text{LZ}\approx 0.02$ (for $\tau=100$). The characteristic scale for $\tau$, set by Eq.~(\ref{eq:LZ}) as $\tau_*=4\ell m_*/(\pi \Delta_c^2)$ is, in this case, $\tau_*\approx 25$.

The same analysis for a larger string ($\ell=9$) shows a rather different behavior: for all values of $\tau$ considered (up to $\tau=100$), the $\tau$-independent linear growth continues well beyond $h_c$. The origin of such discrepancy is the exponential dependence of the gap $\Delta_c$ on $\ell$. For the values considered here, 
the characteristic scale for $\tau$ from Eq.~(\ref{eq:LZ}) is $\tau_*=4\ell m_*/\pi \Delta_c^2\approx 1800$. For the values of $\tau$ in Fig.~\ref{fig:Vdyn}(b), the Landau-Zener probability of ending up in the excited state goes from $P_\text{LZ}\approx 0.997$ (for $\tau=5$) to $P_\text{LZ}\approx 0.973$ (for $\tau=100$). In other words, the evolution is completely diabatic given the small size of the gap, and the probability of string breaking (defined as the probability $P_0$ of ending up in the ground state for $h>h_c$) is extremely small. Figure \ref{fig:Vdyn}(b), nonetheless, reveals that the linear growth of the potential is eventually suppressed in the evolution for values of $h$ significantly larger than $h_c$ (larger than the values considered, for example, in the Fig.~\ref{fig:LZsc}). This suppression, therefore, cannot be attributed to the avoided crossing between the string and the broken-string configuration, and cannot be understood using the simple two-level Landau-Zener approach. As we will show below, the suppression of the potential in this case is due to the other states in the spectrum, which correspond to \emph{bubbles} that nucleate and thus break the string. Extending the evolution to larger values of $h$ (compared to the ones in Fig.~\ref{fig:LZsc} and to the static string-breaking point $h_c$) can therefore lead to string breaking even when the gap $\Delta_c$ is prohibitively small.
\begin{figure}[t!]
    \centering
    \includegraphics[width=\linewidth]{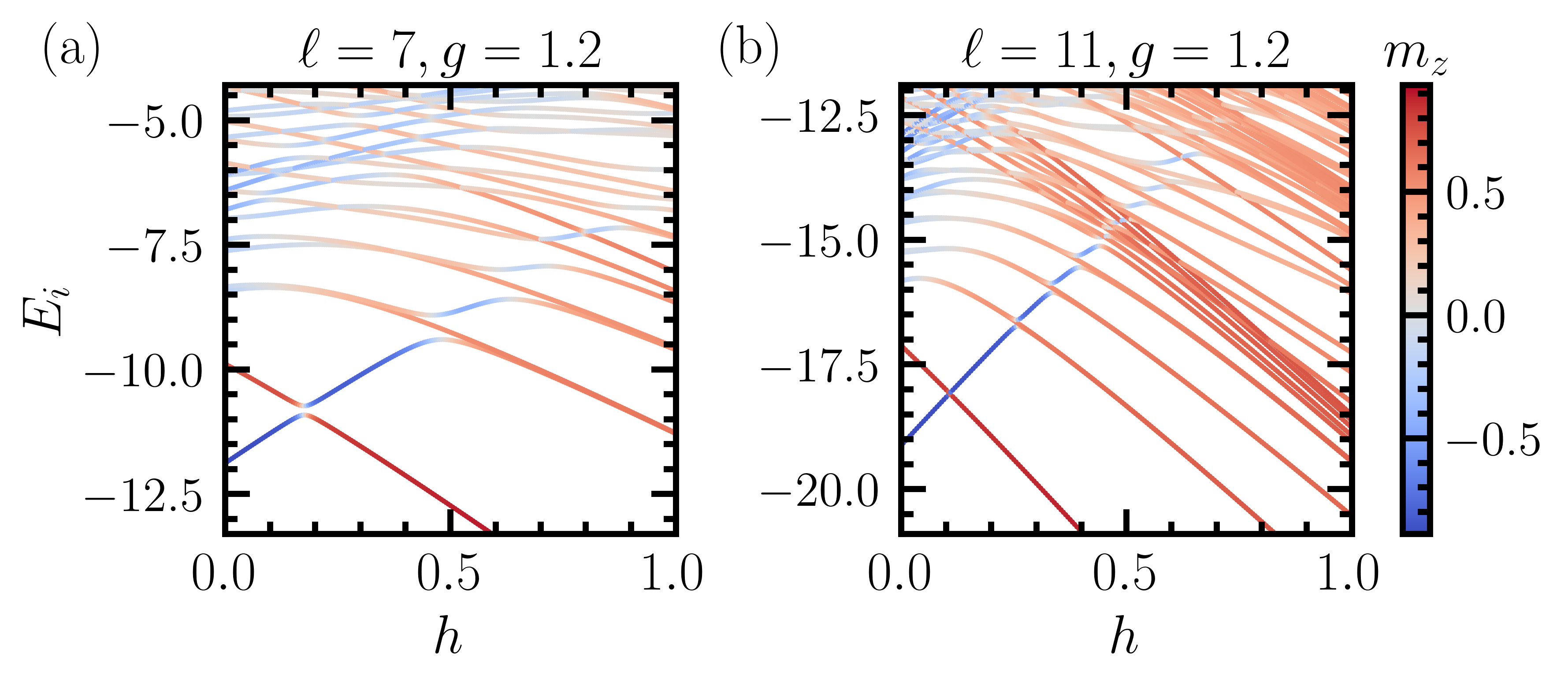}
    \includegraphics[width=\linewidth]{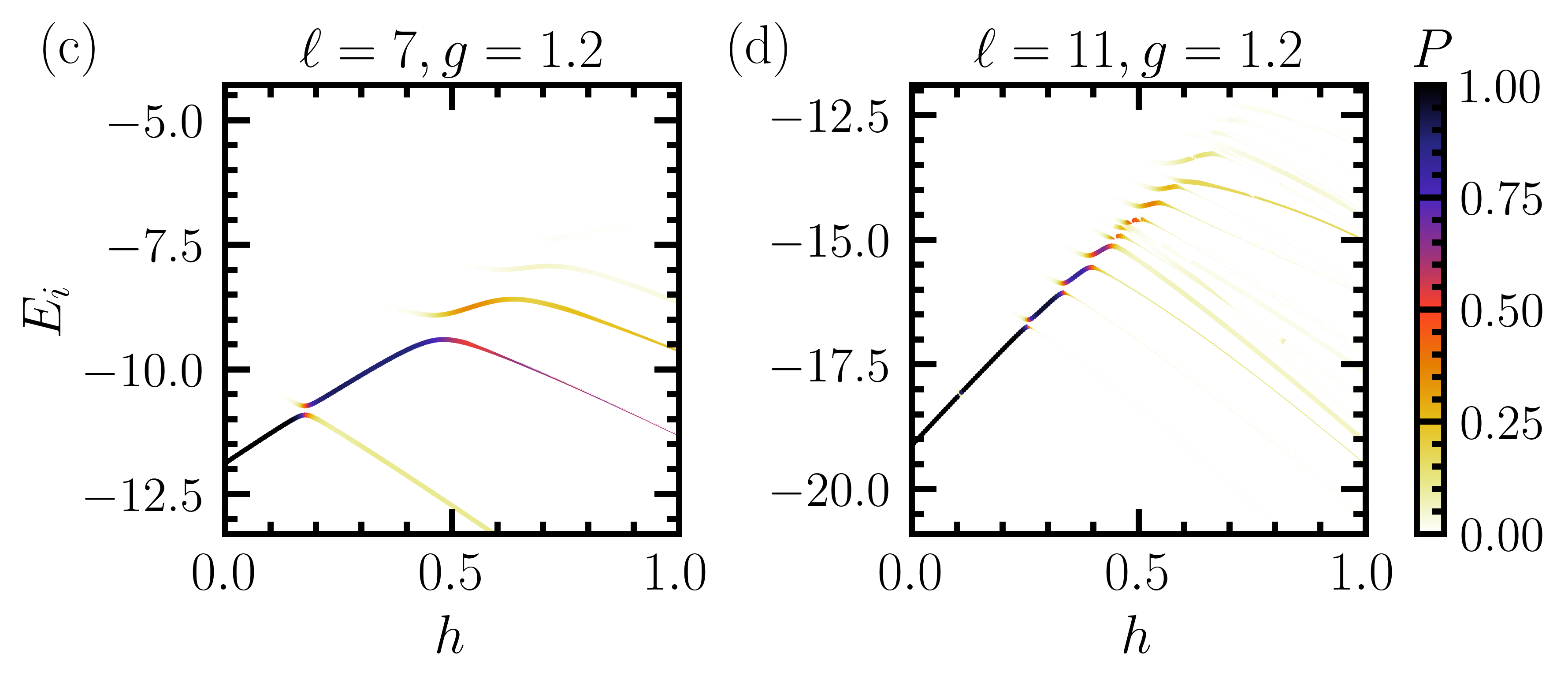}
    \caption{(a,b) Spectrum of $H$ in Eq.~\eqref{eq:H} for the short-range Ising coupling in Eq.~\eqref{eq:Jij} (with $\xi = 1$) as a function of the longitudinal field $h$ for $g=1.2$ and strings of length (a) $\ell=7$ and (b) $\ell=11$. After the first avoided crossing at $h=h_c$, the string state undergoes multiple avoided crossings with other states in the spectrum. (c, d) Population of the eigenstates as the state evolves from the $h=0$ ground state (string state) with a linear $h$ ramp with $\tau=25$. 
    }
    \label{fig:spectrum_multi}
\end{figure}
%

%%%%%%%%%%%%
\subsubsection{Bubble nucleation}
\label{subsec:bubbles}
Our earlier observations demonstrate that, in the dynamical protocol of this work, the Landau-Zener approach and quasi-adiabatic evolution can explain the mechanism of string breaking when short strings are considered. Extending this approach to long strings is challenging due to the exponentially small gaps, which make quasi-adiabatic evolution practically impossible in accessible time scales. To observe string breaking for longer strings, the string tension $h$ needs to be increased further beyond the (static) string-breaking point $h_c$. 
Note that we previously chose $h_f=2h_c$, in which case $h_f$ goes to zero as $\ell^{-1}$ for large $\ell$. We now will evolve from $h=0$ to $h=h_f$, independent of $\ell$. String breaking can then be understood as the effect of transitioning through numerous avoided crossings with other states in the spectrum~\cite{Sinha2021},
as shown in Fig.~\ref{fig:spectrum_multi}. In the limit $g=0$, analyzed in Appendix \ref{app:bubbles}, these states are associated with ``bubbles'' (i.e., domains with opposite magnetization). For $g\neq0$, such bubbles dynamically nucleate in the string. 
Since the size of the flipped domain is smaller than $\ell$, the matrix elements that determine the transition rates from the string state are larger for a bubble state than for the broken string state considered in the previous sections.
These levels can then be populated in the dynamics on practically accessible time scales. As $h$ grows during the evolution, increasingly smaller bubbles become energetically favorable.

\begin{figure}[t!]
    \centering
    \includegraphics[width=\linewidth]{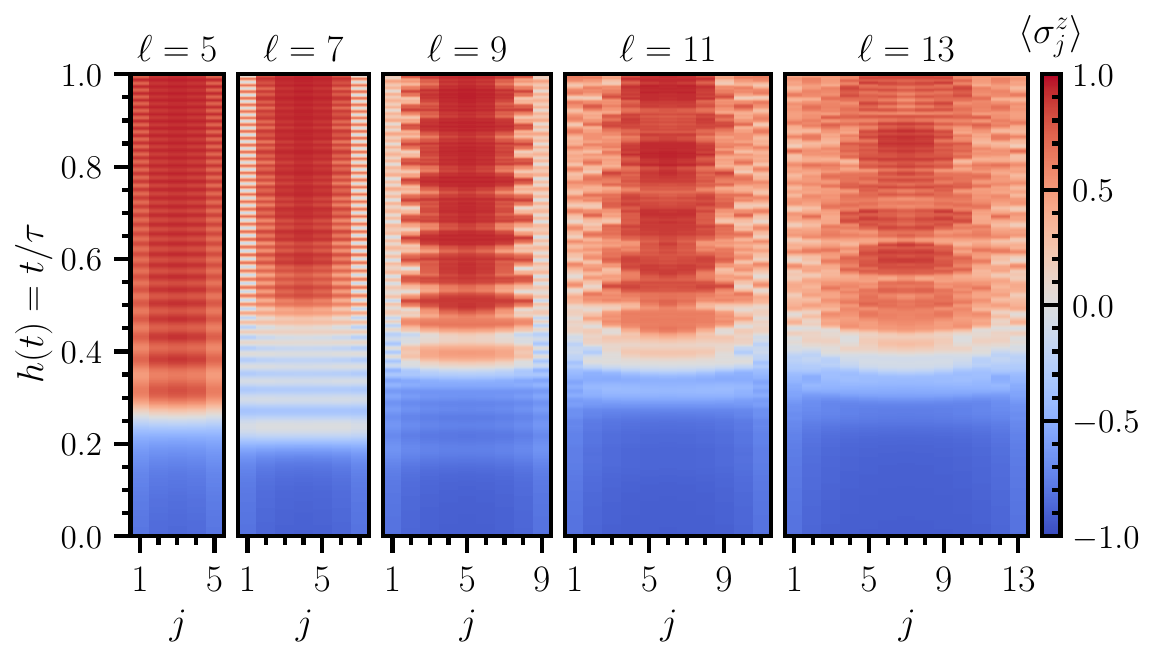}
    \caption{Local magnetization in the time evolution with the linear ramp $h(t)=t/\tau$ for $\tau=100$, $g=1.2$, and string lengths $\ell=5,7,9,11,13$. }
    \label{fig:spatial}
\end{figure}

Interestingly, the inhomogeneous nature of the bubble states has observable effects in the dynamics. In Fig.~\ref{fig:spatial}, we plot the local magnetization as the system evolves from $h(t=0)=0.0$ to $h(t=100)=1.0$ for different lengths $\ell$ of the string. In all the cases considered, the string breaks and magnetization changes sign. For $\ell=5$, the breaking occurs at $h\approx h_c(\ell=5)=0.25$, and the magnetization profile remains spatially homogeneous throughout the evolution. For $\ell=7$, we observe homogeneous oscillations in the magnetization starting at $h\approx h_c(\ell=7)=0.21$: these oscillations demonstrate a partial occupation of the broken string state, but do not completely change the sign of the magnetization, showing that the evolution is far from adiabatic. A more significant growth of the magnetization is observed for $h> 0.5$, after transitioning through the avoided crossing with the next largest bubble [Fig.~\ref{fig:spectrum_multi}(a)], which is of size $6$ (see Appendix~\ref{app:bubbles} for an explanation when $g=0$). As a consequence, since a bubble of size 6 can form either adjacent to the left edge (hence leaving out the rightmost site) or adjacent to the right edge (hence leaving out the leftmost site), the magnetization at the boundary sites 
$j=1$ and $j=7$ differs from that in the bulk.  More complex patterns are observed for $\ell=9,11,13$, suggesting that different types of bubbles are produced. In these cases, the magnetization changes sign at $h\approx 0.4$. We find that this value is almost independent of $\ell$ but changes significantly with the total time of the ramp (the $\tau$ dependence will be studied in Sec.~\ref{sec:scaling}).~\footnote{The $\ell$ independence is expected for large enough strings: there, the local evolution of the magnetization in the bulk (where the effective field $h_j^\text{eff}$ is small) is not affected by the edges (at least, for a certain time, set by the Lieb-Robinson bound \cite{Lieb1972}), and hence does not depend on $\ell$. We note, however, that this argument may not hold for the values of $\ell$ considered here and larger system sizes may be needed to eventually suppress the effect of the boundaries.}

\begin{figure}[t!]
    \centering
    \includegraphics[width=\linewidth]{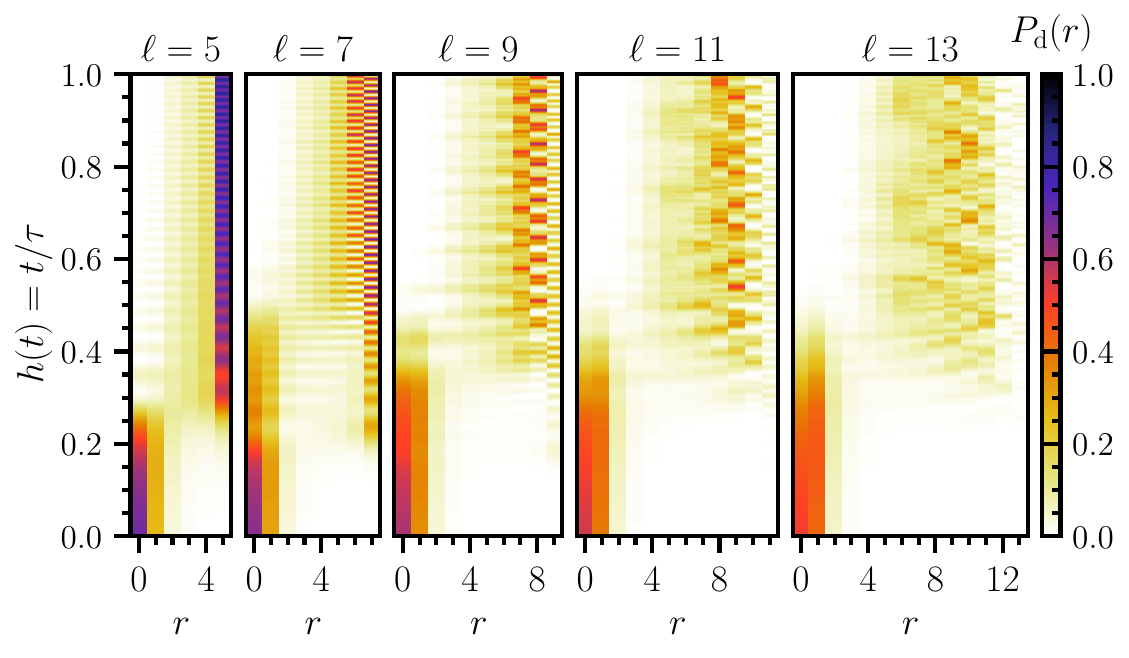}
    \caption{Probability distribution, $P_d(r)$ defined in Eq.~\eqref{eq:Pd}, of the size of the largest domain, $r$, during the time evolution with the linear ramp $h(t)=t/\tau$ for $\tau=100$, $g=1.2$, and string lengths $\ell=5,7,9,11,13$. }
    \label{fig:domain}
\end{figure}

To further characterize the formation of bubbles in the evolution, we analyze the probability distribution of the size of the largest domain. Given a state $\ket{\Psi}$, the probability that the largest domain with positive magnetization in $\ket{\Psi}$ has length $r$ is
\begin{equation}
    P_d(r)=\sum_{\mathbf{s}}\delta_{d(\mathbf{s}), r} |\braket{\mathbf{s}|\Psi}|^2,
    \label{eq:Pd}
\end{equation}
where the sum runs over all bitstring configurations $\mathbf{s}=(s_1,\dots, s_\ell)$ in the $z$ basis and $d(\mathbf{s})$ is the length of the largest domain with positive magnetization in the bitstring $\mathbf{s}$ (i.e., it is the length of the longest sequence of consecutive sites with $s_j=+1$). As shown in Fig.~\ref{fig:domain}, for $\ell=5$, the probability distribution is peaked around $r=0$ ($r=\ell$) before (after) the string breaking. For larger $\ell$, we find that the final probability distribution is much broader and peaked at values $r<\ell$, showing that the string breaks through the formation of smaller bubbles.

We emphasize that, in this regime, bubble nucleation is a genuinely nonperturbative, strongly correlated phenomenon, for which analytical descriptions such as those developed in Ref.~\cite{Sinha2021} break down. In particular, we are far from both the low-density regime of bubbles ($g\tau^2 \ll 1$) and the regime dominated by small bubbles of length $r=1$ ($g\ll 1$) explored in that work. Instead, bubbles are not dilute and span a broad range of sizes, as illustrated in Fig.~\ref{fig:domain}. A key qualitative consequence is that the density of bubbles of length $r=1$  \emph{decreases} with increasing $\tau$ (see Appendix \ref{app:nucleation}), in stark contrast to the behavior found in the regimes discussed in Ref.~\cite{Sinha2021}. In our regime, slower ramps promote the growth of extended bubbles rather than the proliferation of isolated $r=1$ bubbles, which are suppressed at larger $\tau$. 

%%%%%%%%%%%%
\subsubsection{Scaling}
\label{sec:scaling}
For long strings and accessible evolution times, string breaking cannot be understood as a simple Landau-Zener process. The aim in this subsection 
is to characterize the $\tau$ dependence of the observed dynamics in this more complex regime.
\begin{figure}[t!]
    \centering
\includegraphics[width=\linewidth]{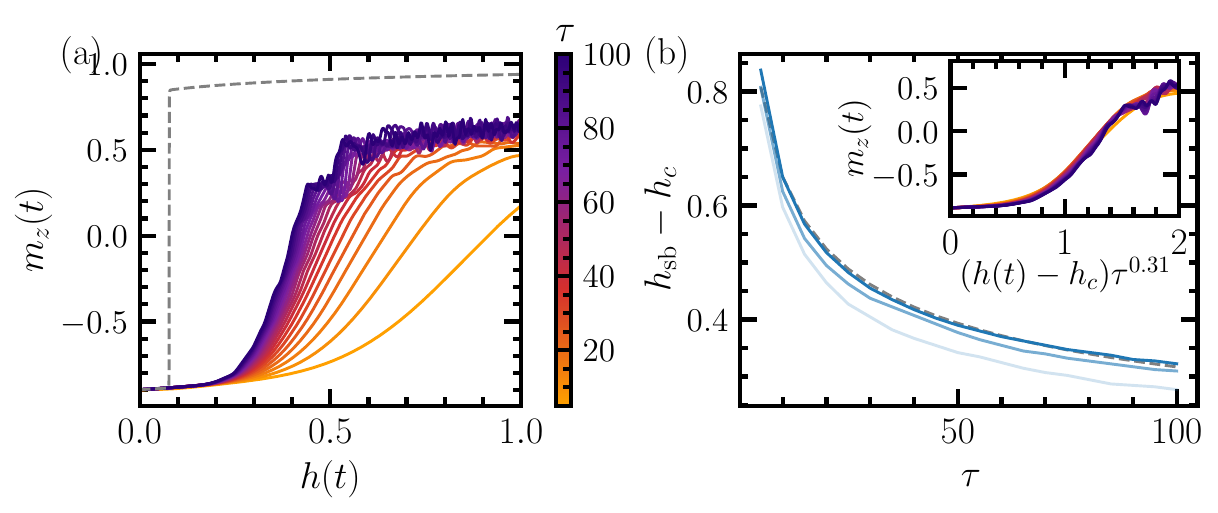}
    \caption{(a) Time evolution of the average magnetization $m_z$ for $g=1.2$ and $\ell=15$. In the adiabatic limit $\tau\rightarrow \infty$ (dashed gray line), the magnetization changes sign abruptly at $h_c=0.078$. (b) $h_{\text{sb}}$ is the value of $h(t)$ where $m_z(t)=0$. The difference $h_{\text{sb}}-h_c$ is plotted here as a function of $\tau$ for $\ell=11,13,15$, with increasing intensity of color (from light to dark) corresponding to larger system sizes. The dotted gray curve is the fit $h_\text{sb}-h_c\approx 1.3\times \tau^{-0.31}$ of the $\ell=15$ data. Inset: collapse of the time evolution of the average magnetization plotted as a function of $(h(t)-h_c)\tau^{0.31}$ for $\ell=15$. }
    \label{fig:collapse}
\end{figure}

Figure~\ref{fig:collapse}(a) displays the magnetization in the evolution with the linear ramp $h(t)=t/\tau$ for different values of $\tau$ for a system of $\ell=15$ dynamical spins. For the range of $\tau$ considered here, the magnetization changes sign at some value $h_\text{sb}(\tau)<h_\text{f}$. In Fig.~\ref{fig:collapse}(b), we plot $h_\text{sb}-h_c$ as a function of $\tau$. The dependence is well approximated by a power law (see Appendix \ref{app:nucleation}): a fit yields $h_\text{sb}-h_c\approx 1.3 \, \tau^{-0.31}$. The inset in Fig.~\ref{fig:collapse}(b) shows that, by appropriately rescaling the 
$x$ axis in Fig.~\ref{fig:collapse}(a), it is possible to find a good collapse of the curves obtained for different $\tau$ onto one curve. We conjecture that such scaling properties may emerge close to the critical point~\cite{Pelissetto2020,pelissetto2023scaling}.  The conjecture is motivated by the following physical picture. At the critical point of the model, the correlation length diverges. It is well known that ramps across a continuous phase transition lead to universal scaling behavior described by the Kibble–Zurek mechanism, where the correlation length cannot grow indefinitely during the dynamics but instead “freezes out” at a value determined by the ramp speed and universal critical exponents.
In our case, the system undergoes a first-order transition that becomes weakly first order close to the critical point. Although the correlation length does not diverge in this regime, it can still become very large. As a result, when the ramp crosses the transition sufficiently close to criticality, the dynamics may be unable to adiabatically follow the instantaneous ground state, leading to a freeze-out mechanism qualitatively similar to that occurring at continuous transitions. This provides the physical basis for our conjecture that scaling behavior may emerge in this regime. Our numerical results indicate that the extracted exponents are not universal, but instead depend smoothly on the transverse field $g$ (although we expect that universal critical exponents will be recovered for $g\rightarrow g_*$). To understand whether the conjecture applies here, we need a more systematic study of the correlation length and more extensive simulations of the dynamics for different values of $g$, which we leave for future work. 

We note that the approach based on describing the dynamics in terms of individual level crossings, as demonstrated in Ref.~\cite{Sinha2021}, is instead expected to hold in the perturbative regime of small transverse field, away from criticality. Bridging these two distinct regimes---the perturbative small-$g$ limit, where a description in terms of individual level crossings is expected to apply, and the near-critical regime characterized by large correlation lengths and collective dynamics---remains an open problem.

%%%%%%%%%%%%%%%%%%%%%%%%%%%%%%%%%%%%%%%%%%%%%%%%%%%%%%%%%%%%%%%%%%%%%%%%%%%%%%%%%%%%%%%%%%%%%%%%%%%%%%%%%%%%%%%%%%%%%%%%%%%%%%%%%%%%%%%%%%%%%%%%%%%%%%%%%%%%%%%%%%%%%%%%%%%%%%%%%%%%%%%%%%%%%%%%%%%%%%%%%%%%%%%%
\section{String breaking dynamics with power-law Interactions}
\label{sec:longrange}
\begin{figure*}[t!]
    \centering
      \includegraphics[width=\linewidth]{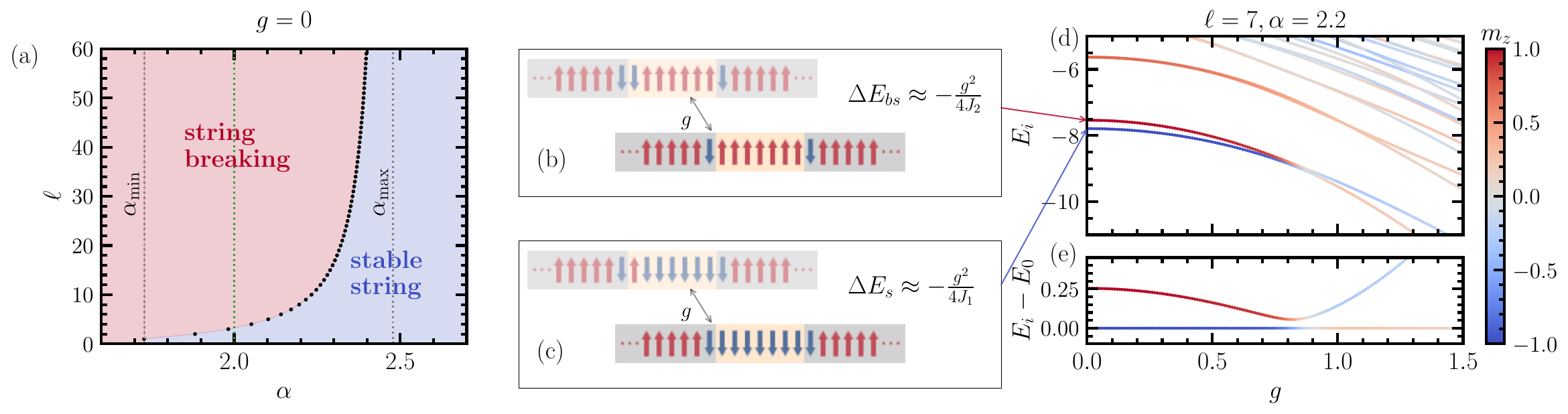}
    \caption{(a) String-breaking length $\ell_c$ as a function of $\alpha$ for $g=0$. The black dots indicate, for each integer $\ell$, the value of $\alpha$ for which $E_\text{bs}=E_\text{s}$. The blue (red) region indicates the values of $\ell$ and $\alpha$ where the string is stable (breaks). The string-breaking length diverges at the vertical dotted line corresponding to $\alpha=\alpha_\text{max}$, and no string breaking occurs for $\alpha>\alpha_{\text{max}}$. For $\alpha<\alpha_\text{min}$ (vertical dotted line) the string is unstable for every $\ell$. The green dotted line $\alpha=2$ marks the boundary between the confined ($\alpha \le 2$) and deconfined ($\alpha>2$) regimes. (b) In the large $\alpha$ limit, the process of flipping the spin $j=1$ gives a correction $\Delta E_{bs}\approx -g^2/(4J_2)$ to the energy of the broken string in second-order perturbation theory. (c) In the same limit, the process of flipping the spin $j=1$ gives a correction $\Delta E_s \approx -g^2/(4J_1)$ to the energy of the string in second-order perturbation theory. (d) Low-energy spectrum with $\alpha=2.2$ and $\ell=7$ as a function of the transverse-field strength $g$. (e) Energy difference 
    $E_i-E_0$ as a function of $g$, where $E_0$ denotes the ground-state energy.
    }
    \label{fig:lr}  
\end{figure*}
We now shift our discussion to the case of power-law decaying interactions. The setup is the same as in Fig.~\ref{fig:1}(a), described by the Hamiltonian in Eq.~(\ref{eq:H}), with the effective field $h_j^\text{eff}$ in Eq.~(\ref{eq:Beff}) that accounts for the interaction with the external, non-dynamical spins. Instead of the exponentially decaying interactions in Eq.~(\ref{eq:Jij}), we now assume a power-law decay of the form
\begin{equation}
\label{eq:pl}
    J_{i,j}\equiv J_{|j-i|}=|j-i|^{-\alpha}
\end{equation}
with $\alpha>1$.
In this case, the effective longitudinal field takes the form
\begin{equation}
\label{eq:BeffLR}
    h_j^\mathrm{eff}=-2\zeta(\alpha)+j^{-\alpha}+(\ell-j+1)^{-\alpha} +\sum_{k=1}^j k^{-\alpha}+\sum_{k=1}^{\ell-j+1} k^{-\alpha},
\end{equation}
where $\zeta(\alpha)=\sum_{n=1}^\infty n^{-\alpha}$ is the Riemann zeta function.

Interactions of the form in Eq.~(\ref{eq:pl}) describe emergent Ising couplings in certain trapped-ions setups~\cite{MonroeReview}, as well as in Rydberg-atom arrays~\cite{browaeys2020many,Saffman2010}, 
magnetic atoms~\cite{norcia21},
NV centers~\cite{cai2013large}, and polar molecules~\cite{Gorshkov2011,yan2013observation}. Systems with long-range interactions are known to exhibit, among other phenomena, a peculiar non-ballistic spreading
of quantum correlations, a counter-intuitive slowdown of entanglement dynamics, suppression
of thermalization and equilibration, anomalous scaling of defects upon traversing criticality (of
relevance here), dynamical phase transitions, and genuinely non-equilibrium phases stabilized
by periodic driving \cite{Defenu2023,DEFENU20241}. 
As was shown in Refs.~\cite{Liu2019,Lerose2019,tan2021domain}, for $1<\alpha \le 2$, the model exhibits domain-wall confinement even in the absence of the homogeneous longitudinal field $h$: the effective potential as a function of distance grows sub-linearly ($V(\ell)\sim \ell^{2-\alpha}$) for $1<\alpha < 2$, and logarithmically ($V(\ell)\sim \log\ell$) for $\alpha=2$
~\footnote{This property underlies the stabilization of finite-temperature order in this model~\cite{dyson1969existence}},~\footnote{For $0\le \alpha\le1$, the notion of locality, and hence domain-wall confinement, is not well-defined~\cite{Liu2019,Lerose2019}.}.
 Note that this kind of sub-linear confinement is associated with a non-standard notion of string tension, that changes significantly with the length.
For $\alpha>2$, on the other hand, the model is deconfined: the potential grows before eventual saturation, with a saturation value that diverges as $\alpha$ approaches $2$ from above. 
As we will show in this section, string breaking occurs in this model even in the absence of $h$, arising solely from the effect of quantum fluctuations. We will therefore set $h=0$ throughout this section.

%%%%%%%%%%%%
\subsection{String-breaking statics}
Similar to Sec.\ \ref{sec:dyn}, we start by analyzing the statics of the string and the broken string in the long-range model, before moving on to a nonequilibrium scenario.

%%%%%%%%%%%%
\subsubsection{Case $g=0$}
To understand the string-breaking effect for the case of power-law interactions, it is useful to first focus on the case $g=0$. The energies $E_s$ and $E_{bs}$ of the string and broken-string states, respectively, can be
straightforwardly computed, yielding
\begin{equation}
    E_{bs}-E_s = 2 \sum_{j=1}^\ell  h_j^\text{eff}=-4\ell \zeta(\alpha)+4\sum_{j=1}^\ell (\ell-j+2)j^{-\alpha}.
\end{equation}
The sign of $E_{bs}-E_s$ depends on $\alpha$ and $\ell$ in a nontrivial way. Figure~\ref{fig:lr}(a) displays this dependence: the blue region indicates the values of $\ell$ and $\alpha$ where the string is stable ($E_{bs}>E_s$), while the red region indicates the values where the string breaks ($E_{bs}<E_s$). In particular, we observe that, for $1 < \alpha<\alpha_\text{min}=1.72865$ [obtained as the solution of $\zeta(\alpha_\text{min})=2$], the string is broken for every $\ell\ge 1$.  For $\alpha>\alpha_\text{max} =2.4787793$ [where $\alpha_\text{max}$ solves $2\zeta(\alpha_\text{max})=\zeta(\alpha_\text{max}-1)$], the string is stable for every $\ell\ge 1$.
In the intermediate regime $\alpha_\text{min}< \alpha < \alpha_\text{max}$, there is an integer $\ell_c$ such that the string is stable for $\ell\le\ell_c$ and breaks for $\ell>\ell_c$.

Note that the domain walls are genuinely confined only for $1<\alpha\le 2$, while for $\alpha>2$, they can be isolated at the price of a finite energy cost \cite{Liu2019,Lerose2019,tan2021domain}. 
%$V(\infty)\sim 1/(\alpha-2)$, 
%which grows unbounded as $\alpha$ approaches $2$ from above.
For $2<\alpha<\alpha_{\rm{max}}$, this finite energy is large enough to make the broken-string state energetically favorable. In particular, the value $\alpha_{\rm{max}}$ is realized when the string-breaking length diverges, i.e., when two isolated domain walls (corresponding to a very long string) have the same energy as two isolated mesons (which can be interpreted as a long broken string). Accordingly, the above analysis shows that string breaking occurs even for those values of $\alpha$ for which domain walls are not confined.

%%%%%%%%%%%%
\subsubsection{Case $g \neq 0$: string breaking from quantum fluctuations \label{sec:quantumfluc}}
Let us now consider the effect of a nonzero transverse field $g$. We will show that the transverse field can destabilize a string and induce string breaking. To understand this phenomenon, it suffices to study a perturbative regime where $g$ is small. We consider values of $\ell$ and $\alpha$ such that the ground state has a string for $g=0$. The transverse field $g\ll 1$ gives a correction to the energy of the string and broken-string states in second-order perturbation theory. 

All processes that flip one spin contribute to the perturbative shift in second-order perturbation theory.
The second-order corrections to the string and broken-string energies from these processes read
\begin{equation}
\label{eq:Es2}
    E_{s}(g)\approx E_{s}(g=0)-\frac{g^2}{2}\sum_{j=1}^L \frac{1}{\tilde h_j+h_j^\text{eff}},
\end{equation}
\begin{equation}
\label{eq:Ebs2}
    E_{bs}(g)\approx E_{bs}(g=0)-\frac{g^2}{2}\sum_{j=1}^L \frac{1}{\tilde h_j-h_j^\text{eff}},
\end{equation}
where we have defined
\begin{equation}
    \tilde h_j = \sum_{k=1,\dots, \ell; k\neq j} J_{j,k}.
\end{equation}
If the condition
\begin{equation}
\label{eq:cond}
\sum_j \left(\tilde h_j-h_j^\text{eff}\right)^{-1}> \sum_j\left(\tilde h_j+h_j^\text{eff}\right)^{-1}
\end{equation}
is satisfied, the energy of the broken string state decreases faster with $g$ than the one of the string state, and the two levels can cross upon increasing $g$. It should be noted that the matrix element connecting the string and broken-string states is much smaller than the second-order corrections computed here, appearing only at order $g^\ell$ in perturbation theory. This strong separation of scales is crucial for the argument, as it ensures that the avoided crossing remains sharp over a wide range of $g$.

While the condition Eq.~(\ref{eq:cond}) can be explicitly checked for any values $\alpha$ and $\ell$ of interest, it is useful to consider a simple argument to verify it for a sufficiently large $\alpha$.
For the string state, every spin flip increases the number of domain walls [Fig.~\ref{fig:lr}(c)], so the energy cost is larger than $4J_1$ for every spin.
 For the broken-string state, on the other hand,  
flipping a spin adjacent to the boundary
does not change the number of domain walls, and the energy cost of flipping one such spin is roughly given by $4J_2$, where $J_2$ is the next-nearest-neighbor interaction. This process has a lower energy cost than $4J_1$ and therefore contributes a larger negative perturbative correction to the energy of the broken-string state. Processes that flip the other spins change the number of domain walls and thus 
cost at least an energy $4J_1$, similarly to the case of the string state.

\begin{figure}
    \centering
    \includegraphics[width=\linewidth]{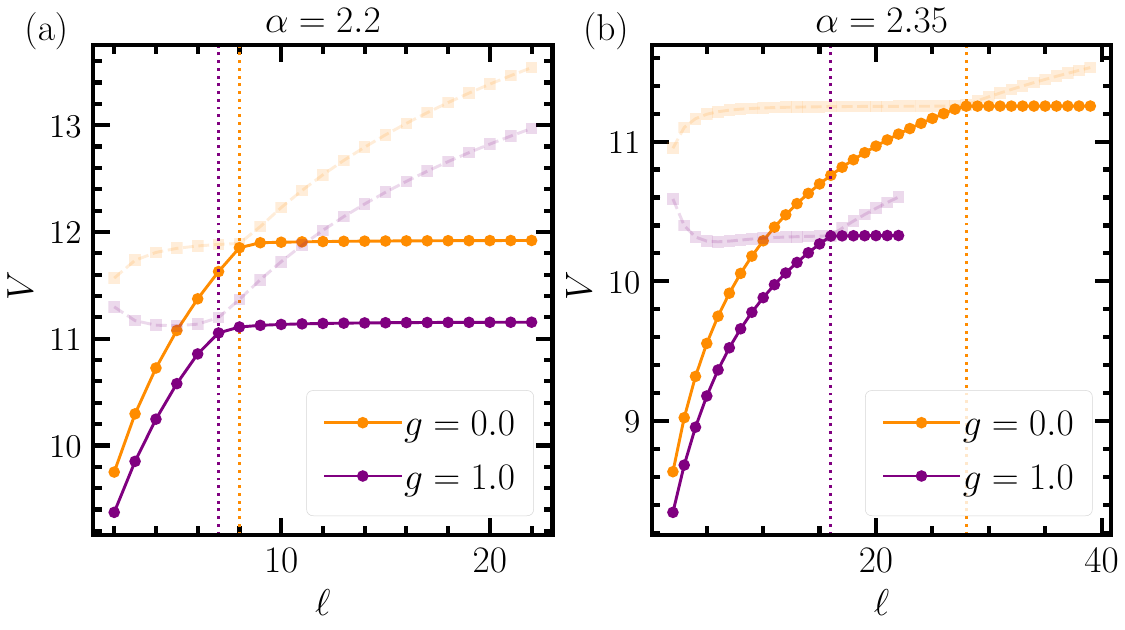}
    \caption{Potential of the string $V(\ell)$, defined as the excess energy $E_0(\ell)-E_0^{\rm{vac}}(\ell)$ between the ground state in the presence and in the absence of static charges, shown by the solid lines for (a) $\alpha=2.2$ and (b) $\alpha=2.35$. 
    The dashed (semi-transparent) lines show the excess energy $E_1(\ell)-E_0^{\rm{vac}}(\ell)$ of the first excited state, demonstrating that the point where the potential's slope discontinuously changes 
    can be understood as a level crossing between the ground and first excited states. The crossings at $\ell_c$ (marked by dotted vertical lines) underlie the crossover from a string-type potential to a meson-meson-type potential, which is the hallmark of string breaking. If $\ell$ is between the dotted vertical lines, then changing $g$ from $0$ to $1$ induces string breaking. }
    \label{fig:LRV}
\end{figure}
This argument suggests 
that the two energy levels can cross upon increasing $g$, i.e., it is possible to have string breaking generated by increased quantum fluctuations. Even for strings that are classically stable, i.e., $E_{bs}(g=0)-E_s(g=0)>0$, it is possible to have string breaking at some value of $g=g_{c}$. This is illustrated in Fig.~\ref{fig:lr}(d), where we plot the low-energy spectrum obtained with exact diagonalization: for $\alpha=2.2$, a string of length $\ell=7$ is classically stable, but an avoided level crossing occurs at $g=g_{c}\approx 0.8$. The energy difference between the energy levels in the spectrum and the ground-state energy is plotted as a function of $g$ in Fig.~\ref{fig:lr}(e). String breaking is manifested in the sudden change of magnetization $m_z$ of the ground state, with $m_z$ going from negative to positive values at the avoided level crossing.
We note that the avoided crossing occurs at $g \approx 0.8$, where $g$ is not particularly small and the validity of perturbation theory is not guaranteed. Nevertheless, we find that the second-order estimates in Eqs.~(\ref{eq:Es2}) and (\ref{eq:Ebs2}) remain good approximations up to $g \approx 1$, allowing the crossing point to be predicted with remarkable accuracy. 

The effect of the transverse field in inducing string breaking can also be observed by plotting the potential of the string as a function of string length, as shown in Fig.~\ref{fig:LRV}. At short distances, the potential grows as $V(\ell) \approx V^\infty_s-c_s/\ell^{\alpha-2}$ (where $c_s$ and $V^\infty_s$ are coefficients depending on $\alpha$ but not on $\ell$). When string breaking occurs, the potential of the string levels off
for $\ell>\ell_c$. We note that, because of the long-range interactions, the potential keeps slowly growing with $\ell$ until saturation, as $V\approx V^\infty_{bs}-c_{bs}/\ell^{\alpha}$ (where $c_{bs}$ and $V^\infty_{bs}$ are $\ell$-independent coefficients). 
The string-breaking point $\ell_c$ (dotted vertical lines) decreases with increasing $g$, confirming the observation that the transverse field can enhance the tendency of the string to break. 

We remark that string breaking induced by quantum fluctuations (that is without any longitudinal field) is possible, in principle, also for the exponentially decaying interactions in Eq.~(\ref{eq:Jij}). A requirement, in that case, is that the difference $E_{bs}-E_s$ in Eq.~(\ref{eq:DeltaE}) is small for $h=0$. Then, the energy difference can be overcome by the perturbative corrections from the transverse field. If the energy difference is too large, the large transverse field needed drives the system into the paramagnetic phase. In order to have a sufficiently small energy difference in Eq.~(\ref{eq:DeltaE}), the range $\xi$ of the interactions should be close to $1/\log(2)$ (but still smaller, otherwise the string is trivially broken for $h=0$ and $g=0$).

\subsection{String-breaking dynamics}
\begin{figure}
    \centering
    \includegraphics[width=\linewidth]{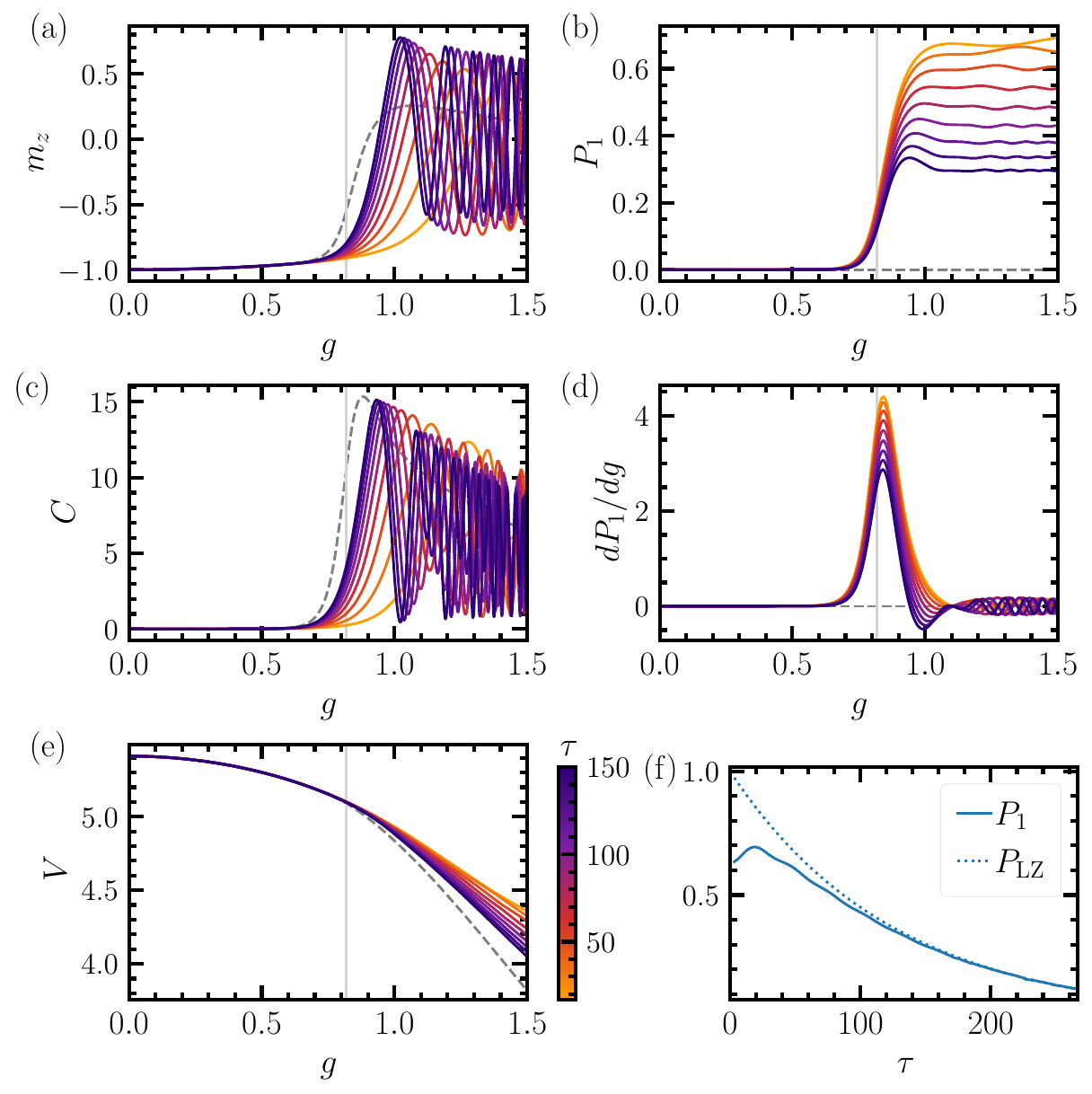}
    \caption{Time evolution as a function of $g(t)=t/\tau$ for different values of $\tau$ (the dashed gray line represents the adiabatic limit $\tau\rightarrow\infty$). The length of the string is $\ell=7$, and interactions decay with a power $\alpha=2.2$. (a) Magnetization along $\hat{z}$ [Eq.~(\ref{eq:mx})]. (b) Population $P_1$ of the (instantaneous) first excited state [Eq.~(\ref{eq:Pn})]. (c) Connected spin-spin correlator [Eq.~(\ref{eq:C})]. (d) $\mathrm{d}P_1/\mathrm{d}g$. (e) Potential, $V$, of the static charges in the dynamical protocol, as defined in Eq.~(\ref{eq:V}). (f) Final values of populations after the ramp from $g=0$ to $g=g_\text{f}=1.5$. The solid line denotes population $P_1$ of the first excited state, while the dotted line denotes the population obtained from the Landau-Zener formula in Eq.~(\ref{eq:LZ}).}
    \label{fig:LR2}
\end{figure}
We now consider a dynamical protocol to probe string breaking, similar to the one considered in Sec.~\ref{sec:dyn}, but in the case where the breaking is induced by the transverse field $g$.
We are interested in observing string breaking as $g$ is linearly increased from $g=0$ to $g>g_c$. The initial state is a product state polarized along $-\hat{z}$. It is evolved with the Hamiltonian in Eq.~(\ref{eq:H}), where $h$ is set to zero, and the effective field $h_j^\text{eff}$ is given in Eq.~(\ref{eq:BeffLR}). The transverse-field strength $g$ is set to vary in time as
\begin{equation}
    g(t) =t/\tau.
\end{equation}
Figure~\ref{fig:LR2} displays the results of the time evolution from time $t=0$ to $t=1.5 \, \tau$ as a function of $g(t)$ for different values of $\tau$. 

The dynamics of the magnetization $m_z$ and of the population of the (instantaneous) first excited state $P_1$ [Fig.~\ref{fig:LR2}(a,b)] show, similarly to the case discussed in Sec.~\ref{sec:dyn}, that the evolution remains almost adiabatic for all values of $\tau$ for $g<g_c$, while significantly different behavior is observed after the avoided level crossing as a function of $\tau$.
The final probabilities at the end of the linear ramp are in good agreement with the ones obtained from a Landau-Zener prediction, similar to the one in Eq.~(\ref{eq:LZ}) [Fig.~\ref{fig:LR2}(f)]. Furthermore, $d P_1/dg$ displays a maximum at the crossing point $g_c$ [Fig.~\ref{fig:LR2}(d)], while $g$ corresponding to the first peak of the correlator $C$ shows more significant deviations from $g_c$ [Fig.~\ref{fig:LR2}(c)] (nevertheless, the value $g_c$ can be obtained from extrapolating the adiabatic limit $\tau\rightarrow \infty$). A qualitative difference with respect to the case studied in Sec.~\ref{sec:dyn} is observed in the evolution of the string potential $V$ [Fig.~\ref{fig:LR2}(e)], as the potential decreases with $g$. Indeed, while $h$ plays the role of a string tension, with the potential of the string growing proportionally to $h\ell$, a similar interpretation does not apply to the transverse field $g$.

%%%%%%%%%%%%%%%%%%%%%%%%%%%%%%%%%%%%%%%%%%%%%%%%%%%%%%%%%%%%%%%%%%%%%%%%%%%%%%%%%%%%%%%%%%%%%%%%%%%%%%%%%%%%%%%%%%%%%%%%%%%%%%%%%%%%%%%%%%%%%%%%%%%%%%%%%%%%%%%%%%%%%%%%%%%%%%%%%%%%%%%%%%%%%%%%%%%%%%%%%%%%%%%%
\section{Conclusions}
\label{sec:conclusions}
In this work, we examine the real-time dynamics of a quantum spin chain to unveil the mechanisms behind string breaking. Our study involves a protocol where the string tension undergoes a gradual increase in time, ultimately exceeding the critical value associated with the celebrated ground-state string breaking. The protocol we propose goes beyond the dynamics following a quantum quench. Quenches have been the most studied protocol to date to induce string breaking in quantum spin chains~\cite{SALA2018,Verdel19_ResonantSB,Verdel2023}. Our non-instantaneous protocol, on the other hand, probes string breaking in a more controlled setting, by drawing connections to the system's spectrum at low energies. 
It further serves as a probe of static string breaking, allowing for the experimental measurement of the string-breaking point and the properties of the crossover.
Concretely, the adiabatic limit recovers ground-state string breaking, while diabatic processes show a rich interplay between energetics and dynamics of allowed transitions throughout the evolution.
For short enough strings, we describe the dynamical breaking phenomenon as a Landau-Zener process, marked by a tunable population of the string and broken-string states as they cross in energy from the ground state to the first excited state (a similar analysis was presented in Ref.~\cite{Pelissetto2020}). In contrast, longer strings exhibit a completely diabatic crossing under realistic ramp time scales. In this case, string breaking occurs at higher string-tension values due to crossings with higher-energy levels, going beyond previous studies based on a single Landau–Zener process \cite{Pelissetto2020}. We establish that this intricate breaking process involves the formation of a superposition of reversed domains (``bubbles''), resulting in the fragmentation of the string into pieces rather than a singular, complete break.

For the case of power-law decaying interactions, we observe dynamical string breaking even in regimes where domain walls are not confined, i.e., for values of $\alpha$ where the static potential does not diverge with the distance. This shows that  string breaking can occur dynamically without the presence of confinement in this setting. We identify string breaking driven purely by quantum fluctuations, in the absence of an external field or thermal effects. This mechanism differs qualitatively from classical or semiclassical intuition and reflects genuine quantum many-body effects.

In the following, we draw a few parallels between string breaking in our work and closely related phenomena in other contexts. We also point out new directions that can result from the present study.
\begin{itemize}
\item[$\circ$]
\emph{False-vacuum decay and bubble nucleation.}
Bubble nucleation, expansion, and collision are a hallmark of first-order phase transitions, and are conjectured to play a role in the early universe~\cite{Hawking:1982ga,Coleman:1977py,Callan:1977pt}. Notably, for long strings, the concept of string breaking in one-dimensional systems parallels a first-order phase transition, and the dynamics is analogous to a false-vacuum decay~\cite{Sinha2021,LagneseFVD,lagnese2023detecting,Darbha2024a,Darbha2024b,vodeb2024stirring}. Similar methods can, therefore, be applied to both problems. For example, bubble nucleation in a false vacuum decay can be examined as a two-level (at low bubble densities) or multilevel (at high densities) Landau-Zener transition~\cite{Sinha2021}. While this approach can quantitatively capture the formation of bubbles of various sizes in some parameter regimes, the hoppings and interactions of the bubbles may play an interesting role in other regimes, as demonstrated, for example, in a recent experiment in a quantum annealer \cite{vodeb2024stirring}. These effects are expected to be especially relevant far from the perturbative limit of small transverse fields and closer to the scaling limit of the Ising field theory. The formation of bubbles from the false vacuum has also been studied for a quench setup~\cite{lagnese2023detecting}. It would be interesting to generalize those results to the case of the linear ramp of this work. 
The complex string-breaking process observed for longer strings in this work provides another unique lens into the dynamics of bubble formation, going beyond previous studies that examined the regime of small transverse fields and long evolution times~\cite{Sinha2021,vodeb2024stirring}. Our analysis reveals that the regime of moderately large transverse field exhibits qualitatively different behavior compared to the perturbative small–transverse-field limit. Although achieving an analytical, even approximate, understanding of this regime is highly challenging, the emergence of clear scaling laws strongly suggests the presence of a distinct and physically rich regime, that is well suited to exploration with quantum simulators. Our study, therefore, paves the way for characterizing the phenomenon of bubble nucleation, including size distributions, bubble positions, and correlations between different bubbles, in novel regimes, with direct experimental relevance. Such studies may be informed by scaling laws inherited from the vicinity of the critical point, resulting in universal behavior. These represent a compelling direction for future work.

\item[$\circ$]
\emph{Parallels to string breaking in particle collisions.} An overarching goal is to shed light on string-breaking dynamics in the much more complex setting of high-energy particle colliders. There, dynamics are primarily governed by quantum chromodynamics, a non-Abelian gauge theory in 3+1 dimensions. The model studied in this work can be shown to have a 1+1-dimensional Abelian gauge-theory dual~\cite{balian1975gauge,zhang2018quantum,Lerose2020,borla2020gauging,Surace_2021,exp1}, hence offering a connection to gauge-theory dynamics, although in a substantially simpler setting. Importantly, time-dependent string tension has parallels to the time-dependent length of an expanding string resulting from a collision~\cite{andersson1983parton},
or to the effect of a time-dependent temperature on the string~\cite{Huntsmith2020}. The formation and evolution of broken-string fragments, both the edge mesonic pairs produced out of a fully adiabatic string breaking or other particle excitations produced in diabatic transitions, may provide insights into inelastic particle production in high-energy collisions~\cite{Surace_2021,belyansky2024high,bennewitz2024simulating,su2024cold,su2024cold}.
In particle colliders, dynamical charges can also form strings. Dynamical probe charges can be studied in a similar setup: an unbroken string remains localized for very long times~\cite{Verdel19_ResonantSB,Lerose2020}, so its evolution is essentially the same as for the case of static charges. Once the string has broken, however, its smaller components are free to move \cite{Verdel19_ResonantSB,ch2019confinement,SALA2018}, leading to qualitatively different dynamics.

\item[$\circ$]
\emph{String breaking as a thermodynamic process and thermalization dynamics.} Quench, diabatic, and adiabatic protocols represent different thermodynamic processes. It is, therefore, interesting to explore the connection between quantum dynamics of string breaking, including bubble formation and evolution, as explored in this work, and the associated thermodynamic quantities such as internal energy, work, heat, and free energy in each process. Steps toward this goal have been taken recently for quench processes in strongly-coupled quantum many-body systems~\cite{davoudi2024quantum}, but need to be generalized to be applicable to the setting of this work. Furthermore, the possible occurrence of thermalization (or prethermalization) during evolution for sufficiently slow ramps can lead to a progression through successive equilibrium (or quasi-equilibrium) states of the instantaneous Hamiltonian, opening up new possibilities for investigating thermalization dynamics of confining theories in a controlled setting.

\item[$\circ$]
\emph{Experimental realization.} Our results find immediate applicability in the exploration of string-breaking dynamics using quantum simulators. Specifically, our protocols are readily suited to trapped-ion experiments, as demonstrated in Ref.~\cite{tan2021domain}, where signatures of composite excitations and confinement dynamics were already observed. The possibility of engineering a range of spin-spin interactions with both exponentially decaying and power-law interactions makes experimental implementations of string breaking in both scenarios feasible in trapped-ion systems. The inhomogeneous profile of the longitudinal field can be engineered with individual addressing beams~\cite{exp1,exp2}. As we have shown, all the key signatures of the physics under discussion are already clearly visible for system sizes up to $L=15$, directly comparable to those accessible in current experiments. Alternatively, a larger system, including the frozen or dynamical external sites, can be realized directly, such that no additional inhomogeneous field is required. 
A trapped-ion experimental demonstration of the protocol studied in this work is, in fact, underway~\cite{exp2}. Other potential experimental platforms include Rydberg-atom arrays and ultracold atoms in optical lattices, where realizations of the Ising model and the PXP model have already been  demonstrated~\cite{Labuhn2016,Bernien2017, yang2020observation}. Notably, confinement and string breaking can be studied in the PXP model in the presence of a staggered detuning, representing a model equivalent to a U$(1)$ lattice gauge theory with a $\theta$ angle where $\theta\neq \pi$~\cite{Surace2020,Halimeh2022,zhang2023observation}.

\end{itemize}

%%%%%%%%%%%%%%%%%%%%%%%%%%%%%%%%%%%%%%%%%%%%%%%%%%%%%%%%%%%%%%%%%%%%%%%%%%%%%%%%%%%%%%%%%%%%%%%%%%%%%%%%%%%%%%%%%%%%%%%%%%%%%%%%%%%%%%%%%%%%%%%%%%%%%%%%%%%%%%%%%%%%%%%%%%%%%%%%%%%%%%%%%%%%%%%%%%%%%%%%%%%%%%%%
\begin{acknowledgements}
We thank Alvise Bastianello, Michael Knap, and John Preskill for insightful discussions.

F.M.S.~acknowledges support provided by the U.S.~Department of Energy (DOE) QuantISED program through the theory consortium ``Intersections of QIS and Theoretical Particle Physics'' at Fermilab, and by Amazon Web Services, AWS Quantum Program.
This research was supported in part by the Munich Institute for {\mbox{Astro-,}} Particle and BioPhysics (MIAPbP), which is funded by the Deutsche Forschungsgemeinschaft (DFG, German Research Foundation) under Germany´s Excellence Strategy–EXC-2094–390783311. This research was also supported in part by the National Science Foundation (NSF) under Grants No.~NSF PHY-1748958 and PHY-2309135.
A.L.~acknowledges funding through a Leverhulme-Peierls Fellowship at the University of Oxford. E.R.B., A.S., B.W., Z.D., and A.V.G.~were supported in part by the NSF Quantum Leap Challenge Institute (QLCI), award No.~OMA-2120757. E.R.B., A.S., B.W., and A.V.G.~were also supported in part by the NSF STAQ program, DOE ASCR Quantum Testbed Pathfinder program (awards No.~DE-SC0019040 and No.~DE-SC0024220), ONR MURI, AFOSR MURI,  DARPA SAVaNT ADVENT, ARL (W911NF-24-2-0107), and NQVL:QSTD:Pilot:FTL. E.R.B.~acknowledges support from the DOE, Office of Science, Office of Advanced Scientific Computing Research (ASCR), Computational Science Graduate Fellowship, award no.~DE-SC0023112. Support is also acknowledged from the U.S.~Department of Energy, Office of Science, National Quantum Information Science Research Centers, Quantum Systems Accelerator (QSA) and from the U.S.~Department of Energy, Office of Science, Accelerated Research in Quantum Computing, Fundamental Algorithmic Research toward Quantum Utility (FAR-Qu). Z.D.~further acknowledges
support by the DOE, Office of Science, Early Career
Award, award no. DE-SC0020271.

\section*{Data Availability Statement}
The data that support the findings of this article are openly available \cite{data}.
\end{acknowledgements}

%%%%%%%%%%%%%%%%%%%%%%%%%%%%%%%%%%%%%%%%%%%%%%%%%%%%%%%%%%%%%%%%%%%%%%%%%%%%%%%%%%%%%%%%%%%%%%%%%%%%%%%%%%%%%%%%%%%%%%%%%%%%%%%%%%%%%%%%%%%%%%%%%%%%%%%%%%%%%%%%%%%%%%%%%%%%%%%%%%%%%%%%%%%%%%%%%%%%%%%%%%%%%%%%

\bibliography{bib}

\begin{thebibliography}{92}%
\makeatletter
\providecommand \@ifxundefined [1]{%
 \@ifx{#1\undefined}
}%
\providecommand \@ifnum [1]{%
 \ifnum #1\expandafter \@firstoftwo
 \else \expandafter \@secondoftwo
 \fi
}%
\providecommand \@ifx [1]{%
 \ifx #1\expandafter \@firstoftwo
 \else \expandafter \@secondoftwo
 \fi
}%
\providecommand \natexlab [1]{#1}%
\providecommand \enquote  [1]{``#1''}%
\providecommand \bibnamefont  [1]{#1}%
\providecommand \bibfnamefont [1]{#1}%
\providecommand \citenamefont [1]{#1}%
\providecommand \href@noop [0]{\@secondoftwo}%
\providecommand \href [0]{\begingroup \@sanitize@url \@href}%
\providecommand \@href[1]{\@@startlink{#1}\@@href}%
\providecommand \@@href[1]{\endgroup#1\@@endlink}%
\providecommand \@sanitize@url [0]{\catcode `\\12\catcode `\$12\catcode
  `\&12\catcode `\#12\catcode `\^12\catcode `\_12\catcode `\%12\relax}%
\providecommand \@@startlink[1]{}%
\providecommand \@@endlink[0]{}%
\providecommand \url  [0]{\begingroup\@sanitize@url \@url }%
\providecommand \@url [1]{\endgroup\@href {#1}{\urlprefix }}%
\providecommand \urlprefix  [0]{URL }%
\providecommand \Eprint [0]{\href }%
\providecommand \doibase [0]{https://doi.org/}%
\providecommand \selectlanguage [0]{\@gobble}%
\providecommand \bibinfo  [0]{\@secondoftwo}%
\providecommand \bibfield  [0]{\@secondoftwo}%
\providecommand \translation [1]{[#1]}%
\providecommand \BibitemOpen [0]{}%
\providecommand \bibitemStop [0]{}%
\providecommand \bibitemNoStop [0]{.\EOS\space}%
\providecommand \EOS [0]{\spacefactor3000\relax}%
\providecommand \BibitemShut  [1]{\csname bibitem#1\endcsname}%
\let\auto@bib@innerbib\@empty
%</preamble>
\bibitem [{\citenamefont {Greensite}(2011)}]{Greensite:2011zz}%
  \BibitemOpen
  \bibfield  {author} {\bibinfo {author} {\bibfnamefont {J.}~\bibnamefont
  {Greensite}},\ }\href {https://doi.org/10.1007/978-3-642-14382-3} {\emph
  {\bibinfo {title} {{An introduction to the confinement problem}}}},\ Vol.\
  \bibinfo {volume} {821}\ (\bibinfo {year} {2011})\BibitemShut {NoStop}%
\bibitem [{\citenamefont {Nambu}(1979)}]{NAMBU1979372}%
  \BibitemOpen
  \bibfield  {author} {\bibinfo {author} {\bibfnamefont {Y.}~\bibnamefont
  {Nambu}},\ }\bibfield  {title} {\bibinfo {title} {{QCD} and the string
  model},\ }\href
  {https://doi.org/https://doi.org/10.1016/0370-2693(79)91193-6} {\bibfield
  {journal} {\bibinfo  {journal} {Physics Letters B}\ }\textbf {\bibinfo
  {volume} {80}},\ \bibinfo {pages} {372} (\bibinfo {year} {1979})}\BibitemShut
  {NoStop}%
\bibitem [{\citenamefont {Bali}\ \emph {et~al.}(1998)\citenamefont {Bali},
  \citenamefont {Eicker}, \citenamefont {Giusti}, \citenamefont {Gl{\"a}ssner},
  \citenamefont {Guesken}, \citenamefont {Hoeber}, \citenamefont {Lacock},
  \citenamefont {Lippert}, \citenamefont {Martinelli}, \citenamefont {Rapuano}
  \emph {et~al.}}]{bali1998glueballs}%
  \BibitemOpen
  \bibfield  {author} {\bibinfo {author} {\bibfnamefont {G.}~\bibnamefont
  {Bali}}, \bibinfo {author} {\bibfnamefont {N.}~\bibnamefont {Eicker}},
  \bibinfo {author} {\bibfnamefont {L.}~\bibnamefont {Giusti}}, \bibinfo
  {author} {\bibfnamefont {U.}~\bibnamefont {Gl{\"a}ssner}}, \bibinfo {author}
  {\bibfnamefont {S.}~\bibnamefont {Guesken}}, \bibinfo {author} {\bibfnamefont
  {H.}~\bibnamefont {Hoeber}}, \bibinfo {author} {\bibfnamefont
  {P.}~\bibnamefont {Lacock}}, \bibinfo {author} {\bibfnamefont
  {T.}~\bibnamefont {Lippert}}, \bibinfo {author} {\bibfnamefont
  {G.}~\bibnamefont {Martinelli}}, \bibinfo {author} {\bibfnamefont
  {F.}~\bibnamefont {Rapuano}}, \emph {et~al.},\ }\bibfield  {title} {\bibinfo
  {title} {Glueballs and string breaking from full {QCD}},\ }\href
  {https://doi.org/https://doi.org/10.1016/S0920-5632(97)00724-X} {\bibfield
  {journal} {\bibinfo  {journal} {Nuclear Physics B-Proceedings Supplements}\
  }\textbf {\bibinfo {volume} {63}},\ \bibinfo {pages} {209} (\bibinfo {year}
  {1998})}\BibitemShut {NoStop}%
\bibitem [{\citenamefont {Aoki}\ \emph {et~al.}(1999)\citenamefont {Aoki},
  \citenamefont {Boyd}, \citenamefont {Burkhalter}, \citenamefont {Ejiri},
  \citenamefont {Fukugita}, \citenamefont {Hashimoto}, \citenamefont {Iwasaki},
  \citenamefont {Kanaya}, \citenamefont {Kaneko}, \citenamefont {Kuramashi}
  \emph {et~al.}}]{aoki1999static}%
  \BibitemOpen
  \bibfield  {author} {\bibinfo {author} {\bibfnamefont {S.}~\bibnamefont
  {Aoki}}, \bibinfo {author} {\bibfnamefont {G.}~\bibnamefont {Boyd}}, \bibinfo
  {author} {\bibfnamefont {R.}~\bibnamefont {Burkhalter}}, \bibinfo {author}
  {\bibfnamefont {S.}~\bibnamefont {Ejiri}}, \bibinfo {author} {\bibfnamefont
  {M.}~\bibnamefont {Fukugita}}, \bibinfo {author} {\bibfnamefont
  {S.}~\bibnamefont {Hashimoto}}, \bibinfo {author} {\bibfnamefont
  {Y.}~\bibnamefont {Iwasaki}}, \bibinfo {author} {\bibfnamefont
  {K.}~\bibnamefont {Kanaya}}, \bibinfo {author} {\bibfnamefont
  {T.}~\bibnamefont {Kaneko}}, \bibinfo {author} {\bibfnamefont
  {Y.}~\bibnamefont {Kuramashi}}, \emph {et~al.},\ }\bibfield  {title}
  {\bibinfo {title} {The static quark potential in full {QCD}},\ }\href
  {https://doi.org/https://doi.org/10.1016/S0920-5632(99)85027-0} {\bibfield
  {journal} {\bibinfo  {journal} {Nuclear Physics B-Proceedings Supplements}\
  }\textbf {\bibinfo {volume} {73}},\ \bibinfo {pages} {216} (\bibinfo {year}
  {1999})}\BibitemShut {NoStop}%
\bibitem [{\citenamefont {Pennanen}\ \emph {et~al.}(2000)\citenamefont
  {Pennanen}, \citenamefont {Michael}, \citenamefont {Collaboration} \emph
  {et~al.}}]{pennanen2000string}%
  \BibitemOpen
  \bibfield  {author} {\bibinfo {author} {\bibfnamefont {P.}~\bibnamefont
  {Pennanen}}, \bibinfo {author} {\bibfnamefont {C.}~\bibnamefont {Michael}},
  \bibinfo {author} {\bibfnamefont {U.}~\bibnamefont {Collaboration}}, \emph
  {et~al.},\ }\bibfield  {title} {\bibinfo {title} {String breaking in
  zero-temperature lattice {QCD}},\ }\href
  {https://doi.org/10.48550/arXiv.hep-lat/0001015} {\bibfield  {journal}
  {\bibinfo  {journal} {arXiv preprint hep-lat/0001015}\ } (\bibinfo {year}
  {2000})}\BibitemShut {NoStop}%
\bibitem [{\citenamefont {Bali}\ \emph {et~al.}(2000)\citenamefont {Bali},
  \citenamefont {Bolder}, \citenamefont {Eicker}, \citenamefont {Lippert},
  \citenamefont {Orth}, \citenamefont {Ueberholz}, \citenamefont {Schilling},
  \citenamefont {Struckmann} \emph {et~al.}}]{bali2000static}%
  \BibitemOpen
  \bibfield  {author} {\bibinfo {author} {\bibfnamefont {G.~S.}\ \bibnamefont
  {Bali}}, \bibinfo {author} {\bibfnamefont {B.}~\bibnamefont {Bolder}},
  \bibinfo {author} {\bibfnamefont {N.}~\bibnamefont {Eicker}}, \bibinfo
  {author} {\bibfnamefont {T.}~\bibnamefont {Lippert}}, \bibinfo {author}
  {\bibfnamefont {B.}~\bibnamefont {Orth}}, \bibinfo {author} {\bibfnamefont
  {P.}~\bibnamefont {Ueberholz}}, \bibinfo {author} {\bibfnamefont
  {K.}~\bibnamefont {Schilling}}, \bibinfo {author} {\bibfnamefont
  {T.}~\bibnamefont {Struckmann}}, \emph {et~al.},\ }\bibfield  {title}
  {\bibinfo {title} {Static potentials and glueball masses from {QCD}
  simulations with {Wilson} sea quarks},\ }\href
  {https://doi.org/https://doi.org/10.1103/PhysRevD.62.054503} {\bibfield
  {journal} {\bibinfo  {journal} {Physical Review D}\ }\textbf {\bibinfo
  {volume} {62}},\ \bibinfo {pages} {054503} (\bibinfo {year}
  {2000})}\BibitemShut {NoStop}%
\bibitem [{\citenamefont {Duncan}\ \emph {et~al.}(2001)\citenamefont {Duncan},
  \citenamefont {Eichten},\ and\ \citenamefont {Thacker}}]{duncan2001string}%
  \BibitemOpen
  \bibfield  {author} {\bibinfo {author} {\bibfnamefont {A.}~\bibnamefont
  {Duncan}}, \bibinfo {author} {\bibfnamefont {E.}~\bibnamefont {Eichten}},\
  and\ \bibinfo {author} {\bibfnamefont {H.}~\bibnamefont {Thacker}},\
  }\bibfield  {title} {\bibinfo {title} {String breaking in four dimensional
  lattice {QCD}},\ }\href
  {https://doi.org/https://doi.org/10.1103/PhysRevD.63.111501} {\bibfield
  {journal} {\bibinfo  {journal} {Physical Review D}\ }\textbf {\bibinfo
  {volume} {63}},\ \bibinfo {pages} {111501} (\bibinfo {year}
  {2001})}\BibitemShut {NoStop}%
\bibitem [{\citenamefont {Bernard}\ \emph {et~al.}(2001)\citenamefont
  {Bernard}, \citenamefont {DeGrand}, \citenamefont {DeTar}, \citenamefont
  {Lacock}, \citenamefont {Gottlieb}, \citenamefont {Heller}, \citenamefont
  {Hetrick}, \citenamefont {Orginos}, \citenamefont {Toussaint},\ and\
  \citenamefont {Sugar}}]{bernard2001zero}%
  \BibitemOpen
  \bibfield  {author} {\bibinfo {author} {\bibfnamefont {C.}~\bibnamefont
  {Bernard}}, \bibinfo {author} {\bibfnamefont {T.}~\bibnamefont {DeGrand}},
  \bibinfo {author} {\bibfnamefont {C.}~\bibnamefont {DeTar}}, \bibinfo
  {author} {\bibfnamefont {P.}~\bibnamefont {Lacock}}, \bibinfo {author}
  {\bibfnamefont {S.}~\bibnamefont {Gottlieb}}, \bibinfo {author}
  {\bibfnamefont {U.~M.}\ \bibnamefont {Heller}}, \bibinfo {author}
  {\bibfnamefont {J.}~\bibnamefont {Hetrick}}, \bibinfo {author} {\bibfnamefont
  {K.}~\bibnamefont {Orginos}}, \bibinfo {author} {\bibfnamefont
  {D.}~\bibnamefont {Toussaint}},\ and\ \bibinfo {author} {\bibfnamefont
  {R.~L.}\ \bibnamefont {Sugar}},\ }\bibfield  {title} {\bibinfo {title} {Zero
  temperature string breaking in lattice quantum chromodynamics},\ }\href
  {https://doi.org/https://doi.org/10.1103/PhysRevD.64.074509} {\bibfield
  {journal} {\bibinfo  {journal} {Physical Review D}\ }\textbf {\bibinfo
  {volume} {64}},\ \bibinfo {pages} {074509} (\bibinfo {year}
  {2001})}\BibitemShut {NoStop}%
\bibitem [{\citenamefont {Kratochvila}\ and\ \citenamefont
  {De~Forcrand}(2003)}]{kratochvila2003observing}%
  \BibitemOpen
  \bibfield  {author} {\bibinfo {author} {\bibfnamefont {S.}~\bibnamefont
  {Kratochvila}}\ and\ \bibinfo {author} {\bibfnamefont {P.}~\bibnamefont
  {De~Forcrand}},\ }\bibfield  {title} {\bibinfo {title} {Observing string
  breaking with {Wilson} loops},\ }\href
  {https://doi.org/https://doi.org/10.1016/j.nuclphysb.2003.08.014} {\bibfield
  {journal} {\bibinfo  {journal} {Nuclear Physics B}\ }\textbf {\bibinfo
  {volume} {671}},\ \bibinfo {pages} {103} (\bibinfo {year}
  {2003})}\BibitemShut {NoStop}%
\bibitem [{\citenamefont {Bali}\ \emph {et~al.}(2005)\citenamefont {Bali},
  \citenamefont {Neff}, \citenamefont {Duessel}, \citenamefont {Lippert},
  \citenamefont {Schilling},\ and\ \citenamefont
  {Collaboration)}}]{bali2005observation}%
  \BibitemOpen
  \bibfield  {author} {\bibinfo {author} {\bibfnamefont {G.~S.}\ \bibnamefont
  {Bali}}, \bibinfo {author} {\bibfnamefont {H.}~\bibnamefont {Neff}}, \bibinfo
  {author} {\bibfnamefont {T.}~\bibnamefont {Duessel}}, \bibinfo {author}
  {\bibfnamefont {T.}~\bibnamefont {Lippert}}, \bibinfo {author} {\bibfnamefont
  {K.}~\bibnamefont {Schilling}},\ and\ \bibinfo {author} {\bibfnamefont
  {S.}~\bibnamefont {Collaboration)}},\ }\bibfield  {title} {\bibinfo {title}
  {Observation of string breaking in {QCD}},\ }\href
  {https://doi.org/https://doi.org/10.1103/PhysRevD.71.114513} {\bibfield
  {journal} {\bibinfo  {journal} {Physical Review D}\ }\textbf {\bibinfo
  {volume} {71}},\ \bibinfo {pages} {114513} (\bibinfo {year}
  {2005})}\BibitemShut {NoStop}%
\bibitem [{\citenamefont {Andersson}\ \emph {et~al.}(1983)\citenamefont
  {Andersson}, \citenamefont {Gustafson}, \citenamefont {Ingelman},\ and\
  \citenamefont {Sj{\"o}strand}}]{andersson1983parton}%
  \BibitemOpen
  \bibfield  {author} {\bibinfo {author} {\bibfnamefont {B.}~\bibnamefont
  {Andersson}}, \bibinfo {author} {\bibfnamefont {G.}~\bibnamefont
  {Gustafson}}, \bibinfo {author} {\bibfnamefont {G.}~\bibnamefont
  {Ingelman}},\ and\ \bibinfo {author} {\bibfnamefont {T.}~\bibnamefont
  {Sj{\"o}strand}},\ }\bibfield  {title} {\bibinfo {title} {Parton
  fragmentation and string dynamics},\ }\href
  {https://doi.org/https://doi.org/10.1016/0370-1573(83)90080-7} {\bibfield
  {journal} {\bibinfo  {journal} {Physics Reports}\ }\textbf {\bibinfo {volume}
  {97}},\ \bibinfo {pages} {31} (\bibinfo {year} {1983})}\BibitemShut {NoStop}%
\bibitem [{\citenamefont {Brambilla}\ \emph {et~al.}(2014)\citenamefont
  {Brambilla}, \citenamefont {Eidelman}, \citenamefont {Foka}, \citenamefont
  {Gardner}, \citenamefont {Kronfeld}, \citenamefont {Alford}, \citenamefont
  {Alkofer}, \citenamefont {Butenschoen}, \citenamefont {Cohen}, \citenamefont
  {Erdmenger} \emph {et~al.}}]{brambilla2014qcd}%
  \BibitemOpen
  \bibfield  {author} {\bibinfo {author} {\bibfnamefont {N.}~\bibnamefont
  {Brambilla}}, \bibinfo {author} {\bibfnamefont {S.}~\bibnamefont {Eidelman}},
  \bibinfo {author} {\bibfnamefont {P.}~\bibnamefont {Foka}}, \bibinfo {author}
  {\bibfnamefont {S.}~\bibnamefont {Gardner}}, \bibinfo {author} {\bibfnamefont
  {A.}~\bibnamefont {Kronfeld}}, \bibinfo {author} {\bibfnamefont
  {M.}~\bibnamefont {Alford}}, \bibinfo {author} {\bibfnamefont
  {R.}~\bibnamefont {Alkofer}}, \bibinfo {author} {\bibfnamefont
  {M.}~\bibnamefont {Butenschoen}}, \bibinfo {author} {\bibfnamefont
  {T.}~\bibnamefont {Cohen}}, \bibinfo {author} {\bibfnamefont
  {J.}~\bibnamefont {Erdmenger}}, \emph {et~al.},\ }\bibfield  {title}
  {\bibinfo {title} {{QCD and strongly coupled gauge theories: challenges and
  perspectives}},\ }\href
  {https://doi.org/https://doi.org/10.1140/epjc/s10052-014-2981-5} {\bibfield
  {journal} {\bibinfo  {journal} {The European Physical Journal C}\ }\textbf
  {\bibinfo {volume} {74}},\ \bibinfo {pages} {2981} (\bibinfo {year}
  {2014})}\BibitemShut {NoStop}%
\bibitem [{\citenamefont {Bauer}\ \emph {et~al.}(2023)\citenamefont {Bauer},
  \citenamefont {Davoudi}, \citenamefont {Balantekin}, \citenamefont
  {Bhattacharya}, \citenamefont {Carena}, \citenamefont {De~Jong},
  \citenamefont {Draper}, \citenamefont {El-Khadra}, \citenamefont {Gemelke},
  \citenamefont {Hanada} \emph {et~al.}}]{bauer2023quantum}%
  \BibitemOpen
  \bibfield  {author} {\bibinfo {author} {\bibfnamefont {C.~W.}\ \bibnamefont
  {Bauer}}, \bibinfo {author} {\bibfnamefont {Z.}~\bibnamefont {Davoudi}},
  \bibinfo {author} {\bibfnamefont {A.~B.}\ \bibnamefont {Balantekin}},
  \bibinfo {author} {\bibfnamefont {T.}~\bibnamefont {Bhattacharya}}, \bibinfo
  {author} {\bibfnamefont {M.}~\bibnamefont {Carena}}, \bibinfo {author}
  {\bibfnamefont {W.~A.}\ \bibnamefont {De~Jong}}, \bibinfo {author}
  {\bibfnamefont {P.}~\bibnamefont {Draper}}, \bibinfo {author} {\bibfnamefont
  {A.}~\bibnamefont {El-Khadra}}, \bibinfo {author} {\bibfnamefont
  {N.}~\bibnamefont {Gemelke}}, \bibinfo {author} {\bibfnamefont
  {M.}~\bibnamefont {Hanada}}, \emph {et~al.},\ }\bibfield  {title} {\bibinfo
  {title} {Quantum simulation for high-energy physics},\ }\href
  {https://doi.org/https://doi.org/10.1103/PRXQuantum.4.027001} {\bibfield
  {journal} {\bibinfo  {journal} {PRX quantum}\ }\textbf {\bibinfo {volume}
  {4}},\ \bibinfo {pages} {027001} (\bibinfo {year} {2023})}\BibitemShut
  {NoStop}%
\bibitem [{\citenamefont {Hebenstreit}\ \emph {et~al.}(2013)\citenamefont
  {Hebenstreit}, \citenamefont {Berges},\ and\ \citenamefont
  {Gelfand}}]{Hebenstreit2013}%
  \BibitemOpen
  \bibfield  {author} {\bibinfo {author} {\bibfnamefont {F.}~\bibnamefont
  {Hebenstreit}}, \bibinfo {author} {\bibfnamefont {J.}~\bibnamefont
  {Berges}},\ and\ \bibinfo {author} {\bibfnamefont {D.}~\bibnamefont
  {Gelfand}},\ }\bibfield  {title} {\bibinfo {title} {Real-time dynamics of
  string breaking},\ }\href {https://doi.org/10.1103/PhysRevLett.111.201601}
  {\bibfield  {journal} {\bibinfo  {journal} {Phys. Rev. Lett.}\ }\textbf
  {\bibinfo {volume} {111}},\ \bibinfo {pages} {201601} (\bibinfo {year}
  {2013})}\BibitemShut {NoStop}%
\bibitem [{\citenamefont {Hebenstreit}\ and\ \citenamefont
  {Berges}(2014)}]{Hebenstreit2014}%
  \BibitemOpen
  \bibfield  {author} {\bibinfo {author} {\bibfnamefont {F.}~\bibnamefont
  {Hebenstreit}}\ and\ \bibinfo {author} {\bibfnamefont {J.}~\bibnamefont
  {Berges}},\ }\bibfield  {title} {\bibinfo {title} {{Connecting real-time
  properties of the massless Schwinger model to the massive case}},\ }\href
  {https://doi.org/10.1103/PhysRevD.90.045034} {\bibfield  {journal} {\bibinfo
  {journal} {Phys. Rev. D}\ }\textbf {\bibinfo {volume} {90}},\ \bibinfo
  {pages} {045034} (\bibinfo {year} {2014})}\BibitemShut {NoStop}%
\bibitem [{\citenamefont {K{\"u}hn}\ \emph {et~al.}(2015)\citenamefont
  {K{\"u}hn}, \citenamefont {Zohar}, \citenamefont {Cirac},\ and\ \citenamefont
  {Ba{\~n}uls}}]{Kuhn2015}%
  \BibitemOpen
  \bibfield  {author} {\bibinfo {author} {\bibfnamefont {S.}~\bibnamefont
  {K{\"u}hn}}, \bibinfo {author} {\bibfnamefont {E.}~\bibnamefont {Zohar}},
  \bibinfo {author} {\bibfnamefont {J.~I.}\ \bibnamefont {Cirac}},\ and\
  \bibinfo {author} {\bibfnamefont {M.~C.}\ \bibnamefont {Ba{\~n}uls}},\
  }\bibfield  {title} {\bibinfo {title} {{Non-Abelian} string breaking
  phenomena with matrix product states},\ }\href
  {https://doi.org/10.1007/JHEP07(2015)130} {\bibfield  {journal} {\bibinfo
  {journal} {Journal of High Energy Physics}\ }\textbf {\bibinfo {volume}
  {2015}},\ \bibinfo {pages} {1} (\bibinfo {year} {2015})}\BibitemShut
  {NoStop}%
\bibitem [{\citenamefont {Buyens}\ \emph {et~al.}(2016)\citenamefont {Buyens},
  \citenamefont {Haegeman}, \citenamefont {Verschelde}, \citenamefont
  {Verstraete},\ and\ \citenamefont {Van~Acoleyen}}]{Buyens2016}%
  \BibitemOpen
  \bibfield  {author} {\bibinfo {author} {\bibfnamefont {B.}~\bibnamefont
  {Buyens}}, \bibinfo {author} {\bibfnamefont {J.}~\bibnamefont {Haegeman}},
  \bibinfo {author} {\bibfnamefont {H.}~\bibnamefont {Verschelde}}, \bibinfo
  {author} {\bibfnamefont {F.}~\bibnamefont {Verstraete}},\ and\ \bibinfo
  {author} {\bibfnamefont {K.}~\bibnamefont {Van~Acoleyen}},\ }\bibfield
  {title} {\bibinfo {title} {{Confinement and String Breaking for
  ${\mathrm{QED}}_{2}$ in the Hamiltonian Picture}},\ }\href
  {https://doi.org/10.1103/PhysRevX.6.041040} {\bibfield  {journal} {\bibinfo
  {journal} {Phys. Rev. X}\ }\textbf {\bibinfo {volume} {6}},\ \bibinfo {pages}
  {041040} (\bibinfo {year} {2016})}\BibitemShut {NoStop}%
\bibitem [{\citenamefont {Pichler}\ \emph {et~al.}(2016)\citenamefont
  {Pichler}, \citenamefont {Dalmonte}, \citenamefont {Rico}, \citenamefont
  {Zoller},\ and\ \citenamefont {Montangero}}]{Pichler2016}%
  \BibitemOpen
  \bibfield  {author} {\bibinfo {author} {\bibfnamefont {T.}~\bibnamefont
  {Pichler}}, \bibinfo {author} {\bibfnamefont {M.}~\bibnamefont {Dalmonte}},
  \bibinfo {author} {\bibfnamefont {E.}~\bibnamefont {Rico}}, \bibinfo {author}
  {\bibfnamefont {P.}~\bibnamefont {Zoller}},\ and\ \bibinfo {author}
  {\bibfnamefont {S.}~\bibnamefont {Montangero}},\ }\bibfield  {title}
  {\bibinfo {title} {Real-time dynamics in {U(1)} lattice gauge theories with
  tensor networks},\ }\href {https://doi.org/10.1103/PhysRevX.6.011023}
  {\bibfield  {journal} {\bibinfo  {journal} {Phys. Rev. X}\ }\textbf {\bibinfo
  {volume} {6}},\ \bibinfo {pages} {011023} (\bibinfo {year}
  {2016})}\BibitemShut {NoStop}%
\bibitem [{\citenamefont {Kormos}\ \emph {et~al.}(2017)\citenamefont {Kormos},
  \citenamefont {Collura}, \citenamefont {Tak{\'a}cs},\ and\ \citenamefont
  {Calabrese}}]{Kormos:2017aa}%
  \BibitemOpen
  \bibfield  {author} {\bibinfo {author} {\bibfnamefont {M.}~\bibnamefont
  {Kormos}}, \bibinfo {author} {\bibfnamefont {M.}~\bibnamefont {Collura}},
  \bibinfo {author} {\bibfnamefont {G.}~\bibnamefont {Tak{\'a}cs}},\ and\
  \bibinfo {author} {\bibfnamefont {P.}~\bibnamefont {Calabrese}},\ }\bibfield
  {title} {\bibinfo {title} {Real time confinement following a quantum quench
  to a non-integrable model},\ }\href {https://doi.org/10.1038/nphys3934}
  {\bibfield  {journal} {\bibinfo  {journal} {Nature Physics}\ }\textbf
  {\bibinfo {volume} {13}},\ \bibinfo {pages} {246} (\bibinfo {year}
  {2017})}\BibitemShut {NoStop}%
\bibitem [{\citenamefont {Sala}\ \emph {et~al.}(2018)\citenamefont {Sala},
  \citenamefont {Shi}, \citenamefont {K\"uhn}, \citenamefont {Ba\~nuls},
  \citenamefont {Demler},\ and\ \citenamefont {Cirac}}]{SALA2018}%
  \BibitemOpen
  \bibfield  {author} {\bibinfo {author} {\bibfnamefont {P.}~\bibnamefont
  {Sala}}, \bibinfo {author} {\bibfnamefont {T.}~\bibnamefont {Shi}}, \bibinfo
  {author} {\bibfnamefont {S.}~\bibnamefont {K\"uhn}}, \bibinfo {author}
  {\bibfnamefont {M.~C.}\ \bibnamefont {Ba\~nuls}}, \bibinfo {author}
  {\bibfnamefont {E.}~\bibnamefont {Demler}},\ and\ \bibinfo {author}
  {\bibfnamefont {J.~I.}\ \bibnamefont {Cirac}},\ }\bibfield  {title} {\bibinfo
  {title} {{Variational study of U(1) and SU(2) lattice gauge theories with
  Gaussian states in $1+1$ dimensions}},\ }\href
  {https://doi.org/10.1103/PhysRevD.98.034505} {\bibfield  {journal} {\bibinfo
  {journal} {Phys. Rev. D}\ }\textbf {\bibinfo {volume} {98}},\ \bibinfo
  {pages} {034505} (\bibinfo {year} {2018})}\BibitemShut {NoStop}%
\bibitem [{\citenamefont {Spitz}\ and\ \citenamefont
  {Berges}(2019)}]{Spitz2019}%
  \BibitemOpen
  \bibfield  {author} {\bibinfo {author} {\bibfnamefont {D.}~\bibnamefont
  {Spitz}}\ and\ \bibinfo {author} {\bibfnamefont {J.}~\bibnamefont {Berges}},\
  }\bibfield  {title} {\bibinfo {title} {{Schwinger pair production and string
  breaking in non-Abelian gauge theory from real-time lattice improved
  Hamiltonians}},\ }\href {https://doi.org/10.1103/PhysRevD.99.036020}
  {\bibfield  {journal} {\bibinfo  {journal} {Phys. Rev. D}\ }\textbf {\bibinfo
  {volume} {99}},\ \bibinfo {pages} {036020} (\bibinfo {year}
  {2019})}\BibitemShut {NoStop}%
\bibitem [{\citenamefont {Chanda}\ \emph {et~al.}(2020)\citenamefont {Chanda},
  \citenamefont {Zakrzewski}, \citenamefont {Lewenstein},\ and\ \citenamefont
  {Tagliacozzo}}]{ch2019confinement}%
  \BibitemOpen
  \bibfield  {author} {\bibinfo {author} {\bibfnamefont {T.}~\bibnamefont
  {Chanda}}, \bibinfo {author} {\bibfnamefont {J.}~\bibnamefont {Zakrzewski}},
  \bibinfo {author} {\bibfnamefont {M.}~\bibnamefont {Lewenstein}},\ and\
  \bibinfo {author} {\bibfnamefont {L.}~\bibnamefont {Tagliacozzo}},\
  }\bibfield  {title} {\bibinfo {title} {Confinement and lack of thermalization
  after quenches in the bosonic {Schwinger} model},\ }\href
  {https://doi.org/10.1103/PhysRevLett.124.180602} {\bibfield  {journal}
  {\bibinfo  {journal} {Phys. Rev. Lett.}\ }\textbf {\bibinfo {volume} {124}},\
  \bibinfo {pages} {180602} (\bibinfo {year} {2020})}\BibitemShut {NoStop}%
\bibitem [{\citenamefont {Lerose}\ \emph {et~al.}(2020)\citenamefont {Lerose},
  \citenamefont {Surace}, \citenamefont {Mazza}, \citenamefont {Perfetto},
  \citenamefont {Collura},\ and\ \citenamefont {Gambassi}}]{Lerose2020}%
  \BibitemOpen
  \bibfield  {author} {\bibinfo {author} {\bibfnamefont {A.}~\bibnamefont
  {Lerose}}, \bibinfo {author} {\bibfnamefont {F.~M.}\ \bibnamefont {Surace}},
  \bibinfo {author} {\bibfnamefont {P.~P.}\ \bibnamefont {Mazza}}, \bibinfo
  {author} {\bibfnamefont {G.}~\bibnamefont {Perfetto}}, \bibinfo {author}
  {\bibfnamefont {M.}~\bibnamefont {Collura}},\ and\ \bibinfo {author}
  {\bibfnamefont {A.}~\bibnamefont {Gambassi}},\ }\bibfield  {title} {\bibinfo
  {title} {Quasilocalized dynamics from confinement of quantum excitations},\
  }\href {https://doi.org/10.1103/PhysRevB.102.041118} {\bibfield  {journal}
  {\bibinfo  {journal} {Phys. Rev. B}\ }\textbf {\bibinfo {volume} {102}},\
  \bibinfo {pages} {041118} (\bibinfo {year} {2020})}\BibitemShut {NoStop}%
\bibitem [{\citenamefont {Magnifico}\ \emph {et~al.}(2020)\citenamefont
  {Magnifico}, \citenamefont {Dalmonte}, \citenamefont {Facchi}, \citenamefont
  {Pascazio}, \citenamefont {Pepe},\ and\ \citenamefont
  {Ercolessi}}]{Magnifico2020realtimedynamics}%
  \BibitemOpen
  \bibfield  {author} {\bibinfo {author} {\bibfnamefont {G.}~\bibnamefont
  {Magnifico}}, \bibinfo {author} {\bibfnamefont {M.}~\bibnamefont {Dalmonte}},
  \bibinfo {author} {\bibfnamefont {P.}~\bibnamefont {Facchi}}, \bibinfo
  {author} {\bibfnamefont {S.}~\bibnamefont {Pascazio}}, \bibinfo {author}
  {\bibfnamefont {F.~V.}\ \bibnamefont {Pepe}},\ and\ \bibinfo {author}
  {\bibfnamefont {E.}~\bibnamefont {Ercolessi}},\ }\bibfield  {title} {\bibinfo
  {title} {Real {T}ime {D}ynamics and {C}onfinement in the {$\mathbb{Z}_{n}$}
  {S}chwinger-{W}eyl lattice model for 1+1 {QED}},\ }\href
  {https://doi.org/10.22331/q-2020-06-15-281} {\bibfield  {journal} {\bibinfo
  {journal} {{Quantum}}\ }\textbf {\bibinfo {volume} {4}},\ \bibinfo {pages}
  {281} (\bibinfo {year} {2020})}\BibitemShut {NoStop}%
\bibitem [{\citenamefont {Milsted}\ \emph {et~al.}(2022)\citenamefont
  {Milsted}, \citenamefont {Liu}, \citenamefont {Preskill},\ and\ \citenamefont
  {Vidal}}]{Milsted22}%
  \BibitemOpen
  \bibfield  {author} {\bibinfo {author} {\bibfnamefont {A.}~\bibnamefont
  {Milsted}}, \bibinfo {author} {\bibfnamefont {J.}~\bibnamefont {Liu}},
  \bibinfo {author} {\bibfnamefont {J.}~\bibnamefont {Preskill}},\ and\
  \bibinfo {author} {\bibfnamefont {G.}~\bibnamefont {Vidal}},\ }\bibfield
  {title} {\bibinfo {title} {Collisions of false-vacuum bubble walls in a
  quantum spin chain},\ }\href {https://doi.org/10.1103/PRXQuantum.3.020316}
  {\bibfield  {journal} {\bibinfo  {journal} {PRX Quantum}\ }\textbf {\bibinfo
  {volume} {3}},\ \bibinfo {pages} {020316} (\bibinfo {year}
  {2022})}\BibitemShut {NoStop}%
\bibitem [{\citenamefont {Lee}\ \emph {et~al.}(2023)\citenamefont {Lee},
  \citenamefont {Mulligan}, \citenamefont {Ringer},\ and\ \citenamefont
  {Yao}}]{Lee2023}%
  \BibitemOpen
  \bibfield  {author} {\bibinfo {author} {\bibfnamefont {K.}~\bibnamefont
  {Lee}}, \bibinfo {author} {\bibfnamefont {J.}~\bibnamefont {Mulligan}},
  \bibinfo {author} {\bibfnamefont {F.}~\bibnamefont {Ringer}},\ and\ \bibinfo
  {author} {\bibfnamefont {X.}~\bibnamefont {Yao}},\ }\bibfield  {title}
  {\bibinfo {title} {Liouvillian dynamics of the open {Schwinger} model: String
  breaking and kinetic dissipation in a thermal medium},\ }\href
  {https://doi.org/10.1103/PhysRevD.108.094518} {\bibfield  {journal} {\bibinfo
   {journal} {Phys. Rev. D}\ }\textbf {\bibinfo {volume} {108}},\ \bibinfo
  {pages} {094518} (\bibinfo {year} {2023})}\BibitemShut {NoStop}%
\bibitem [{\citenamefont {Verdel}\ \emph {et~al.}(2020)\citenamefont {Verdel},
  \citenamefont {Liu}, \citenamefont {Whitsitt}, \citenamefont {Gorshkov},\
  and\ \citenamefont {Heyl}}]{Verdel19_ResonantSB}%
  \BibitemOpen
  \bibfield  {author} {\bibinfo {author} {\bibfnamefont {R.}~\bibnamefont
  {Verdel}}, \bibinfo {author} {\bibfnamefont {F.}~\bibnamefont {Liu}},
  \bibinfo {author} {\bibfnamefont {S.}~\bibnamefont {Whitsitt}}, \bibinfo
  {author} {\bibfnamefont {A.~V.}\ \bibnamefont {Gorshkov}},\ and\ \bibinfo
  {author} {\bibfnamefont {M.}~\bibnamefont {Heyl}},\ }\bibfield  {title}
  {\bibinfo {title} {Real-time dynamics of string breaking in quantum spin
  chains},\ }\href {https://doi.org/10.1103/PhysRevB.102.014308} {\bibfield
  {journal} {\bibinfo  {journal} {Phys. Rev. B}\ }\textbf {\bibinfo {volume}
  {102}},\ \bibinfo {pages} {014308} (\bibinfo {year} {2020})}\BibitemShut
  {NoStop}%
\bibitem [{\citenamefont {Surace}\ \emph {et~al.}(2020)\citenamefont {Surace},
  \citenamefont {Mazza}, \citenamefont {Giudici}, \citenamefont {Lerose},
  \citenamefont {Gambassi},\ and\ \citenamefont {Dalmonte}}]{Surace2020}%
  \BibitemOpen
  \bibfield  {author} {\bibinfo {author} {\bibfnamefont {F.~M.}\ \bibnamefont
  {Surace}}, \bibinfo {author} {\bibfnamefont {P.~P.}\ \bibnamefont {Mazza}},
  \bibinfo {author} {\bibfnamefont {G.}~\bibnamefont {Giudici}}, \bibinfo
  {author} {\bibfnamefont {A.}~\bibnamefont {Lerose}}, \bibinfo {author}
  {\bibfnamefont {A.}~\bibnamefont {Gambassi}},\ and\ \bibinfo {author}
  {\bibfnamefont {M.}~\bibnamefont {Dalmonte}},\ }\bibfield  {title} {\bibinfo
  {title} {Lattice gauge theories and string dynamics in {Rydberg} atom quantum
  simulators},\ }\href {https://doi.org/10.1103/PhysRevX.10.021041} {\bibfield
  {journal} {\bibinfo  {journal} {Phys. Rev. X}\ }\textbf {\bibinfo {volume}
  {10}},\ \bibinfo {pages} {021041} (\bibinfo {year} {2020})}\BibitemShut
  {NoStop}%
\bibitem [{\citenamefont {Verdel}\ \emph {et~al.}(2023)\citenamefont {Verdel},
  \citenamefont {Zhu},\ and\ \citenamefont {Heyl}}]{Verdel2023}%
  \BibitemOpen
  \bibfield  {author} {\bibinfo {author} {\bibfnamefont {R.}~\bibnamefont
  {Verdel}}, \bibinfo {author} {\bibfnamefont {G.-Y.}\ \bibnamefont {Zhu}},\
  and\ \bibinfo {author} {\bibfnamefont {M.}~\bibnamefont {Heyl}},\ }\bibfield
  {title} {\bibinfo {title} {Dynamical localization transition of string
  breaking in quantum spin chains},\ }\href
  {https://doi.org/10.1103/PhysRevLett.131.230402} {\bibfield  {journal}
  {\bibinfo  {journal} {Phys. Rev. Lett.}\ }\textbf {\bibinfo {volume} {131}},\
  \bibinfo {pages} {230402} (\bibinfo {year} {2023})}\BibitemShut {NoStop}%
\bibitem [{\citenamefont {Batini}\ \emph {et~al.}(2024)\citenamefont {Batini},
  \citenamefont {Kuhn}, \citenamefont {Berges},\ and\ \citenamefont
  {Floerchinger}}]{batini2024particle}%
  \BibitemOpen
  \bibfield  {author} {\bibinfo {author} {\bibfnamefont {L.}~\bibnamefont
  {Batini}}, \bibinfo {author} {\bibfnamefont {L.}~\bibnamefont {Kuhn}},
  \bibinfo {author} {\bibfnamefont {J.}~\bibnamefont {Berges}},\ and\ \bibinfo
  {author} {\bibfnamefont {S.}~\bibnamefont {Floerchinger}},\ }\bibfield
  {title} {\bibinfo {title} {Particle production and hadronization temperature
  in the massive {Schwinger} model},\ }\href
  {https://doi.org/10.1103/PhysRevD.110.045017} {\bibfield  {journal} {\bibinfo
   {journal} {Phys. Rev. D}\ }\textbf {\bibinfo {volume} {110}},\ \bibinfo
  {pages} {045017} (\bibinfo {year} {2024})}\BibitemShut {NoStop}%
\bibitem [{\citenamefont {McCoy}\ and\ \citenamefont {Wu}(1978)}]{McCoyWu}%
  \BibitemOpen
  \bibfield  {author} {\bibinfo {author} {\bibfnamefont {B.~M.}\ \bibnamefont
  {McCoy}}\ and\ \bibinfo {author} {\bibfnamefont {T.~T.}\ \bibnamefont {Wu}},\
  }\bibfield  {title} {\bibinfo {title} {Two-dimensional {Ising} field theory
  in a magnetic field: Breakup of the cut in the two-point function},\ }\href
  {https://doi.org/10.1103/PhysRevD.18.1259} {\bibfield  {journal} {\bibinfo
  {journal} {Phys. Rev. D}\ }\textbf {\bibinfo {volume} {18}},\ \bibinfo
  {pages} {1259} (\bibinfo {year} {1978})}\BibitemShut {NoStop}%
\bibitem [{\citenamefont {Delfino}\ \emph {et~al.}(1996)\citenamefont
  {Delfino}, \citenamefont {Mussardo},\ and\ \citenamefont
  {Simonetti}}]{DELFINO1996469}%
  \BibitemOpen
  \bibfield  {author} {\bibinfo {author} {\bibfnamefont {G.}~\bibnamefont
  {Delfino}}, \bibinfo {author} {\bibfnamefont {G.}~\bibnamefont {Mussardo}},\
  and\ \bibinfo {author} {\bibfnamefont {P.}~\bibnamefont {Simonetti}},\
  }\bibfield  {title} {\bibinfo {title} {Non-integrable quantum field theories
  as perturbations of certain integrable models},\ }\href
  {https://doi.org/https://doi.org/10.1016/0550-3213(96)00265-9} {\bibfield
  {journal} {\bibinfo  {journal} {Nucl. Phys. B}\ }\textbf {\bibinfo {volume}
  {473}},\ \bibinfo {pages} {469} (\bibinfo {year} {1996})}\BibitemShut
  {NoStop}%
\bibitem [{\citenamefont {Liu}\ \emph {et~al.}(2019)\citenamefont {Liu},
  \citenamefont {Lundgren}, \citenamefont {Titum}, \citenamefont {Pagano},
  \citenamefont {Zhang}, \citenamefont {Monroe},\ and\ \citenamefont
  {Gorshkov}}]{Liu2019}%
  \BibitemOpen
  \bibfield  {author} {\bibinfo {author} {\bibfnamefont {F.}~\bibnamefont
  {Liu}}, \bibinfo {author} {\bibfnamefont {R.}~\bibnamefont {Lundgren}},
  \bibinfo {author} {\bibfnamefont {P.}~\bibnamefont {Titum}}, \bibinfo
  {author} {\bibfnamefont {G.}~\bibnamefont {Pagano}}, \bibinfo {author}
  {\bibfnamefont {J.}~\bibnamefont {Zhang}}, \bibinfo {author} {\bibfnamefont
  {C.}~\bibnamefont {Monroe}},\ and\ \bibinfo {author} {\bibfnamefont {A.~V.}\
  \bibnamefont {Gorshkov}},\ }\bibfield  {title} {\bibinfo {title} {Confined
  quasiparticle dynamics in long-range interacting quantum spin chains},\
  }\href {https://doi.org/10.1103/PhysRevLett.122.150601} {\bibfield  {journal}
  {\bibinfo  {journal} {Phys. Rev. Lett.}\ }\textbf {\bibinfo {volume} {122}},\
  \bibinfo {pages} {150601} (\bibinfo {year} {2019})}\BibitemShut {NoStop}%
\bibitem [{Note1()}]{Note1}%
  \BibitemOpen
  \bibinfo {note} {The term ``charge'' can be understood from the operation of
  ``gauging'' the global $\protect \mathbb {Z}_2$ symmetry of the Ising model,
  i.e., promoting the global symmetry to a local one by introducing redundant
  degrees of freedom. In the lattice gauge theory obtained from this procedure,
  domain walls play the role of charges. See, for example, Refs.~\cite
  {balian1975gauge,zhang2018quantum,Lerose2020,borla2020gauging,Surace_2021,exp1}
  for an explicit derivation of the mapping between the Ising chain and the
  Ising lattice gauge theory.}\BibitemShut {Stop}%
\bibitem [{\citenamefont {Tan}\ \emph {et~al.}(2021)\citenamefont {Tan},
  \citenamefont {Becker}, \citenamefont {Liu}, \citenamefont {Pagano},
  \citenamefont {Collins}, \citenamefont {De}, \citenamefont {Feng},
  \citenamefont {Kaplan}, \citenamefont {Kyprianidis}, \citenamefont {Lundgren}
  \emph {et~al.}}]{tan2021domain}%
  \BibitemOpen
  \bibfield  {author} {\bibinfo {author} {\bibfnamefont {W.~L.}\ \bibnamefont
  {Tan}}, \bibinfo {author} {\bibfnamefont {P.}~\bibnamefont {Becker}},
  \bibinfo {author} {\bibfnamefont {F.}~\bibnamefont {Liu}}, \bibinfo {author}
  {\bibfnamefont {G.}~\bibnamefont {Pagano}}, \bibinfo {author} {\bibfnamefont
  {K.}~\bibnamefont {Collins}}, \bibinfo {author} {\bibfnamefont
  {A.}~\bibnamefont {De}}, \bibinfo {author} {\bibfnamefont {L.}~\bibnamefont
  {Feng}}, \bibinfo {author} {\bibfnamefont {H.}~\bibnamefont {Kaplan}},
  \bibinfo {author} {\bibfnamefont {A.}~\bibnamefont {Kyprianidis}}, \bibinfo
  {author} {\bibfnamefont {R.}~\bibnamefont {Lundgren}}, \emph {et~al.},\
  }\bibfield  {title} {\bibinfo {title} {Domain-wall confinement and dynamics
  in a quantum simulator},\ }\href {https://doi.org/10.1038/s41567-021-01194-3}
  {\bibfield  {journal} {\bibinfo  {journal} {Nature Physics}\ }\textbf
  {\bibinfo {volume} {17}},\ \bibinfo {pages} {742} (\bibinfo {year}
  {2021})}\BibitemShut {NoStop}%
\bibitem [{\citenamefont {De}\ \emph {et~al.}(2024)\citenamefont {De},
  \citenamefont {Lerose}, \citenamefont {Luo}, \citenamefont {Surace},
  \citenamefont {Schuckert}, \citenamefont {Bennewitz}, \citenamefont {Ware},
  \citenamefont {Morong}, \citenamefont {Collins}, \citenamefont {Davoudi}
  \emph {et~al.}}]{exp1}%
  \BibitemOpen
  \bibfield  {author} {\bibinfo {author} {\bibfnamefont {A.}~\bibnamefont
  {De}}, \bibinfo {author} {\bibfnamefont {A.}~\bibnamefont {Lerose}}, \bibinfo
  {author} {\bibfnamefont {D.}~\bibnamefont {Luo}}, \bibinfo {author}
  {\bibfnamefont {F.~M.}\ \bibnamefont {Surace}}, \bibinfo {author}
  {\bibfnamefont {A.}~\bibnamefont {Schuckert}}, \bibinfo {author}
  {\bibfnamefont {E.~R.}\ \bibnamefont {Bennewitz}}, \bibinfo {author}
  {\bibfnamefont {B.}~\bibnamefont {Ware}}, \bibinfo {author} {\bibfnamefont
  {W.}~\bibnamefont {Morong}}, \bibinfo {author} {\bibfnamefont {K.~S.}\
  \bibnamefont {Collins}}, \bibinfo {author} {\bibfnamefont {Z.}~\bibnamefont
  {Davoudi}}, \emph {et~al.},\ }\bibfield  {title} {\bibinfo {title}
  {Observation of string-breaking dynamics in a quantum simulator},\ }\href
  {https://doi.org/10.48550/arXiv.2410.13815} {\bibfield  {journal} {\bibinfo
  {journal} {arXiv preprint arXiv:2410.13815}\ } (\bibinfo {year}
  {2024})}\BibitemShut {NoStop}%
\bibitem [{\citenamefont {Ciavarella}(2024)}]{Ciavarella2024}%
  \BibitemOpen
  \bibfield  {author} {\bibinfo {author} {\bibfnamefont {A.~N.}\ \bibnamefont
  {Ciavarella}},\ }\bibfield  {title} {\bibinfo {title} {String breaking in the
  heavy quark limit with scalable circuits},\ }\href
  {https://doi.org/10.48550/arXiv.2411.05915} {\bibfield  {journal} {\bibinfo
  {journal} {arXiv preprint arXiv:2411.05915}\ } (\bibinfo {year}
  {2024})}\BibitemShut {NoStop}%
\bibitem [{\citenamefont {Cochran}\ \emph {et~al.}(2024)\citenamefont
  {Cochran}, \citenamefont {Jobst}, \citenamefont {Rosenberg}, \citenamefont
  {Lensky}, \citenamefont {Gyawali}, \citenamefont {Eassa}, \citenamefont
  {Will}, \citenamefont {Abanin}, \citenamefont {Acharya}, \citenamefont {Beni}
  \emph {et~al.}}]{Cochran2024}%
  \BibitemOpen
  \bibfield  {author} {\bibinfo {author} {\bibfnamefont {T.~A.}\ \bibnamefont
  {Cochran}}, \bibinfo {author} {\bibfnamefont {B.}~\bibnamefont {Jobst}},
  \bibinfo {author} {\bibfnamefont {E.}~\bibnamefont {Rosenberg}}, \bibinfo
  {author} {\bibfnamefont {Y.~D.}\ \bibnamefont {Lensky}}, \bibinfo {author}
  {\bibfnamefont {G.}~\bibnamefont {Gyawali}}, \bibinfo {author} {\bibfnamefont
  {N.}~\bibnamefont {Eassa}}, \bibinfo {author} {\bibfnamefont
  {M.}~\bibnamefont {Will}}, \bibinfo {author} {\bibfnamefont {D.}~\bibnamefont
  {Abanin}}, \bibinfo {author} {\bibfnamefont {R.}~\bibnamefont {Acharya}},
  \bibinfo {author} {\bibfnamefont {L.~A.}\ \bibnamefont {Beni}}, \emph
  {et~al.},\ }\bibfield  {title} {\bibinfo {title} {Visualizing dynamics of
  charges and strings in (2+ 1) {D} lattice gauge theories},\ }\href
  {https://doi.org/10.48550/arXiv.2409.17142} {\bibfield  {journal} {\bibinfo
  {journal} {arXiv preprint arXiv:2409.17142}\ } (\bibinfo {year}
  {2024})}\BibitemShut {NoStop}%
\bibitem [{\citenamefont {Gonzalez-Cuadra}\ \emph {et~al.}(2024)\citenamefont
  {Gonzalez-Cuadra}, \citenamefont {Hamdan}, \citenamefont {Zache},
  \citenamefont {Braverman}, \citenamefont {Kornjaca}, \citenamefont {Lukin},
  \citenamefont {Cantu}, \citenamefont {Liu}, \citenamefont {Wang},
  \citenamefont {Keesling} \emph {et~al.}}]{Gonzalez2024}%
  \BibitemOpen
  \bibfield  {author} {\bibinfo {author} {\bibfnamefont {D.}~\bibnamefont
  {Gonzalez-Cuadra}}, \bibinfo {author} {\bibfnamefont {M.}~\bibnamefont
  {Hamdan}}, \bibinfo {author} {\bibfnamefont {T.~V.}\ \bibnamefont {Zache}},
  \bibinfo {author} {\bibfnamefont {B.}~\bibnamefont {Braverman}}, \bibinfo
  {author} {\bibfnamefont {M.}~\bibnamefont {Kornjaca}}, \bibinfo {author}
  {\bibfnamefont {A.}~\bibnamefont {Lukin}}, \bibinfo {author} {\bibfnamefont
  {S.~H.}\ \bibnamefont {Cantu}}, \bibinfo {author} {\bibfnamefont
  {F.}~\bibnamefont {Liu}}, \bibinfo {author} {\bibfnamefont {S.-T.}\
  \bibnamefont {Wang}}, \bibinfo {author} {\bibfnamefont {A.}~\bibnamefont
  {Keesling}}, \emph {et~al.},\ }\bibfield  {title} {\bibinfo {title}
  {Observation of string breaking on a (2+1) {D} {Rydberg} quantum simulator},\
  }\href {https://doi.org/10.48550/arXiv.2410.16558} {\bibfield  {journal}
  {\bibinfo  {journal} {arXiv preprint arXiv:2410.16558}\ } (\bibinfo {year}
  {2024})}\BibitemShut {NoStop}%
\bibitem [{\citenamefont {Luo}\ \emph {et~al.}(2025)\citenamefont {Luo},
  \citenamefont {Surace}, \citenamefont {De}, \citenamefont {Lerose},
  \citenamefont {Bennewitz}, \citenamefont {Ware}, \citenamefont {Schuckert},
  \citenamefont {Davoudi}, \citenamefont {Gorshkov}, \citenamefont {Katz} \emph
  {et~al.}}]{exp2}%
  \BibitemOpen
  \bibfield  {author} {\bibinfo {author} {\bibfnamefont {D.}~\bibnamefont
  {Luo}}, \bibinfo {author} {\bibfnamefont {F.~M.}\ \bibnamefont {Surace}},
  \bibinfo {author} {\bibfnamefont {A.}~\bibnamefont {De}}, \bibinfo {author}
  {\bibfnamefont {A.}~\bibnamefont {Lerose}}, \bibinfo {author} {\bibfnamefont
  {E.~R.}\ \bibnamefont {Bennewitz}}, \bibinfo {author} {\bibfnamefont
  {B.}~\bibnamefont {Ware}}, \bibinfo {author} {\bibfnamefont {A.}~\bibnamefont
  {Schuckert}}, \bibinfo {author} {\bibfnamefont {Z.}~\bibnamefont {Davoudi}},
  \bibinfo {author} {\bibfnamefont {A.~V.}\ \bibnamefont {Gorshkov}}, \bibinfo
  {author} {\bibfnamefont {O.}~\bibnamefont {Katz}}, \emph {et~al.},\
  }\bibfield  {title} {\bibinfo {title} {Quantum simulation of bubble
  nucleation across a quantum phase transition},\ }\href
  {https://doi.org/10.48550/arXiv.2505.09607} {\bibfield  {journal} {\bibinfo
  {journal} {arXiv preprint arXiv:2505.09607}\ } (\bibinfo {year}
  {2025})}\BibitemShut {NoStop}%
\bibitem [{\citenamefont {Sinha}\ \emph {et~al.}(2021)\citenamefont {Sinha},
  \citenamefont {Chanda},\ and\ \citenamefont {Dziarmaga}}]{Sinha2021}%
  \BibitemOpen
  \bibfield  {author} {\bibinfo {author} {\bibfnamefont {A.}~\bibnamefont
  {Sinha}}, \bibinfo {author} {\bibfnamefont {T.}~\bibnamefont {Chanda}},\ and\
  \bibinfo {author} {\bibfnamefont {J.}~\bibnamefont {Dziarmaga}},\ }\bibfield
  {title} {\bibinfo {title} {Nonadiabatic dynamics across a first-order quantum
  phase transition: Quantized bubble nucleation},\ }\href
  {https://doi.org/10.1103/PhysRevB.103.L220302} {\bibfield  {journal}
  {\bibinfo  {journal} {Phys. Rev. B}\ }\textbf {\bibinfo {volume} {103}},\
  \bibinfo {pages} {L220302} (\bibinfo {year} {2021})}\BibitemShut {NoStop}%
\bibitem [{\citenamefont {Dutta}\ \emph {et~al.}(2015)\citenamefont {Dutta},
  \citenamefont {Aeppli}, \citenamefont {Chakrabarti}, \citenamefont
  {Divakaran}, \citenamefont {Rosenbaum},\ and\ \citenamefont
  {Sen}}]{Dutta_2015}%
  \BibitemOpen
  \bibfield  {author} {\bibinfo {author} {\bibfnamefont {A.}~\bibnamefont
  {Dutta}}, \bibinfo {author} {\bibfnamefont {G.}~\bibnamefont {Aeppli}},
  \bibinfo {author} {\bibfnamefont {B.~K.}\ \bibnamefont {Chakrabarti}},
  \bibinfo {author} {\bibfnamefont {U.}~\bibnamefont {Divakaran}}, \bibinfo
  {author} {\bibfnamefont {T.~F.}\ \bibnamefont {Rosenbaum}},\ and\ \bibinfo
  {author} {\bibfnamefont {D.}~\bibnamefont {Sen}},\ }\bibinfo {title}
  {Experimental realizations of transverse field ising systems},\ in\
  \href@noop {} {\emph {\bibinfo {booktitle} {Quantum Phase Transitions in
  Transverse Field Spin Models: From Statistical Physics to Quantum
  Information}}}\ (\bibinfo  {publisher} {Cambridge University Press},\
  \bibinfo {year} {2015})\ p.\ \bibinfo {pages} {231–244}\BibitemShut
  {NoStop}%
\bibitem [{\citenamefont {Nevado}\ and\ \citenamefont
  {Porras}(2016)}]{Nevado2016}%
  \BibitemOpen
  \bibfield  {author} {\bibinfo {author} {\bibfnamefont {P.}~\bibnamefont
  {Nevado}}\ and\ \bibinfo {author} {\bibfnamefont {D.}~\bibnamefont
  {Porras}},\ }\bibfield  {title} {\bibinfo {title} {Hidden frustrated
  interactions and quantum annealing in trapped-ion spin-phonon chains},\
  }\href {https://doi.org/10.1103/PhysRevA.93.013625} {\bibfield  {journal}
  {\bibinfo  {journal} {Phys. Rev. A}\ }\textbf {\bibinfo {volume} {93}},\
  \bibinfo {pages} {013625} (\bibinfo {year} {2016})}\BibitemShut {NoStop}%
\bibitem [{\citenamefont {Schuckert}\ \emph {et~al.}(2023)\citenamefont
  {Schuckert}, \citenamefont {Katz}, \citenamefont {Feng}, \citenamefont
  {Crane}, \citenamefont {De}, \citenamefont {Hafezi}, \citenamefont
  {Gorshkov},\ and\ \citenamefont {Monroe}}]{schuckert2023observation}%
  \BibitemOpen
  \bibfield  {author} {\bibinfo {author} {\bibfnamefont {A.}~\bibnamefont
  {Schuckert}}, \bibinfo {author} {\bibfnamefont {O.}~\bibnamefont {Katz}},
  \bibinfo {author} {\bibfnamefont {L.}~\bibnamefont {Feng}}, \bibinfo {author}
  {\bibfnamefont {E.}~\bibnamefont {Crane}}, \bibinfo {author} {\bibfnamefont
  {A.}~\bibnamefont {De}}, \bibinfo {author} {\bibfnamefont {M.}~\bibnamefont
  {Hafezi}}, \bibinfo {author} {\bibfnamefont {A.~V.}\ \bibnamefont
  {Gorshkov}},\ and\ \bibinfo {author} {\bibfnamefont {C.}~\bibnamefont
  {Monroe}},\ }\bibfield  {title} {\bibinfo {title} {Observation of a
  finite-energy phase transition in a one-dimensional quantum simulator},\
  }\href {https://doi.org/10.48550/arXiv.2310.19869} {\bibfield  {journal}
  {\bibinfo  {journal} {arXiv preprint arXiv:2310.19869}\ } (\bibinfo {year}
  {2023})}\BibitemShut {NoStop}%
\bibitem [{\citenamefont {Katz}\ \emph {et~al.}(2024)\citenamefont {Katz},
  \citenamefont {Feng}, \citenamefont {Porras},\ and\ \citenamefont
  {Monroe}}]{katz2024observing}%
  \BibitemOpen
  \bibfield  {author} {\bibinfo {author} {\bibfnamefont {O.}~\bibnamefont
  {Katz}}, \bibinfo {author} {\bibfnamefont {L.}~\bibnamefont {Feng}}, \bibinfo
  {author} {\bibfnamefont {D.}~\bibnamefont {Porras}},\ and\ \bibinfo {author}
  {\bibfnamefont {C.}~\bibnamefont {Monroe}},\ }\bibfield  {title} {\bibinfo
  {title} {Observing topological insulator phases with a programmable quantum
  simulator},\ }\href {https://doi.org/10.48550/arXiv.2401.10362} {\bibfield
  {journal} {\bibinfo  {journal} {arXiv preprint arXiv:2401.10362}\ } (\bibinfo
  {year} {2024})}\BibitemShut {NoStop}%
\bibitem [{\citenamefont {Zhang}\ \emph
  {et~al.}(2023{\natexlab{a}})\citenamefont {Zhang}, \citenamefont {Kim},
  \citenamefont {Mark}, \citenamefont {Choi},\ and\ \citenamefont
  {Painter}}]{Zhang2023}%
  \BibitemOpen
  \bibfield  {author} {\bibinfo {author} {\bibfnamefont {X.}~\bibnamefont
  {Zhang}}, \bibinfo {author} {\bibfnamefont {E.}~\bibnamefont {Kim}}, \bibinfo
  {author} {\bibfnamefont {D.~K.}\ \bibnamefont {Mark}}, \bibinfo {author}
  {\bibfnamefont {S.}~\bibnamefont {Choi}},\ and\ \bibinfo {author}
  {\bibfnamefont {O.}~\bibnamefont {Painter}},\ }\bibfield  {title} {\bibinfo
  {title} {A superconducting quantum simulator based on a photonic-bandgap
  metamaterial},\ }\href {https://doi.org/10.1126/science.ade7651} {\bibfield
  {journal} {\bibinfo  {journal} {Science}\ }\textbf {\bibinfo {volume}
  {379}},\ \bibinfo {pages} {278} (\bibinfo {year}
  {2023}{\natexlab{a}})}\BibitemShut {NoStop}%
\bibitem [{\citenamefont {Periwal}\ \emph {et~al.}(2021)\citenamefont
  {Periwal}, \citenamefont {Cooper}, \citenamefont {Kunkel}, \citenamefont
  {Wienand}, \citenamefont {Davis},\ and\ \citenamefont
  {Schleier-Smith}}]{periwal2021programmable}%
  \BibitemOpen
  \bibfield  {author} {\bibinfo {author} {\bibfnamefont {A.}~\bibnamefont
  {Periwal}}, \bibinfo {author} {\bibfnamefont {E.~S.}\ \bibnamefont {Cooper}},
  \bibinfo {author} {\bibfnamefont {P.}~\bibnamefont {Kunkel}}, \bibinfo
  {author} {\bibfnamefont {J.~F.}\ \bibnamefont {Wienand}}, \bibinfo {author}
  {\bibfnamefont {E.~J.}\ \bibnamefont {Davis}},\ and\ \bibinfo {author}
  {\bibfnamefont {M.}~\bibnamefont {Schleier-Smith}},\ }\bibfield  {title}
  {\bibinfo {title} {Programmable interactions and emergent geometry in an
  array of atom clouds},\ }\href
  {https://doi.org/https://doi.org/10.1038/s41586-021-04156-0} {\bibfield
  {journal} {\bibinfo  {journal} {Nature}\ }\textbf {\bibinfo {volume} {600}},\
  \bibinfo {pages} {630} (\bibinfo {year} {2021})}\BibitemShut {NoStop}%
\bibitem [{\citenamefont {Lee}(2016)}]{Lee2016}%
  \BibitemOpen
  \bibfield  {author} {\bibinfo {author} {\bibfnamefont {T.~E.}\ \bibnamefont
  {Lee}},\ }\bibfield  {title} {\bibinfo {title} {Floquet engineering from
  long-range to short-range interactions},\ }\href
  {https://doi.org/10.1103/PhysRevA.94.040701} {\bibfield  {journal} {\bibinfo
  {journal} {Phys. Rev. A}\ }\textbf {\bibinfo {volume} {94}},\ \bibinfo
  {pages} {040701} (\bibinfo {year} {2016})}\BibitemShut {NoStop}%
\bibitem [{Note2()}]{Note2}%
  \BibitemOpen
  \bibinfo {note} {Following Appendix \ref {app:bubbles}, we can show that at
  the string-breaking point $h=h_c$, one has $E_s=E_{bs} < E_\protect \mathrm
  {other}$, where $E_\protect \mathrm {other}$ is the energy of an arbitrary
  configuration with $n<\ell $ spins polarized along $-\protect \hat z$. For a
  larger $h>h_c$, the difference $E_\protect \mathrm {other}-E_{bs}$ can only
  be larger, since it differs from the $h=h_c$ value by a quantity
  $2n(h-h_c)$.}\BibitemShut {Stop}%
\bibitem [{Note3()}]{Note3}%
  \BibitemOpen
  \bibinfo {note} {The initial state considered here is the ground state at
  finite transverse field $g$. Experimentally, this can be prepared with high
  fidelity with a quasi-adiabatic ramp of the transverse field from $g=0$,
  since the gap remains rather large along the ramp.}\BibitemShut {Stop}%
\bibitem [{Note4()}]{Note4}%
  \BibitemOpen
  \bibinfo {note} {To make the time dimensionality of $\tau $ explicit, one can
  reintroduce the energy scale $J_{1}$, set to $1$ earlier for simplicity, and
  define $h(t)=J_{1}\protect \, t/\tau $.}\BibitemShut {Stop}%
\bibitem [{\citenamefont {Pelissetto}\ \emph {et~al.}(2020)\citenamefont
  {Pelissetto}, \citenamefont {Rossini},\ and\ \citenamefont
  {Vicari}}]{Pelissetto2020}%
  \BibitemOpen
  \bibfield  {author} {\bibinfo {author} {\bibfnamefont {A.}~\bibnamefont
  {Pelissetto}}, \bibinfo {author} {\bibfnamefont {D.}~\bibnamefont
  {Rossini}},\ and\ \bibinfo {author} {\bibfnamefont {E.}~\bibnamefont
  {Vicari}},\ }\bibfield  {title} {\bibinfo {title} {Scaling properties of the
  dynamics at first-order quantum transitions when boundary conditions favor
  one of the two phases},\ }\href {https://doi.org/10.1103/PhysRevE.102.012143}
  {\bibfield  {journal} {\bibinfo  {journal} {Phys. Rev. E}\ }\textbf {\bibinfo
  {volume} {102}},\ \bibinfo {pages} {012143} (\bibinfo {year}
  {2020})}\BibitemShut {NoStop}%
\bibitem [{\citenamefont {Zener}\ and\ \citenamefont
  {Fowler}(1932)}]{Zener1932}%
  \BibitemOpen
  \bibfield  {author} {\bibinfo {author} {\bibfnamefont {C.}~\bibnamefont
  {Zener}}\ and\ \bibinfo {author} {\bibfnamefont {R.~H.}\ \bibnamefont
  {Fowler}},\ }\bibfield  {title} {\bibinfo {title} {Non-adiabatic crossing of
  energy levels},\ }\href {https://doi.org/10.1098/rspa.1932.0165} {\bibfield
  {journal} {\bibinfo  {journal} {Proceedings of the Royal Society of London.
  Series A, Containing Papers of a Mathematical and Physical Character}\
  }\textbf {\bibinfo {volume} {137}},\ \bibinfo {pages} {696} (\bibinfo {year}
  {1932})}\BibitemShut {NoStop}%
\bibitem [{Note5()}]{Note5}%
  \BibitemOpen
  \bibinfo {note} {The $\ell $ independence is expected for large enough
  strings: there, the local evolution of the magnetization in the bulk (where
  the effective field $h_j^\protect \text {eff}$ is small) is not affected by
  the edges (at least, for a certain time, set by the Lieb-Robinson bound \cite
  {Lieb1972}), and hence does not depend on $\ell $. We note, however, that
  this argument may not hold for the values of $\ell $ considered here and
  larger system sizes may be needed to eventually suppress the effect of the
  boundaries.}\BibitemShut {Stop}%
\bibitem [{\citenamefont {Pelissetto}\ and\ \citenamefont
  {Vicari}(2023)}]{pelissetto2023scaling}%
  \BibitemOpen
  \bibfield  {author} {\bibinfo {author} {\bibfnamefont {A.}~\bibnamefont
  {Pelissetto}}\ and\ \bibinfo {author} {\bibfnamefont {E.}~\bibnamefont
  {Vicari}},\ }\bibfield  {title} {\bibinfo {title} {Scaling behaviors at
  quantum and classical first-order transitions},\ }\href
  {https://doi.org/10.48550/arXiv.2302.08238} {\bibfield  {journal} {\bibinfo
  {journal} {arXiv preprint arXiv:2302.08238}\ } (\bibinfo {year}
  {2023})}\BibitemShut {NoStop}%
\bibitem [{\citenamefont {Monroe}\ \emph {et~al.}(2021)\citenamefont {Monroe},
  \citenamefont {Campbell}, \citenamefont {Duan}, \citenamefont {Gong},
  \citenamefont {Gorshkov}, \citenamefont {Hess}, \citenamefont {Islam},
  \citenamefont {Kim}, \citenamefont {Linke}, \citenamefont {Pagano},
  \citenamefont {Richerme}, \citenamefont {Senko},\ and\ \citenamefont
  {Yao}}]{MonroeReview}%
  \BibitemOpen
  \bibfield  {author} {\bibinfo {author} {\bibfnamefont {C.}~\bibnamefont
  {Monroe}}, \bibinfo {author} {\bibfnamefont {W.~C.}\ \bibnamefont
  {Campbell}}, \bibinfo {author} {\bibfnamefont {L.-M.}\ \bibnamefont {Duan}},
  \bibinfo {author} {\bibfnamefont {Z.-X.}\ \bibnamefont {Gong}}, \bibinfo
  {author} {\bibfnamefont {A.~V.}\ \bibnamefont {Gorshkov}}, \bibinfo {author}
  {\bibfnamefont {P.~W.}\ \bibnamefont {Hess}}, \bibinfo {author}
  {\bibfnamefont {R.}~\bibnamefont {Islam}}, \bibinfo {author} {\bibfnamefont
  {K.}~\bibnamefont {Kim}}, \bibinfo {author} {\bibfnamefont {N.~M.}\
  \bibnamefont {Linke}}, \bibinfo {author} {\bibfnamefont {G.}~\bibnamefont
  {Pagano}}, \bibinfo {author} {\bibfnamefont {P.}~\bibnamefont {Richerme}},
  \bibinfo {author} {\bibfnamefont {C.}~\bibnamefont {Senko}},\ and\ \bibinfo
  {author} {\bibfnamefont {N.~Y.}\ \bibnamefont {Yao}},\ }\bibfield  {title}
  {\bibinfo {title} {Programmable quantum simulations of spin systems with
  trapped ions},\ }\href {https://doi.org/10.1103/RevModPhys.93.025001}
  {\bibfield  {journal} {\bibinfo  {journal} {Rev. Mod. Phys.}\ }\textbf
  {\bibinfo {volume} {93}},\ \bibinfo {pages} {025001} (\bibinfo {year}
  {2021})}\BibitemShut {NoStop}%
\bibitem [{\citenamefont {Browaeys}\ and\ \citenamefont
  {Lahaye}(2020)}]{browaeys2020many}%
  \BibitemOpen
  \bibfield  {author} {\bibinfo {author} {\bibfnamefont {A.}~\bibnamefont
  {Browaeys}}\ and\ \bibinfo {author} {\bibfnamefont {T.}~\bibnamefont
  {Lahaye}},\ }\bibfield  {title} {\bibinfo {title} {Many-body physics with
  individually controlled {Rydberg} atoms},\ }\href
  {https://doi.org/https://doi.org/10.1038/s41567-019-0733-z} {\bibfield
  {journal} {\bibinfo  {journal} {Nature Physics}\ }\textbf {\bibinfo {volume}
  {16}},\ \bibinfo {pages} {132} (\bibinfo {year} {2020})}\BibitemShut
  {NoStop}%
\bibitem [{\citenamefont {Saffman}\ \emph {et~al.}(2010)\citenamefont
  {Saffman}, \citenamefont {Walker},\ and\ \citenamefont
  {M\o{}lmer}}]{Saffman2010}%
  \BibitemOpen
  \bibfield  {author} {\bibinfo {author} {\bibfnamefont {M.}~\bibnamefont
  {Saffman}}, \bibinfo {author} {\bibfnamefont {T.~G.}\ \bibnamefont
  {Walker}},\ and\ \bibinfo {author} {\bibfnamefont {K.}~\bibnamefont
  {M\o{}lmer}},\ }\bibfield  {title} {\bibinfo {title} {Quantum information
  with {Rydberg} atoms},\ }\href {https://doi.org/10.1103/RevModPhys.82.2313}
  {\bibfield  {journal} {\bibinfo  {journal} {Rev. Mod. Phys.}\ }\textbf
  {\bibinfo {volume} {82}},\ \bibinfo {pages} {2313} (\bibinfo {year}
  {2010})}\BibitemShut {NoStop}%
\bibitem [{\citenamefont {Norcia}\ and\ \citenamefont
  {Ferlaino}(2021)}]{norcia21}%
  \BibitemOpen
  \bibfield  {author} {\bibinfo {author} {\bibfnamefont {M.~A.}\ \bibnamefont
  {Norcia}}\ and\ \bibinfo {author} {\bibfnamefont {F.}~\bibnamefont
  {Ferlaino}},\ }\bibfield  {title} {\bibinfo {title} {Developments in atomic
  control using ultracold magnetic lanthanides},\ }\href
  {https://doi.org/https://doi.org/10.1038/s41567-021-01398-7} {\bibfield
  {journal} {\bibinfo  {journal} {Nature Physics}\ }\textbf {\bibinfo {volume}
  {17}},\ \bibinfo {pages} {1349} (\bibinfo {year} {2021})}\BibitemShut
  {NoStop}%
\bibitem [{\citenamefont {Cai}\ \emph {et~al.}(2013)\citenamefont {Cai},
  \citenamefont {Retzker}, \citenamefont {Jelezko},\ and\ \citenamefont
  {Plenio}}]{cai2013large}%
  \BibitemOpen
  \bibfield  {author} {\bibinfo {author} {\bibfnamefont {J.}~\bibnamefont
  {Cai}}, \bibinfo {author} {\bibfnamefont {A.}~\bibnamefont {Retzker}},
  \bibinfo {author} {\bibfnamefont {F.}~\bibnamefont {Jelezko}},\ and\ \bibinfo
  {author} {\bibfnamefont {M.~B.}\ \bibnamefont {Plenio}},\ }\bibfield  {title}
  {\bibinfo {title} {A large-scale quantum simulator on a diamond surface at
  room temperature},\ }\href
  {https://doi.org/https://doi.org/10.1038/nphys2519} {\bibfield  {journal}
  {\bibinfo  {journal} {Nature Physics}\ }\textbf {\bibinfo {volume} {9}},\
  \bibinfo {pages} {168} (\bibinfo {year} {2013})}\BibitemShut {NoStop}%
\bibitem [{\citenamefont {Gorshkov}\ \emph {et~al.}(2011)\citenamefont
  {Gorshkov}, \citenamefont {Manmana}, \citenamefont {Chen}, \citenamefont
  {Ye}, \citenamefont {Demler}, \citenamefont {Lukin},\ and\ \citenamefont
  {Rey}}]{Gorshkov2011}%
  \BibitemOpen
  \bibfield  {author} {\bibinfo {author} {\bibfnamefont {A.~V.}\ \bibnamefont
  {Gorshkov}}, \bibinfo {author} {\bibfnamefont {S.~R.}\ \bibnamefont
  {Manmana}}, \bibinfo {author} {\bibfnamefont {G.}~\bibnamefont {Chen}},
  \bibinfo {author} {\bibfnamefont {J.}~\bibnamefont {Ye}}, \bibinfo {author}
  {\bibfnamefont {E.}~\bibnamefont {Demler}}, \bibinfo {author} {\bibfnamefont
  {M.~D.}\ \bibnamefont {Lukin}},\ and\ \bibinfo {author} {\bibfnamefont
  {A.~M.}\ \bibnamefont {Rey}},\ }\bibfield  {title} {\bibinfo {title} {Tunable
  superfluidity and quantum magnetism with ultracold polar molecules},\ }\href
  {https://doi.org/10.1103/PhysRevLett.107.115301} {\bibfield  {journal}
  {\bibinfo  {journal} {Phys. Rev. Lett.}\ }\textbf {\bibinfo {volume} {107}},\
  \bibinfo {pages} {115301} (\bibinfo {year} {2011})}\BibitemShut {NoStop}%
\bibitem [{\citenamefont {Yan}\ \emph {et~al.}(2013)\citenamefont {Yan},
  \citenamefont {Moses}, \citenamefont {Gadway}, \citenamefont {Covey},
  \citenamefont {Hazzard}, \citenamefont {Rey}, \citenamefont {Jin},\ and\
  \citenamefont {Ye}}]{yan2013observation}%
  \BibitemOpen
  \bibfield  {author} {\bibinfo {author} {\bibfnamefont {B.}~\bibnamefont
  {Yan}}, \bibinfo {author} {\bibfnamefont {S.~A.}\ \bibnamefont {Moses}},
  \bibinfo {author} {\bibfnamefont {B.}~\bibnamefont {Gadway}}, \bibinfo
  {author} {\bibfnamefont {J.~P.}\ \bibnamefont {Covey}}, \bibinfo {author}
  {\bibfnamefont {K.~R.}\ \bibnamefont {Hazzard}}, \bibinfo {author}
  {\bibfnamefont {A.~M.}\ \bibnamefont {Rey}}, \bibinfo {author} {\bibfnamefont
  {D.~S.}\ \bibnamefont {Jin}},\ and\ \bibinfo {author} {\bibfnamefont
  {J.}~\bibnamefont {Ye}},\ }\bibfield  {title} {\bibinfo {title} {Observation
  of dipolar spin-exchange interactions with lattice-confined polar
  molecules},\ }\href {https://doi.org/https://doi.org/10.1038/nature12483}
  {\bibfield  {journal} {\bibinfo  {journal} {Nature}\ }\textbf {\bibinfo
  {volume} {501}},\ \bibinfo {pages} {521} (\bibinfo {year}
  {2013})}\BibitemShut {NoStop}%
\bibitem [{\citenamefont {Defenu}\ \emph {et~al.}(2023)\citenamefont {Defenu},
  \citenamefont {Donner}, \citenamefont {Macr\`{\i}}, \citenamefont {Pagano},
  \citenamefont {Ruffo},\ and\ \citenamefont {Trombettoni}}]{Defenu2023}%
  \BibitemOpen
  \bibfield  {author} {\bibinfo {author} {\bibfnamefont {N.}~\bibnamefont
  {Defenu}}, \bibinfo {author} {\bibfnamefont {T.}~\bibnamefont {Donner}},
  \bibinfo {author} {\bibfnamefont {T.}~\bibnamefont {Macr\`{\i}}}, \bibinfo
  {author} {\bibfnamefont {G.}~\bibnamefont {Pagano}}, \bibinfo {author}
  {\bibfnamefont {S.}~\bibnamefont {Ruffo}},\ and\ \bibinfo {author}
  {\bibfnamefont {A.}~\bibnamefont {Trombettoni}},\ }\bibfield  {title}
  {\bibinfo {title} {Long-range interacting quantum systems},\ }\href
  {https://doi.org/10.1103/RevModPhys.95.035002} {\bibfield  {journal}
  {\bibinfo  {journal} {Rev. Mod. Phys.}\ }\textbf {\bibinfo {volume} {95}},\
  \bibinfo {pages} {035002} (\bibinfo {year} {2023})}\BibitemShut {NoStop}%
\bibitem [{\citenamefont {Defenu}\ \emph {et~al.}(2024)\citenamefont {Defenu},
  \citenamefont {Lerose},\ and\ \citenamefont {Pappalardi}}]{DEFENU20241}%
  \BibitemOpen
  \bibfield  {author} {\bibinfo {author} {\bibfnamefont {N.}~\bibnamefont
  {Defenu}}, \bibinfo {author} {\bibfnamefont {A.}~\bibnamefont {Lerose}},\
  and\ \bibinfo {author} {\bibfnamefont {S.}~\bibnamefont {Pappalardi}},\
  }\bibfield  {title} {\bibinfo {title} {Out-of-equilibrium dynamics of quantum
  many-body systems with long-range interactions},\ }\href
  {https://doi.org/https://doi.org/10.1016/j.physrep.2024.04.005} {\bibfield
  {journal} {\bibinfo  {journal} {Physics Reports}\ }\textbf {\bibinfo {volume}
  {1074}},\ \bibinfo {pages} {1} (\bibinfo {year} {2024})},\ \bibinfo {note}
  {out-of-equilibrium dynamics of quantum many-body systems with long-range
  interactions}\BibitemShut {NoStop}%
\bibitem [{\citenamefont {Lerose}\ \emph {et~al.}(2019)\citenamefont {Lerose},
  \citenamefont {\ifmmode \check{Z}\else
  \v{Z}\fi{}unkovi\ifmmode~\check{c}\else \v{c}\fi{}}, \citenamefont {Silva},\
  and\ \citenamefont {Gambassi}}]{Lerose2019}%
  \BibitemOpen
  \bibfield  {author} {\bibinfo {author} {\bibfnamefont {A.}~\bibnamefont
  {Lerose}}, \bibinfo {author} {\bibfnamefont {B.}~\bibnamefont {\ifmmode
  \check{Z}\else \v{Z}\fi{}unkovi\ifmmode~\check{c}\else \v{c}\fi{}}}, \bibinfo
  {author} {\bibfnamefont {A.}~\bibnamefont {Silva}},\ and\ \bibinfo {author}
  {\bibfnamefont {A.}~\bibnamefont {Gambassi}},\ }\bibfield  {title} {\bibinfo
  {title} {Quasilocalized excitations induced by long-range interactions in
  translationally invariant quantum spin chains},\ }\href
  {https://doi.org/10.1103/PhysRevB.99.121112} {\bibfield  {journal} {\bibinfo
  {journal} {Phys. Rev. B}\ }\textbf {\bibinfo {volume} {99}},\ \bibinfo
  {pages} {121112} (\bibinfo {year} {2019})}\BibitemShut {NoStop}%
\bibitem [{Note6()}]{Note6}%
  \BibitemOpen
  \bibinfo {note} {This property underlies the stabilization of
  finite-temperature order in this model~\cite
  {dyson1969existence}}\BibitemShut {NoStop}%
\bibitem [{Note7()}]{Note7}%
  \BibitemOpen
  \bibinfo {note} {For $0\le \alpha \le 1$, the notion of locality, and hence
  domain-wall confinement, is not well-defined~\cite
  {Liu2019,Lerose2019}.}\BibitemShut {Stop}%
\bibitem [{\citenamefont {Hawking}\ \emph {et~al.}(1982)\citenamefont
  {Hawking}, \citenamefont {Moss},\ and\ \citenamefont
  {Stewart}}]{Hawking:1982ga}%
  \BibitemOpen
  \bibfield  {author} {\bibinfo {author} {\bibfnamefont {S.~W.}\ \bibnamefont
  {Hawking}}, \bibinfo {author} {\bibfnamefont {I.~G.}\ \bibnamefont {Moss}},\
  and\ \bibinfo {author} {\bibfnamefont {J.~M.}\ \bibnamefont {Stewart}},\
  }\bibfield  {title} {\bibinfo {title} {{Bubble Collisions in the Very Early
  Universe}},\ }\href {https://doi.org/10.1103/PhysRevD.26.2681} {\bibfield
  {journal} {\bibinfo  {journal} {Phys. Rev. D}\ }\textbf {\bibinfo {volume}
  {26}},\ \bibinfo {pages} {2681} (\bibinfo {year} {1982})}\BibitemShut
  {NoStop}%
\bibitem [{\citenamefont {Coleman}(1977)}]{Coleman:1977py}%
  \BibitemOpen
  \bibfield  {author} {\bibinfo {author} {\bibfnamefont {S.~R.}\ \bibnamefont
  {Coleman}},\ }\bibfield  {title} {\bibinfo {title} {{The Fate of the False
  Vacuum. 1. Semiclassical Theory}},\ }\href
  {https://doi.org/10.1103/PhysRevD.16.1248} {\bibfield  {journal} {\bibinfo
  {journal} {Phys. Rev. D}\ }\textbf {\bibinfo {volume} {15}},\ \bibinfo
  {pages} {2929} (\bibinfo {year} {1977})},\ \bibinfo {note} {[Erratum:
  Phys.Rev.D 16, 1248 (1977)]}\BibitemShut {NoStop}%
\bibitem [{\citenamefont {Callan}\ and\ \citenamefont
  {Coleman}(1977)}]{Callan:1977pt}%
  \BibitemOpen
  \bibfield  {author} {\bibinfo {author} {\bibfnamefont {C.~G.}\ \bibnamefont
  {Callan}, \bibfnamefont {Jr.}}\ and\ \bibinfo {author} {\bibfnamefont
  {S.~R.}\ \bibnamefont {Coleman}},\ }\bibfield  {title} {\bibinfo {title}
  {{The Fate of the False Vacuum. 2. First Quantum Corrections}},\ }\href
  {https://doi.org/10.1103/PhysRevD.16.1762} {\bibfield  {journal} {\bibinfo
  {journal} {Phys. Rev. D}\ }\textbf {\bibinfo {volume} {16}},\ \bibinfo
  {pages} {1762} (\bibinfo {year} {1977})}\BibitemShut {NoStop}%
\bibitem [{\citenamefont {Lagnese}\ \emph {et~al.}(2021)\citenamefont
  {Lagnese}, \citenamefont {Surace}, \citenamefont {Kormos},\ and\
  \citenamefont {Calabrese}}]{LagneseFVD}%
  \BibitemOpen
  \bibfield  {author} {\bibinfo {author} {\bibfnamefont {G.}~\bibnamefont
  {Lagnese}}, \bibinfo {author} {\bibfnamefont {F.~M.}\ \bibnamefont {Surace}},
  \bibinfo {author} {\bibfnamefont {M.}~\bibnamefont {Kormos}},\ and\ \bibinfo
  {author} {\bibfnamefont {P.}~\bibnamefont {Calabrese}},\ }\bibfield  {title}
  {\bibinfo {title} {False vacuum decay in quantum spin chains},\ }\href
  {https://doi.org/10.1103/PhysRevB.104.L201106} {\bibfield  {journal}
  {\bibinfo  {journal} {Phys. Rev. B}\ }\textbf {\bibinfo {volume} {104}},\
  \bibinfo {pages} {L201106} (\bibinfo {year} {2021})}\BibitemShut {NoStop}%
\bibitem [{\citenamefont {Lagnese}\ \emph {et~al.}(2023)\citenamefont
  {Lagnese}, \citenamefont {Surace}, \citenamefont {Morampudi},\ and\
  \citenamefont {Wilczek}}]{lagnese2023detecting}%
  \BibitemOpen
  \bibfield  {author} {\bibinfo {author} {\bibfnamefont {G.}~\bibnamefont
  {Lagnese}}, \bibinfo {author} {\bibfnamefont {F.~M.}\ \bibnamefont {Surace}},
  \bibinfo {author} {\bibfnamefont {S.}~\bibnamefont {Morampudi}},\ and\
  \bibinfo {author} {\bibfnamefont {F.}~\bibnamefont {Wilczek}},\ }\bibfield
  {title} {\bibinfo {title} {Detecting a long lived false vacuum with quantum
  quenches},\ }\href {https://doi.org/10.48550/arXiv.2308.08340} {\bibfield
  {journal} {\bibinfo  {journal} {arXiv preprint arXiv:2308.08340}\ } (\bibinfo
  {year} {2023})}\BibitemShut {NoStop}%
\bibitem [{\citenamefont {Darbha}\ \emph
  {et~al.}(2024{\natexlab{a}})\citenamefont {Darbha}, \citenamefont
  {Kornja\ifmmode~\check{c}\else \v{c}\fi{}a}, \citenamefont {Liu},
  \citenamefont {Balewski}, \citenamefont {Hirsbrunner}, \citenamefont {Lopes},
  \citenamefont {Wang}, \citenamefont {Van~Beeumen}, \citenamefont {Camps},\
  and\ \citenamefont {Klymko}}]{Darbha2024a}%
  \BibitemOpen
  \bibfield  {author} {\bibinfo {author} {\bibfnamefont {S.}~\bibnamefont
  {Darbha}}, \bibinfo {author} {\bibfnamefont {M.}~\bibnamefont
  {Kornja\ifmmode~\check{c}\else \v{c}\fi{}a}}, \bibinfo {author}
  {\bibfnamefont {F.}~\bibnamefont {Liu}}, \bibinfo {author} {\bibfnamefont
  {J.}~\bibnamefont {Balewski}}, \bibinfo {author} {\bibfnamefont {M.~R.}\
  \bibnamefont {Hirsbrunner}}, \bibinfo {author} {\bibfnamefont {P.~L.~S.}\
  \bibnamefont {Lopes}}, \bibinfo {author} {\bibfnamefont {S.-T.}\ \bibnamefont
  {Wang}}, \bibinfo {author} {\bibfnamefont {R.}~\bibnamefont {Van~Beeumen}},
  \bibinfo {author} {\bibfnamefont {D.}~\bibnamefont {Camps}},\ and\ \bibinfo
  {author} {\bibfnamefont {K.}~\bibnamefont {Klymko}},\ }\bibfield  {title}
  {\bibinfo {title} {False vacuum decay and nucleation dynamics in neutral atom
  systems},\ }\href {https://doi.org/10.1103/PhysRevB.110.155103} {\bibfield
  {journal} {\bibinfo  {journal} {Phys. Rev. B}\ }\textbf {\bibinfo {volume}
  {110}},\ \bibinfo {pages} {155103} (\bibinfo {year}
  {2024}{\natexlab{a}})}\BibitemShut {NoStop}%
\bibitem [{\citenamefont {Darbha}\ \emph
  {et~al.}(2024{\natexlab{b}})\citenamefont {Darbha}, \citenamefont
  {Kornja\ifmmode~\check{c}\else \v{c}\fi{}a}, \citenamefont {Liu},
  \citenamefont {Balewski}, \citenamefont {Hirsbrunner}, \citenamefont {Lopes},
  \citenamefont {Wang}, \citenamefont {Van~Beeumen}, \citenamefont {Klymko},\
  and\ \citenamefont {Camps}}]{Darbha2024b}%
  \BibitemOpen
  \bibfield  {author} {\bibinfo {author} {\bibfnamefont {S.}~\bibnamefont
  {Darbha}}, \bibinfo {author} {\bibfnamefont {M.}~\bibnamefont
  {Kornja\ifmmode~\check{c}\else \v{c}\fi{}a}}, \bibinfo {author}
  {\bibfnamefont {F.}~\bibnamefont {Liu}}, \bibinfo {author} {\bibfnamefont
  {J.}~\bibnamefont {Balewski}}, \bibinfo {author} {\bibfnamefont {M.~R.}\
  \bibnamefont {Hirsbrunner}}, \bibinfo {author} {\bibfnamefont {P.~L.~S.}\
  \bibnamefont {Lopes}}, \bibinfo {author} {\bibfnamefont {S.-T.}\ \bibnamefont
  {Wang}}, \bibinfo {author} {\bibfnamefont {R.}~\bibnamefont {Van~Beeumen}},
  \bibinfo {author} {\bibfnamefont {K.}~\bibnamefont {Klymko}},\ and\ \bibinfo
  {author} {\bibfnamefont {D.}~\bibnamefont {Camps}},\ }\bibfield  {title}
  {\bibinfo {title} {Long-lived oscillations of metastable states in neutral
  atom systems},\ }\href {https://doi.org/10.1103/PhysRevB.110.155114}
  {\bibfield  {journal} {\bibinfo  {journal} {Phys. Rev. B}\ }\textbf {\bibinfo
  {volume} {110}},\ \bibinfo {pages} {155114} (\bibinfo {year}
  {2024}{\natexlab{b}})}\BibitemShut {NoStop}%
\bibitem [{\citenamefont {Vodeb}\ \emph {et~al.}(2024)\citenamefont {Vodeb},
  \citenamefont {Desaules}, \citenamefont {Hallam}, \citenamefont {Rava},
  \citenamefont {Humar}, \citenamefont {Willsch}, \citenamefont {Jin},
  \citenamefont {Willsch}, \citenamefont {Michielsen},\ and\ \citenamefont
  {Papi{\'c}}}]{vodeb2024stirring}%
  \BibitemOpen
  \bibfield  {author} {\bibinfo {author} {\bibfnamefont {J.}~\bibnamefont
  {Vodeb}}, \bibinfo {author} {\bibfnamefont {J.-Y.}\ \bibnamefont {Desaules}},
  \bibinfo {author} {\bibfnamefont {A.}~\bibnamefont {Hallam}}, \bibinfo
  {author} {\bibfnamefont {A.}~\bibnamefont {Rava}}, \bibinfo {author}
  {\bibfnamefont {G.}~\bibnamefont {Humar}}, \bibinfo {author} {\bibfnamefont
  {D.}~\bibnamefont {Willsch}}, \bibinfo {author} {\bibfnamefont
  {F.}~\bibnamefont {Jin}}, \bibinfo {author} {\bibfnamefont {M.}~\bibnamefont
  {Willsch}}, \bibinfo {author} {\bibfnamefont {K.}~\bibnamefont
  {Michielsen}},\ and\ \bibinfo {author} {\bibfnamefont {Z.}~\bibnamefont
  {Papi{\'c}}},\ }\bibfield  {title} {\bibinfo {title} {Stirring the false
  vacuum via interacting quantized bubbles on a 5564-qubit quantum annealer},\
  }\href {https://doi.org/10.48550/arXiv.2406.14718} {\bibfield  {journal}
  {\bibinfo  {journal} {arXiv preprint arXiv:2406.14718}\ } (\bibinfo {year}
  {2024})}\BibitemShut {NoStop}%
\bibitem [{\citenamefont {Balian}\ \emph {et~al.}(1975)\citenamefont {Balian},
  \citenamefont {Drouffe},\ and\ \citenamefont {Itzykson}}]{balian1975gauge}%
  \BibitemOpen
  \bibfield  {author} {\bibinfo {author} {\bibfnamefont {R.}~\bibnamefont
  {Balian}}, \bibinfo {author} {\bibfnamefont {J.~M.}\ \bibnamefont
  {Drouffe}},\ and\ \bibinfo {author} {\bibfnamefont {C.}~\bibnamefont
  {Itzykson}},\ }\bibfield  {title} {\bibinfo {title} {Gauge fields on a
  lattice. {II.} gauge-invariant {I}sing model},\ }\href
  {https://doi.org/10.1103/PhysRevD.11.2098} {\bibfield  {journal} {\bibinfo
  {journal} {Phys. Rev. D}\ }\textbf {\bibinfo {volume} {11}},\ \bibinfo
  {pages} {2098} (\bibinfo {year} {1975})}\BibitemShut {NoStop}%
\bibitem [{\citenamefont {Zhang}\ \emph {et~al.}(2018)\citenamefont {Zhang}
  \emph {et~al.}}]{zhang2018quantum}%
  \BibitemOpen
  \bibfield  {author} {\bibinfo {author} {\bibfnamefont {J.}~\bibnamefont
  {Zhang}} \emph {et~al.},\ }\bibfield  {title} {\bibinfo {title} {Quantum
  simulation of the universal features of the {Polyakov} loop},\ }\href
  {https://doi.org/10.1103/PhysRevLett.121.223201} {\bibfield  {journal}
  {\bibinfo  {journal} {Phys. Rev. Lett.}\ }\textbf {\bibinfo {volume} {121}},\
  \bibinfo {pages} {223201} (\bibinfo {year} {2018})}\BibitemShut {NoStop}%
\bibitem [{\citenamefont {Borla}\ \emph {et~al.}(2021)\citenamefont {Borla},
  \citenamefont {Verresen}, \citenamefont {Shah},\ and\ \citenamefont
  {Moroz}}]{borla2020gauging}%
  \BibitemOpen
  \bibfield  {author} {\bibinfo {author} {\bibfnamefont {U.}~\bibnamefont
  {Borla}}, \bibinfo {author} {\bibfnamefont {R.}~\bibnamefont {Verresen}},
  \bibinfo {author} {\bibfnamefont {J.}~\bibnamefont {Shah}},\ and\ \bibinfo
  {author} {\bibfnamefont {S.}~\bibnamefont {Moroz}},\ }\bibfield  {title}
  {\bibinfo {title} {{Gauging the Kitaev chain}},\ }\href
  {https://doi.org/10.21468/SciPostPhys.10.6.148} {\bibfield  {journal}
  {\bibinfo  {journal} {SciPost Phys.}\ }\textbf {\bibinfo {volume} {10}},\
  \bibinfo {pages} {148} (\bibinfo {year} {2021})}\BibitemShut {NoStop}%
\bibitem [{\citenamefont {Surace}\ and\ \citenamefont
  {Lerose}(2021)}]{Surace_2021}%
  \BibitemOpen
  \bibfield  {author} {\bibinfo {author} {\bibfnamefont {F.~M.}\ \bibnamefont
  {Surace}}\ and\ \bibinfo {author} {\bibfnamefont {A.}~\bibnamefont
  {Lerose}},\ }\bibfield  {title} {\bibinfo {title} {Scattering of mesons in
  quantum simulators},\ }\href {https://doi.org/10.1088/1367-2630/abfc40}
  {\bibfield  {journal} {\bibinfo  {journal} {New J. Phys.}\ }\textbf {\bibinfo
  {volume} {23}},\ \bibinfo {pages} {062001} (\bibinfo {year}
  {2021})}\BibitemShut {NoStop}%
\bibitem [{\citenamefont {Hunt-Smith}\ and\ \citenamefont
  {Skands}(2020)}]{Huntsmith2020}%
  \BibitemOpen
  \bibfield  {author} {\bibinfo {author} {\bibfnamefont {N.}~\bibnamefont
  {Hunt-Smith}}\ and\ \bibinfo {author} {\bibfnamefont {P.}~\bibnamefont
  {Skands}},\ }\bibfield  {title} {\bibinfo {title} {String fragmentation with
  a time-dependent tension},\ }\href
  {https://doi.org/10.1140/epjc/s10052-020-08654-9} {\bibfield  {journal}
  {\bibinfo  {journal} {European Physical Journal C}\ }\textbf {\bibinfo
  {volume} {80}} (\bibinfo {year} {2020})}\BibitemShut {NoStop}%
\bibitem [{\citenamefont {Belyansky}\ \emph {et~al.}(2024)\citenamefont
  {Belyansky}, \citenamefont {Whitsitt}, \citenamefont {Mueller}, \citenamefont
  {Fahimniya}, \citenamefont {Bennewitz}, \citenamefont {Davoudi},\ and\
  \citenamefont {Gorshkov}}]{belyansky2024high}%
  \BibitemOpen
  \bibfield  {author} {\bibinfo {author} {\bibfnamefont {R.}~\bibnamefont
  {Belyansky}}, \bibinfo {author} {\bibfnamefont {S.}~\bibnamefont {Whitsitt}},
  \bibinfo {author} {\bibfnamefont {N.}~\bibnamefont {Mueller}}, \bibinfo
  {author} {\bibfnamefont {A.}~\bibnamefont {Fahimniya}}, \bibinfo {author}
  {\bibfnamefont {E.~R.}\ \bibnamefont {Bennewitz}}, \bibinfo {author}
  {\bibfnamefont {Z.}~\bibnamefont {Davoudi}},\ and\ \bibinfo {author}
  {\bibfnamefont {A.~V.}\ \bibnamefont {Gorshkov}},\ }\bibfield  {title}
  {\bibinfo {title} {{High-energy collision of quarks and mesons in the
  Schwinger model: From tensor networks to circuit QED}},\ }\href
  {https://doi.org/https://doi.org/10.1103/PhysRevLett.132.091903} {\bibfield
  {journal} {\bibinfo  {journal} {Physical Review Letters}\ }\textbf {\bibinfo
  {volume} {132}},\ \bibinfo {pages} {091903} (\bibinfo {year}
  {2024})}\BibitemShut {NoStop}%
\bibitem [{\citenamefont {Bennewitz}\ \emph {et~al.}(2024)\citenamefont
  {Bennewitz}, \citenamefont {Ware}, \citenamefont {Schuckert}, \citenamefont
  {Lerose}, \citenamefont {Surace}, \citenamefont {Belyansky}, \citenamefont
  {Morong}, \citenamefont {Luo}, \citenamefont {De}, \citenamefont {Collins},
  \citenamefont {Katz}, \citenamefont {Monroe}, \citenamefont {Davoudi},\ and\
  \citenamefont {Gorshkov}}]{bennewitz2024simulating}%
  \BibitemOpen
  \bibfield  {author} {\bibinfo {author} {\bibfnamefont {E.~R.}\ \bibnamefont
  {Bennewitz}}, \bibinfo {author} {\bibfnamefont {B.}~\bibnamefont {Ware}},
  \bibinfo {author} {\bibfnamefont {A.}~\bibnamefont {Schuckert}}, \bibinfo
  {author} {\bibfnamefont {A.}~\bibnamefont {Lerose}}, \bibinfo {author}
  {\bibfnamefont {F.}~\bibnamefont {Surace}}, \bibinfo {author} {\bibfnamefont
  {R.}~\bibnamefont {Belyansky}}, \bibinfo {author} {\bibfnamefont
  {W.}~\bibnamefont {Morong}}, \bibinfo {author} {\bibfnamefont
  {D.}~\bibnamefont {Luo}}, \bibinfo {author} {\bibfnamefont {A.}~\bibnamefont
  {De}}, \bibinfo {author} {\bibfnamefont {K.}~\bibnamefont {Collins}},
  \bibinfo {author} {\bibfnamefont {O.}~\bibnamefont {Katz}}, \bibinfo {author}
  {\bibfnamefont {C.}~\bibnamefont {Monroe}}, \bibinfo {author} {\bibfnamefont
  {Z.}~\bibnamefont {Davoudi}},\ and\ \bibinfo {author} {\bibfnamefont {A.~G.}\
  \bibnamefont {Gorshkov}},\ }\bibfield  {title} {\bibinfo {title} {Simulating
  meson scattering on spin quantum simulators},\ }\href
  {https://doi.org/10.48550/arXiv.2403.07061} {\bibfield  {journal} {\bibinfo
  {journal} {arXiv preprint arXiv:2403.07061}\ } (\bibinfo {year}
  {2024})}\BibitemShut {NoStop}%
\bibitem [{\citenamefont {Su}\ \emph {et~al.}(2024)\citenamefont {Su},
  \citenamefont {Osborne},\ and\ \citenamefont {Halimeh}}]{su2024cold}%
  \BibitemOpen
  \bibfield  {author} {\bibinfo {author} {\bibfnamefont {G.-X.}\ \bibnamefont
  {Su}}, \bibinfo {author} {\bibfnamefont {J.~J.}\ \bibnamefont {Osborne}},\
  and\ \bibinfo {author} {\bibfnamefont {J.~C.}\ \bibnamefont {Halimeh}},\
  }\bibfield  {title} {\bibinfo {title} {Cold-atom particle collider},\
  }\href@noop {} {\bibfield  {journal} {\bibinfo  {journal} {PRX Quantum}\
  }\textbf {\bibinfo {volume} {5}},\ \bibinfo {pages} {040310} (\bibinfo {year}
  {2024})}\BibitemShut {NoStop}%
\bibitem [{\citenamefont {Davoudi}\ \emph {et~al.}(2024)\citenamefont
  {Davoudi}, \citenamefont {Jarzynski}, \citenamefont {Mueller}, \citenamefont
  {Oruganti}, \citenamefont {Powers},\ and\ \citenamefont
  {Halpern}}]{davoudi2024quantum}%
  \BibitemOpen
  \bibfield  {author} {\bibinfo {author} {\bibfnamefont {Z.}~\bibnamefont
  {Davoudi}}, \bibinfo {author} {\bibfnamefont {C.}~\bibnamefont {Jarzynski}},
  \bibinfo {author} {\bibfnamefont {N.}~\bibnamefont {Mueller}}, \bibinfo
  {author} {\bibfnamefont {G.}~\bibnamefont {Oruganti}}, \bibinfo {author}
  {\bibfnamefont {C.}~\bibnamefont {Powers}},\ and\ \bibinfo {author}
  {\bibfnamefont {N.~Y.}\ \bibnamefont {Halpern}},\ }\bibfield  {title}
  {\bibinfo {title} {Quantum thermodynamics of nonequilibrium processes in
  lattice gauge theories},\ }\href {https://doi.org/10.48550/arXiv.2404.02965}
  {\bibfield  {journal} {\bibinfo  {journal} {arXiv preprint arXiv:2404.02965}\
  } (\bibinfo {year} {2024})}\BibitemShut {NoStop}%
\bibitem [{\citenamefont {Labuhn}\ \emph {et~al.}(2016)\citenamefont {Labuhn},
  \citenamefont {Barredo}, \citenamefont {Ravets}, \citenamefont
  {de~Léséleuc}, \citenamefont {Macrì}, \citenamefont {Lahaye},\ and\
  \citenamefont {Browaeys}}]{Labuhn2016}%
  \BibitemOpen
  \bibfield  {author} {\bibinfo {author} {\bibfnamefont {H.}~\bibnamefont
  {Labuhn}}, \bibinfo {author} {\bibfnamefont {D.}~\bibnamefont {Barredo}},
  \bibinfo {author} {\bibfnamefont {S.}~\bibnamefont {Ravets}}, \bibinfo
  {author} {\bibfnamefont {S.}~\bibnamefont {de~Léséleuc}}, \bibinfo {author}
  {\bibfnamefont {T.}~\bibnamefont {Macrì}}, \bibinfo {author} {\bibfnamefont
  {T.}~\bibnamefont {Lahaye}},\ and\ \bibinfo {author} {\bibfnamefont
  {A.}~\bibnamefont {Browaeys}},\ }\bibfield  {title} {\bibinfo {title}
  {{Tunable two-dimensional arrays of single Rydberg atoms for realizing
  quantum Ising models}},\ }\href {https://doi.org/10.1038/nature18274}
  {\bibfield  {journal} {\bibinfo  {journal} {Nature}\ }\textbf {\bibinfo
  {volume} {534}},\ \bibinfo {pages} {667–670} (\bibinfo {year}
  {2016})}\BibitemShut {NoStop}%
\bibitem [{\citenamefont {{Bernien}}\ \emph {et~al.}(2017)\citenamefont
  {{Bernien}}, \citenamefont {{Schwartz}}, \citenamefont {{Keesling}},
  \citenamefont {{Levine}}, \citenamefont {{Omran}}, \citenamefont {{Pichler}},
  \citenamefont {{Choi}}, \citenamefont {{Zibrov}}, \citenamefont {{Endres}},
  \citenamefont {{Greiner}}, \citenamefont {{Vuleti{\'c}}},\ and\ \citenamefont
  {{Lukin}}}]{Bernien2017}%
  \BibitemOpen
  \bibfield  {author} {\bibinfo {author} {\bibfnamefont {H.}~\bibnamefont
  {{Bernien}}}, \bibinfo {author} {\bibfnamefont {S.}~\bibnamefont
  {{Schwartz}}}, \bibinfo {author} {\bibfnamefont {A.}~\bibnamefont
  {{Keesling}}}, \bibinfo {author} {\bibfnamefont {H.}~\bibnamefont
  {{Levine}}}, \bibinfo {author} {\bibfnamefont {A.}~\bibnamefont {{Omran}}},
  \bibinfo {author} {\bibfnamefont {H.}~\bibnamefont {{Pichler}}}, \bibinfo
  {author} {\bibfnamefont {S.}~\bibnamefont {{Choi}}}, \bibinfo {author}
  {\bibfnamefont {A.~S.}\ \bibnamefont {{Zibrov}}}, \bibinfo {author}
  {\bibfnamefont {M.}~\bibnamefont {{Endres}}}, \bibinfo {author}
  {\bibfnamefont {M.}~\bibnamefont {{Greiner}}}, \bibinfo {author}
  {\bibfnamefont {V.}~\bibnamefont {{Vuleti{\'c}}}},\ and\ \bibinfo {author}
  {\bibfnamefont {M.~D.}\ \bibnamefont {{Lukin}}},\ }\bibfield  {title}
  {\bibinfo {title} {{Probing many-body dynamics on a 51-atom quantum
  simulator}},\ }\href {https://doi.org/10.1038/nature24622} {\bibfield
  {journal} {\bibinfo  {journal} {Nature}\ }\textbf {\bibinfo {volume} {551}},\
  \bibinfo {pages} {579} (\bibinfo {year} {2017})}\BibitemShut {NoStop}%
\bibitem [{\citenamefont {Yang}\ \emph {et~al.}(2020)\citenamefont {Yang},
  \citenamefont {Sun}, \citenamefont {Ott}, \citenamefont {Wang}, \citenamefont
  {Zache}, \citenamefont {Halimeh}, \citenamefont {Yuan}, \citenamefont
  {Hauke},\ and\ \citenamefont {Pan}}]{yang2020observation}%
  \BibitemOpen
  \bibfield  {author} {\bibinfo {author} {\bibfnamefont {B.}~\bibnamefont
  {Yang}}, \bibinfo {author} {\bibfnamefont {H.}~\bibnamefont {Sun}}, \bibinfo
  {author} {\bibfnamefont {R.}~\bibnamefont {Ott}}, \bibinfo {author}
  {\bibfnamefont {H.-Y.}\ \bibnamefont {Wang}}, \bibinfo {author}
  {\bibfnamefont {T.~V.}\ \bibnamefont {Zache}}, \bibinfo {author}
  {\bibfnamefont {J.~C.}\ \bibnamefont {Halimeh}}, \bibinfo {author}
  {\bibfnamefont {Z.-S.}\ \bibnamefont {Yuan}}, \bibinfo {author}
  {\bibfnamefont {P.}~\bibnamefont {Hauke}},\ and\ \bibinfo {author}
  {\bibfnamefont {J.-W.}\ \bibnamefont {Pan}},\ }\bibfield  {title} {\bibinfo
  {title} {Observation of gauge invariance in a 71-site {Bose--Hubbard} quantum
  simulator},\ }\href
  {https://doi.org/https://doi.org/10.1038/s41586-020-2910-8} {\bibfield
  {journal} {\bibinfo  {journal} {Nature}\ }\textbf {\bibinfo {volume} {587}},\
  \bibinfo {pages} {392} (\bibinfo {year} {2020})}\BibitemShut {NoStop}%
\bibitem [{\citenamefont {Halimeh}\ \emph {et~al.}(2022)\citenamefont
  {Halimeh}, \citenamefont {McCulloch}, \citenamefont {Yang},\ and\
  \citenamefont {Hauke}}]{Halimeh2022}%
  \BibitemOpen
  \bibfield  {author} {\bibinfo {author} {\bibfnamefont {J.~C.}\ \bibnamefont
  {Halimeh}}, \bibinfo {author} {\bibfnamefont {I.~P.}\ \bibnamefont
  {McCulloch}}, \bibinfo {author} {\bibfnamefont {B.}~\bibnamefont {Yang}},\
  and\ \bibinfo {author} {\bibfnamefont {P.}~\bibnamefont {Hauke}},\ }\bibfield
   {title} {\bibinfo {title} {Tuning the topological $\theta$-angle in
  cold-atom quantum simulators of gauge theories},\ }\href
  {https://doi.org/10.1103/PRXQuantum.3.040316} {\bibfield  {journal} {\bibinfo
   {journal} {PRX Quantum}\ }\textbf {\bibinfo {volume} {3}},\ \bibinfo {pages}
  {040316} (\bibinfo {year} {2022})}\BibitemShut {NoStop}%
\bibitem [{\citenamefont {Zhang}\ \emph
  {et~al.}(2023{\natexlab{b}})\citenamefont {Zhang}, \citenamefont {Liu},
  \citenamefont {Cheng}, \citenamefont {He}, \citenamefont {Wang},
  \citenamefont {Wang}, \citenamefont {Zhu}, \citenamefont {Su}, \citenamefont
  {Zhou}, \citenamefont {Zheng} \emph {et~al.}}]{zhang2023observation}%
  \BibitemOpen
  \bibfield  {author} {\bibinfo {author} {\bibfnamefont {W.-Y.}\ \bibnamefont
  {Zhang}}, \bibinfo {author} {\bibfnamefont {Y.}~\bibnamefont {Liu}}, \bibinfo
  {author} {\bibfnamefont {Y.}~\bibnamefont {Cheng}}, \bibinfo {author}
  {\bibfnamefont {M.-G.}\ \bibnamefont {He}}, \bibinfo {author} {\bibfnamefont
  {H.-Y.}\ \bibnamefont {Wang}}, \bibinfo {author} {\bibfnamefont {T.-Y.}\
  \bibnamefont {Wang}}, \bibinfo {author} {\bibfnamefont {Z.-H.}\ \bibnamefont
  {Zhu}}, \bibinfo {author} {\bibfnamefont {G.-X.}\ \bibnamefont {Su}},
  \bibinfo {author} {\bibfnamefont {Z.-Y.}\ \bibnamefont {Zhou}}, \bibinfo
  {author} {\bibfnamefont {Y.-G.}\ \bibnamefont {Zheng}}, \emph {et~al.},\
  }\bibfield  {title} {\bibinfo {title} {Observation of microscopic confinement
  dynamics by a tunable topological $\theta$-angle},\ }\href
  {https://doi.org/10.48550/arXiv.2306.11794} {\bibfield  {journal} {\bibinfo
  {journal} {arXiv preprint arXiv:2306.11794}\ } (\bibinfo {year}
  {2023}{\natexlab{b}})}\BibitemShut {NoStop}%
\bibitem [{\citenamefont {Surace}\ \emph {et~al.}(2026)\citenamefont {Surace},
  \citenamefont {Lerose}, \citenamefont {Katz}, \citenamefont {Bennewitz},
  \citenamefont {Schuckert}, \citenamefont {Luo}, \citenamefont {De},
  \citenamefont {Ware}, \citenamefont {Morong}, \citenamefont {Collins},
  \citenamefont {Monroe}, \citenamefont {Davoudi},\ and\ \citenamefont
  {Gorshkov}}]{data}%
  \BibitemOpen
  \bibfield  {author} {\bibinfo {author} {\bibfnamefont {F.~M.}\ \bibnamefont
  {Surace}}, \bibinfo {author} {\bibfnamefont {A.}~\bibnamefont {Lerose}},
  \bibinfo {author} {\bibfnamefont {O.}~\bibnamefont {Katz}}, \bibinfo {author}
  {\bibfnamefont {E.}~\bibnamefont {Bennewitz}}, \bibinfo {author}
  {\bibfnamefont {A.}~\bibnamefont {Schuckert}}, \bibinfo {author}
  {\bibfnamefont {D.}~\bibnamefont {Luo}}, \bibinfo {author} {\bibfnamefont
  {A.}~\bibnamefont {De}}, \bibinfo {author} {\bibfnamefont {B.}~\bibnamefont
  {Ware}}, \bibinfo {author} {\bibfnamefont {W.}~\bibnamefont {Morong}},
  \bibinfo {author} {\bibfnamefont {K.}~\bibnamefont {Collins}}, \bibinfo
  {author} {\bibfnamefont {C.}~\bibnamefont {Monroe}}, \bibinfo {author}
  {\bibfnamefont {Z.}~\bibnamefont {Davoudi}},\ and\ \bibinfo {author}
  {\bibfnamefont {A.}~\bibnamefont {Gorshkov}},\ }\bibfield  {title} {\bibinfo
  {title} {String-breaking dynamics in quantum adiabatic and diabatic
  processes},\ }\href {https://doi.org/10.5281/zenodo.19165051}
  {10.5281/zenodo.19165051} (\bibinfo {year} {2026})\BibitemShut {NoStop}%
\bibitem [{\citenamefont {Lieb}\ and\ \citenamefont
  {Robinson}(1972)}]{Lieb1972}%
  \BibitemOpen
  \bibfield  {author} {\bibinfo {author} {\bibfnamefont {E.~H.}\ \bibnamefont
  {Lieb}}\ and\ \bibinfo {author} {\bibfnamefont {D.~W.}\ \bibnamefont
  {Robinson}},\ }\bibfield  {title} {\bibinfo {title} {The finite group
  velocity of quantum spin systems},\ }\href
  {https://doi.org/10.1007/BF01645779} {\bibfield  {journal} {\bibinfo
  {journal} {Commun. Math. Phys.}\ }\textbf {\bibinfo {volume} {28}},\ \bibinfo
  {pages} {251} (\bibinfo {year} {1972})}\BibitemShut {NoStop}%
\bibitem [{\citenamefont {{Dyson}}(1969)}]{dyson1969existence}%
  \BibitemOpen
  \bibfield  {author} {\bibinfo {author} {\bibfnamefont {F.~J.}\ \bibnamefont
  {{Dyson}}},\ }\bibfield  {title} {\bibinfo {title} {{Existence of a
  phase-transition in a one-dimensional Ising ferromagnet}},\ }\href
  {https://doi.org/10.1007/BF01645907} {\bibfield  {journal} {\bibinfo
  {journal} {Comm. Math. Phys.}\ }\textbf {\bibinfo {volume} {12}},\ \bibinfo
  {pages} {91} (\bibinfo {year} {1969})}\BibitemShut {NoStop}%
\end{thebibliography}%

%%%%%%%%%%%%%%%%%%%%%%%%%%%%%%%%%%%%%%%%%%%%%%%%%%%%%%%%%%%%%%%%%%%%%%%%%%%%%%%%%%%%%%%%%%%%%%%%%%%%%%%%%%%%%%%%%%%%%%%%%%%%%%%%%%%%%%%%%%%%%%%%%%%%%%%%%%%%%%%%%%%%%%%%%%%%%%%%%%%%%%%%%%%%%%%%%%%%%%%%%%%%%%%%

\appendix

%%%%%%%%%%%%%%%%%%%%%%%%%%%%%%%%%%%%%%%%%%%%%%%%%%%%%%%%%%%%%%%%%%%%%%%%%%%%%%%%%%%%%%%%%%%%%%%%%%%%%%%%%%%%%%%%%%%%%%%%%%%%%%%%%%%%%%%%%%%%%%%%%%%%%%%%%%%%%%%%%%%%%%%%%%%%%%%%%%%%%%%%%%%%%%%%%%%%%%%%%%%%%%%%
\section{External spins}
\label{app:ext}
This Appendix offers a comparison of spin-chain evolution with static and dynamic external spins. The results of this Appendix 
justify the 
 implementation of static external spins introduced in Sec.~\ref{sec:sb}.

Consider a string of length $\ell=5$ evolving under the Hamiltonian in Eq.~\eqref{eq:H} with exponentially decaying interactions ($\xi = 1$) and with time-dependent $h(t)=t/\tau$.
For the case of dynamical external spins, the chain consists of 15 sites ($j=1,\dots 15$) with two domain walls pinned at a separation of  $\ell+2$: the spins at sites $j=4$ and $j=12$ are polarized along $+\hat z$, while the ones at $j=5$ and $j=11$ are polarized along $-\hat z$. These four spins are static and do not evolve in time, while both the external spins $j=1,2,3,13,14,15$ and the internal spins $j=6,\dots 10$ are dynamical. Additionally, we consider an effective longitudinal field generated by the other external sites $j=-\infty, \dots ,0$ and $j=16, \dots, \infty$, polarized along $+\hat z$. For the scenario involving  static external spins, spins $j=-\infty, \dots , 4$ and $j=12, \dots, \infty$ are static and polarized along $\hat{z}$, spins $j=5$ and $j=11$ are static and polarized along $-\hat{z}$, and spins $j=6,\cdots,10$ are dynamical.

\begin{figure}[h]
    \centering
    \includegraphics[width=\linewidth]{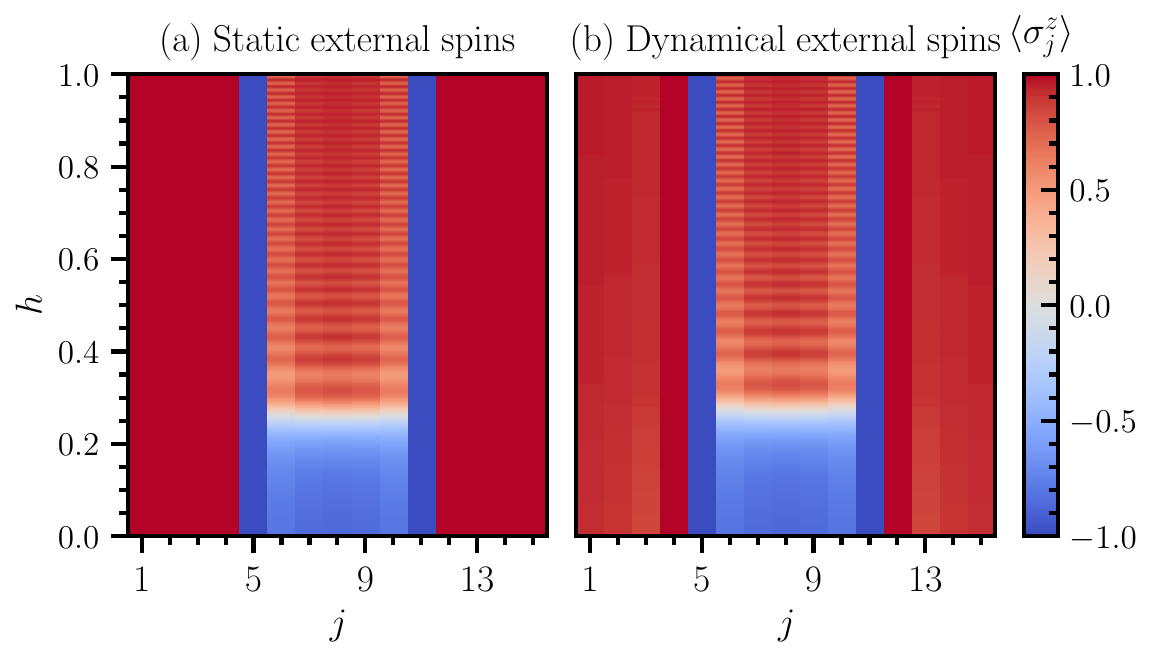}
    \caption{Evolution of a string of length $\ell=5$ in the presence of (a) static external spins and (b) dynamical external spins. The parameters of the evolution are as in Fig.~\ref{fig:spatial} ($\xi=1$, $\tau=100$, and $g=1.2$). }
    \label{fig:ext}
\end{figure}

\begin{figure*}[t]
    \centering
    \includegraphics[width=\linewidth]{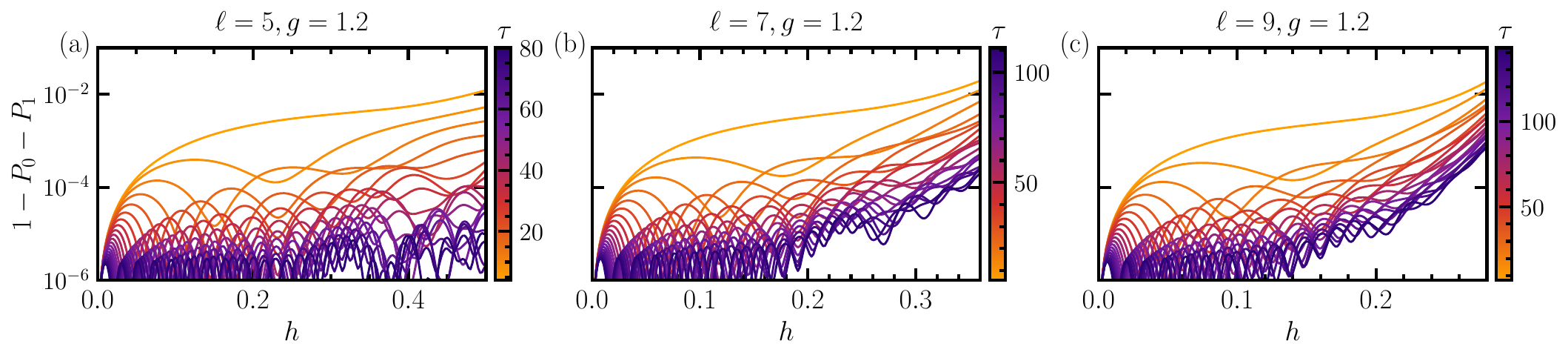}
    \caption{Population of higher levels (beyond the ground state and the first excited state) as a function of $h(t)=t/\tau$ for different values of $\tau$ and different system sizes, (a) $\ell=5$, (b) $\ell=7$, and 
 (c) $\ell=9$.}
    \label{fig:p_other}
\end{figure*}

Figure~\ref{fig:ext} displays the spatiotemporal profile of average magnetization along $\hat{z}$ for both scenarios. No significant difference between the two cases is observed. This can be understood at a ``mean-field" level: the external dynamical spins are surrounded by spins with positive magnetization 
and the effective field they experience polarizes them along $+\hat z$. Adding the additional $h$ field stabilizes this positive magnetization and further enhances this effect. Therefore, the external spins are approximately static throughout the evolution. The internal spins, on the other hand, especially the ones close to the static charges, are subject to the competing effect  of positively and negatively polarized spins and of the $h$ field that makes the string unstable. These are the spins that exhibit the nontrivial time evolution~\cite{exp1}.

For power-law decaying interactions, we expect that the external dynamical spins might be more strongly affected by the internal spins, but this effect is still rather suppressed for the value of $\alpha$ considered in Sec.~\ref{sec:longrange}.

%%%%%%%%%%%%%%%%%%%%%%%%%%%%%%%%%%%%%%%%%%%%%%%%%%%%%%%%%%%%%%%%%%%%%%%%%%%%%%%%%%%%%%%%%%%%%%%%%%%%%%%%%%%%%%%%%%%%%%%%%%%%%%%%%%%%%%%%%%%%%%%%%%%%%%%%%%%%%%%%%%%%%%%%%%%%%%%%%%%%%%%%%%%%%%%%%%%%%%%%%%%%%%%%
\section{Two-level approximation}
\label{app:data}

This Appendix provides evidence supporting the two-level approximation in the real-time dynamics described in Sec.~\ref{subsec:LZ}. The Landau-Zener approach is justified as long as the $h$-field ramp is sufficiently slow such that the dynamics are restricted to the two lowest levels in the spectrum, shown in Fig.~\ref{fig:spectrum_sc}(a,b,c) for different system sizes.

In Fig.~\ref{fig:p_other}, we plot the population of the levels beyond the lowest two levels as $h$ grows in time with the linear ramp $h(t)=t/\tau$, for different values of $\tau$.
For the smallest values of $\tau$ considered here ($\tau=4.0$ for $\ell=5$, $\tau=5.6$ for $\ell=7$, and $\tau=7.1$ for $\ell=9$), the combined population of these levels remains smaller than $\sim 10^{-2}$ for the entire evolution, and it rapidly decreases by many orders of magnitudes at larger values of $\tau$. This confirms the validity of the two-level approximation in the ranges of parameters considered in Secs.~\ref{subsec:obsramp} and \ref{subsec:LZ}.

%%%%%%%%%%%%%%%%%%%%%%%%%%%%%%%%%%%%%%%%%%%%%%%%%%%%%%%%%%%%%%%%%%%%%%%%%%%%%%%%%%%%%%%%%%%%%%%%%%%%%%%%%%%%%%%%%%%%%%%%%%%%%%%%%%%%%%%%%%%%%%%%%%%%%%%%%%%%%%%%%%%%%%%%%%%%%%%%%%%%%%%%%%%%%%%%%%%%%%%%%%%%%%%%
\section{Lowest-energy configurations for $g=0$}
\label{app:bubbles}
This Appendix expands the discussion of static properties of the Ising spin chain with exponentially decaying interactions in the limit of $g=0$ beyond what was presented in Sec.~\ref{subsec:h0}. We are, in particular, interested in characterizing the energy of different bitstring configurations, beyond the string and broken string states considered in the main text. We will also show that the lowest energy always corresponds to either the string state or the broken string state.

It is convenient to label the bitstring configurations (in the $z$ basis) by the positions $\{i_1, i_2,\dots i_n\}$ of the spins polarized along $-\hat z$, with $i_1<i_2<\dots i_n$. We use the convention that $i_1=0$ and $i_n=\ell+1$ are the static external spins located next to the edges, and we refer to the other spins as the dynamical spins. The energy corresponding to one such configuration can be written as
\begin{multline}
\label{eq:Econfig}
    E(\{i_1,\dots, i_n\})= 2n\left(\frac{1}{e^{1/\xi}-1}+h\right)\\
    -2\sum_{j=1}^{n-1}\sum_{k=j+1}^n e^{-(i_k-i_j)/\xi},
\end{multline}
where the energy is computed with respect to a reference zero-energy state having all the spins in the $+\hat z$ direction. For a fixed number $n$ of $\downarrow$ spins, the lowest-energy configuration is obtained when the number of domain walls is minimal, i.e., when all the $\downarrow$ dynamical spins are adjacent to one of the two edges: these are the configurations $\{0,1,\dots n-2, \ell+1\}$ and $\{0, \ell-n+3, \dots, \ell+1\}$. This statement can be proved by contradiction, assuming that the lowest-energy configuration contains a domain of $\downarrow$ spins that are not adjacent to one of the edges. One can split the interaction energy of such a configuration [i.e., the double sum in Eq.~(\ref{eq:Econfig})] into four parts: (i) the interaction energy $E_d$ obtained for $i_j$ and $ i_k$ contained in the domain; (ii) the interaction energy $E_l$ obtained for $i_j$ located to the left of the domain and $i_k$ in the domain; (iii) the interaction energy $E_r$ obtained for $i_j$ in the domain and $i_k$ located to the right of the domain; and (iv) the interaction energy $E_o$ for $i_j$ and $i_k$ both outside of the domain. One can now imagine moving the entire domain one site to the left or right: the new interaction energies are $E_d+ e^{-1/\xi}E_l+e^{1/\xi} E_r +E_o$ or $E_d+ e^{1/\xi}E_l+e^{-1/\xi} E_r +E_o$, respectively. Stating that the original configuration was (one of) the lowest-energy state(s) implies that $E_l+E_r \le e^{-1/\xi}E_l+e^{1/\xi} E_r$, which gives $ |E_l|>e^{1/\xi}|E_r|$, and $E_l+E_r \le e^{1/\xi}E_l+e^{-1/\xi} E_r$, which gives $ |E_r|>e^{1/\xi}|E_l|$, leading to a contradiction.

\begin{figure}[t]
    \centering
    \includegraphics[width=0.65\linewidth]{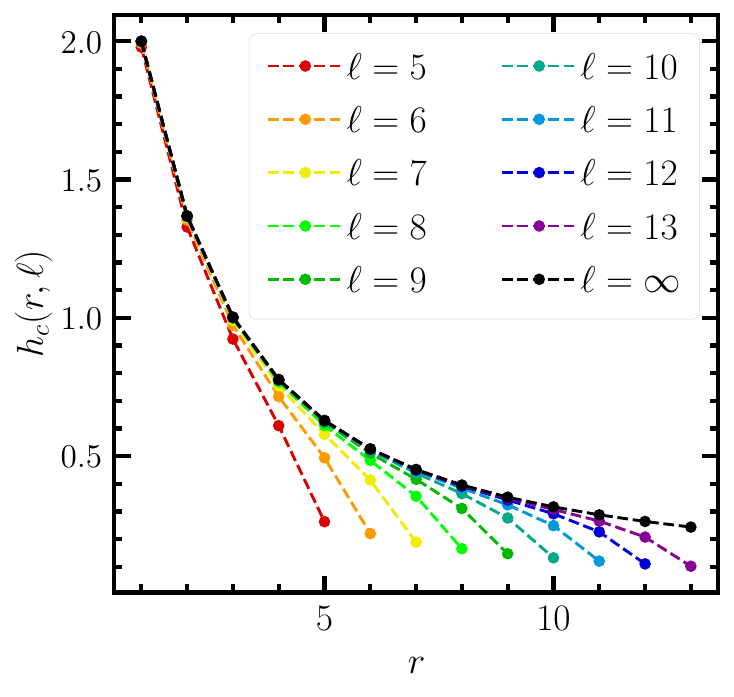}
    \caption{Value $h_c(r,\ell)$ of the longitudinal field for which $E_*(\ell-r,\ell)=E_*(\ell,\ell)$, as a function of $r$ for different lengths $\ell$ of the string and with $g=0$. For any $\ell$, the smallest $h_c(r,\ell)$ value occurs at $r=\ell$.}
    \label{fig:bubbles}
\end{figure}

Having ruled out the possibility of other lowest-energy configurations, we can focus on the state with $n-2$ dynamical spins in the $\downarrow$ state all adjacent to one edge, with an energy denoted by $E_*(n-2,\ell)$. With this notation, the energy of the string state is $E_*(\ell,\ell)$, while the energy of the broken string state is $E_*(0,\ell)$. For each $r$ and $\ell$, there is a critical longitudinal field $h_c(r,\ell)$ at which $E_*(\ell-r,\ell)=E_*(\ell,\ell)$. The critical field $h_c(r,\ell)$, hence, represents the value of $h$ of the first level crossing between the string state and a bubble state of size $r$.  The values of $h_c(r,\ell)$ as a function of $r$ for different values of $\ell$ are shown in Fig.~\ref{fig:bubbles}. 
The value of $h_c$ is smallest for $r=\ell$, confirming that the lowest energy state is the string for $h<h_c(\ell, \ell)$ and the broken string for $h>h_c(\ell, \ell)$. Nevertheless, the other crossings with smaller bubbles play an important role in the dynamical observation of string breaking through a non-adiabatic protocol, as shown in Sec.~\ref{subsec:bubbles}.

\section{Analysis of bubble-nucleation dynamics}
\label{app:nucleation}
In this appendix, we provide additional information about the density of bubbles in the complex process of bubble nucleation and the numerically extracted scaling laws of Sec.~\ref{sec:dyn}B.4 and \ref{sec:dyn}B.5.

\subsection{Bubble densities} During the dynamics induced by the linear ramp, bubbles of various sizes are nucleated. We characterize this process by introducing the average density $\lambda_r$ of bubbles of size $r$, defined as
\begin{equation}
\lambda_r = \frac{1}{\ell - r - 1}\sum_{j=1}^{\ell - r - 1}
P^\downarrow_j\, P^\uparrow_{j+1} P^\uparrow_{j+2} \dots P^\uparrow_{j+r}\, P^\downarrow_{j+r+1},
\end{equation}
where $P_j^{\uparrow,\downarrow}$ denote the projectors onto the two spin states in the $\hat z$ basis at site $j$.
We compute the expectation value of $\lambda_r$ at the end of the linear ramp from $h=0$ to $h=1$. Since these expectation values display residual temporal oscillations, we further average them over the interval $h\in[0.8,1]$ along the ramp.

\begin{figure}[h!]
    \centering
    \includegraphics[width=\linewidth]{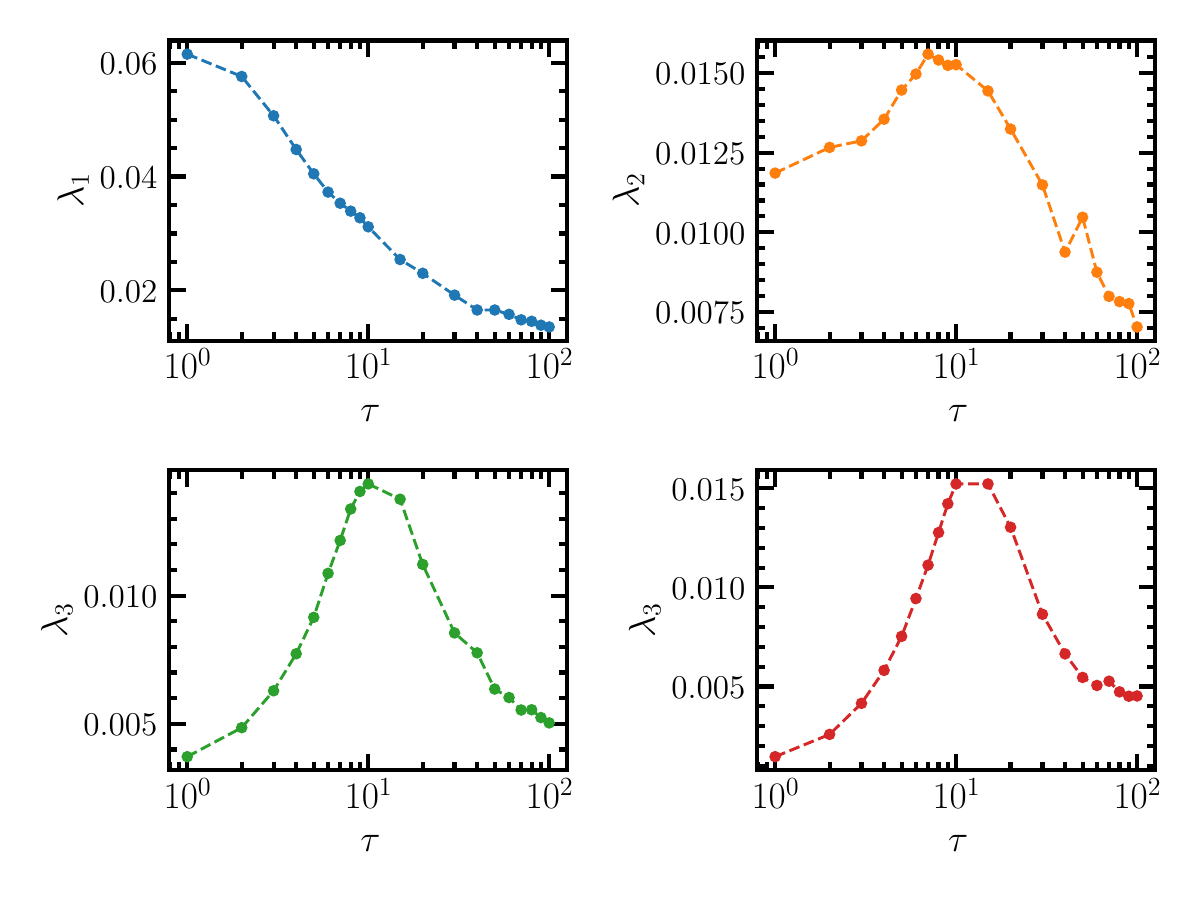}
    \caption{Expectation value of the average bubble density $\lambda_r$ for bubbles of sizes $r=1,2,3,4$. The data are obtained by time-averaging the expectation values over the interval $h\in[0.8,1]$ along the ramp. The transverse field is set to $g=1.2$, and the total string length is $\ell=13$.}
    \label{fig:bubble_densities}
\end{figure}

As shown in Fig.~\ref{fig:bubble_densities}, the density of bubbles of a given size $r$ exhibits a nontrivial dependence on the ramp time $\tau$, with a maximum at a characteristic value of $\tau$ that systematically increases with $r$. This is the result of two competing effects. On the one hand, increasing $\tau$ enhances the probability of bubble nucleation, as expected from Landau–Zener theory. On the other hand, for sufficiently large $\tau$, the nucleation of small bubbles is suppressed, since the formation of larger bubbles becomes increasingly favorable.

\subsection{Fit of scaling exponent} 
\begin{figure}[h]
    \centering
    \includegraphics[width=\linewidth]{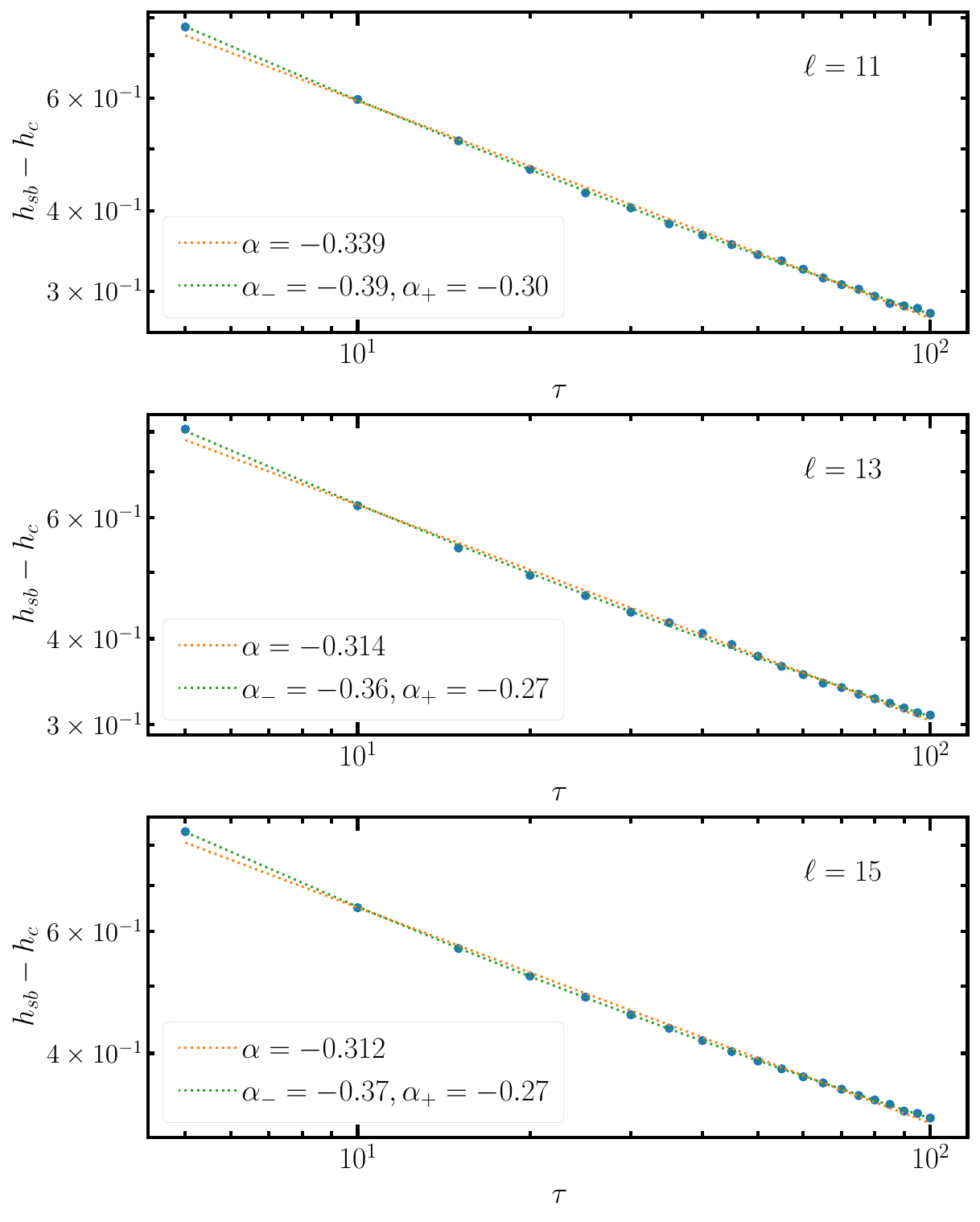}
    \caption{Same data as in Fig.~\ref{fig:collapse}(b), plotted in log-log scale, for $\ell=11,13,15$. The dotted orange and green lines represent a linear and a quadratic fit (in log-log scale), respectively. The parameter $\alpha$ is the slope extracted from the linear fit, while $\alpha_\pm$ are the slopes at the two values $\tau_-=5$, $\tau_+=100$, extracted from the quadratic fit.  }
    \label{fig:fit}
\end{figure}

The data from Fig.~\ref{fig:collapse}(b) are replotted in Fig.~\ref{fig:fit} as a log–log plot. The exponent $\alpha$ is initially extracted from a linear fit in log-log scale (dotted orange line). As shown, the linear fit exhibits a slight systematic curvature, suggesting that a quadratic fit (green dotted line) may be more appropriate. From the quadratic fit, we extract the slopes $\alpha_{\pm}$ at the two extreme values of $\tau$ considered, $\tau_- = 5$ and $\tau_+ = 100$. The values of $\alpha$ and $\alpha_\pm$ are reported in the legends. The slopes $\alpha_{\pm}$ provide an estimate of the variation of $\alpha$ over the interval considered, which is roughly $30\%$. In contrast, the dependence of the exponent on $\ell$ is much weaker, with a variation of less than $1\%$ between $\ell = 13$ and $\ell = 15$.

\end{document}